\documentclass[A4,english, APS,  PRD, nofootinbib,superscriptaddress]{revtex4}
\usepackage[latin9]{inputenc}
\usepackage{color}
\usepackage{bm}
\usepackage{amsmath}
\usepackage{graphicx}
\usepackage[unicode=true,
 bookmarks=true,bookmarksnumbered=false,bookmarksopen=false,
 breaklinks=false,pdfborder={0 0 0},pdfborderstyle={},backref=false,colorlinks=true]
 {hyperref}
\hypersetup{
 linkcolor=blue, citecolor=cyan}

\makeatletter
\@ifundefined{textcolor}{}
{%
 \definecolor{BLACK}{gray}{0}
 \definecolor{WHITE}{gray}{1}
 \definecolor{RED}{rgb}{1,0,0}
 \definecolor{GREEN}{rgb}{0,1,0}
 \definecolor{BLUE}{rgb}{0,0,1}
 \definecolor{CYAN}{cmyk}{1,0,0,0}
 \definecolor{MAGENTA}{cmyk}{0,1,0,0}
 \definecolor{YELLOW}{cmyk}{0,0,1,0}
}

\usepackage{subfigure}
\usepackage{slashed}
\usepackage{bm}
\usepackage{xspace}

\newcommand{\lam}{\mbox{$\Lambda$}\xspace}

\def\wrho{{\widehat{\rho}}}

\def\wrho{{\widehat{\rho}}}

\usepackage{amssymb}

\makeatother

\begin{document}
\title{Spin polarization in relativistic heavy-ion collisions}
\author{Francesco Becattini}
\affiliation{University of Florence and INFN, Via G. Sansone 1, I-50019, Sesto
Fiorentino (Firenze), Italy}
\author{Matteo Buzzegoli}
\affiliation{Department of Physics and Astronomy, Iowa State University, 2323 Osborn
Drive, Ames, Iowa 50011, USA}
\author{Takafumi Niida}
\affiliation{Department of Physics, Institute of Pure and Applied Sciences, University
of Tsukuba, 1-1-1 Tennoudai, Tsukuba, Ibaraki 305-8571, Japan}
\author{Shi Pu}
\affiliation{Department of Modern Physics, University of Science and Technology
of China, Hefei, Anhui 230026, China}
\author{Ai-Hong Tang}
\affiliation{Brookhaven National Laboratory, Upton, New York 11973, USA}
\author{Qun Wang}
\affiliation{Department of Modern Physics, University of Science and Technology
of China, Hefei, Anhui 230026, China}
\begin{abstract}
Polarization has opened a new physics chapter in relativistic heavy
ion collisions. Since the first prediction and experimental observation
of global spin polarization, a lot of progress has been made in understanding
its features, both at experimental and theoretical level. In this
paper, we give an overview on the recent advances in this field. The
covered topics include a review of measurements of global and local
spin polarization of hyperons and the global spin alignment of vector
mesons. We account for the basic theoretical framework to describe
spin polarization in a relativistic fluid such as the Quark Gluon
Plasma, including statistical quantum field theory and local thermodynamic
equilibrium, spin hydrodynamics, relativistic kinetic theory with
spin and coalescence models.
\end{abstract}
\maketitle
\tableofcontents{}

\section{Introduction}

\label{sec:general_spin}For more than thirty years, all theoretical
and experimental investigations in relativistic heavy ion physics
have been based on the measurement of particles momenta. The main
goal of theoretical models was to predict momentum distributions and
correlations, while experimental works aimed at measuring the momentum
spectra of identified particles to test these models. In 2017, the
observation of spin polarization of $\Lambda$ hyperons demonstrated
the possibility to study the physics of the QCD matter formed in these
collisions by using a completely new tool. The spin physics in relativistic
heavy-ion collisions has since quickly developed and becomed one of
the most promising lines of research in this field, whose potential
is yet to be fully explored.


Around 2004 it was proposed \citep{Liang:2004ph,Voloshin:2004ha}
that particles produced in heavy ion collisions at finite impact parameter
could be globally polarized along the direction of the total orbital
angular momentum. Quantitative theoretical predictions \citep{Liang:2004ph}
were based upon the spin-orbit coupling in a perturbative QCD-inspired
model, leading to large values of polarization (around 30\%, which
were corrected thereafter to be less than 4\% \citep{Gao:2007bc}).
Besides the apparent large uncertainty, one of the key problems in
this original approach is the difficulty of reconciling a perturbative-collisional
approach with the consolidated evidence that the Quark Gluon Plasma
(QGP) is a very strongly interacting system. About the time when the
first measurement of global $\Lambda$ polarization at RHIC was released
\citep{STAR:2007ccu} setting an upper limit of few percent, the idea
of a polarization related to hydrodynamic motion, and particularly
vorticity, was put forward \citep{Becattini:2007sr}. The idea is
as follows: if the QGP achieves and is able to maintain local thermodynamic
equilibrium until it decouples into freely streaming non-interacting
hadrons, as it was established by the study of particle spectra, spin
degrees of freedom should likewise be at, or near, local thermodynamic
equilibrium. The consequence is that a mean spin polarization of produced
hadrons can be calculated at the freeze-out just like their momentum
spectrum. Specifically, this model implies that spin polarization
is driven by hydrodynamic vorticity \citep{Becattini:2007nd,Becattini:2013fla}
(more precisely the thermal vorticity, that is the anti-symmetric
gradient of four-temperature, as it will become clear later) and makes
it possible to obtain quantitative predictions in relativistic heavy
ion collisions as hydrodynamic vorticity can be calculated with the
hydrodynamic model. Such predictions were released circa 2015 \citep{Becattini:2013vja,Becattini:2015ska,Karpenko:2016jyx}
and provided a global polarization of $\Lambda$ hyperons of about
1\% at $\sqrt{s_{{\rm NN}}}=200$ GeV, i.e., less than the lower bound
set by the first measurement of STAR experiment.

Spurred by these new predictions, the experiment STAR carried out
an improved measurement with larger statistics at $\sqrt{s_{{\rm NN}}}=200$
GeV and new measurements at lower energies, reporting a positive evidence
of global spin polarization of $\Lambda$ hyperons in Au+Au collisions
\citep{STAR:2017ckg}. The data turned out to be in very good agreement
with the predictions based on hydrodynamics and local equilibrium
and the result was interpreted as a confirmation, at the relativistic
and subatomic level, of the link between spin and rotation which was
predicted more than a century ago and experimentally observed in the
Barnett and Einstein-De Haas effects \citep{Barnett:1915uqc,Einstein1915}.
This finding triggered a lot of enthusiasm and many new developments
both at experimental and theoretical levels.

Indeed, there has been a considerable progress since. Over the past
few years, the experiments confirmed the first observations and demonstrated
the capability of measuring spin polarization as a function of momentum
(so-called local spin polarization) \citep{STAR:2019erd}. Besides,
the measurement of the global polarization of $\Xi$ hyperons \citep{STAR:2020xbm},
in good agreement with hydrodynamic predictions, confirmed that this
phenomenon is not driven by specific hadron-dependent couplings or
properties, like in $pp$ collisions, but by collective properties
of the system. Spin polarization has been observed at very low energy
\citep{STAR:2021beb,Kornas:2020qzi} and at the highest energy of
the LHC \citep{ALICE:2019onw,ALICE:2021pzu}.

Unlike for global polarization, local spin polarization as a function
of the azimuthal angle of emission in the transverse plane turned
out to be starkly different from the prediction of the combined hydrodynamic
model and local equilibrium assumption \citep{Becattini:2020ngo}.
This discrepancy has provided a strong motivation for theoretical
studies in different directions which has led to a remarkable advance
in the understanding of spin thermodynamics and kinetics in a relativistic
fluid. Much work has been devoted to the devolepment of the kinetic
theory with spin \citep{Gao:2012ix,Chen:2012ca,Hidaka:2016yjf,Hidaka:2017auj,Gao:2019znl,Weickgenannt:2019dks,Hattori:2019ahi,Wang:2019moi,Gao:2020pfu,Weickgenannt:2020aaf,Yang:2020hri,Liu:2020flb,Weickgenannt:2021cuo,Sheng:2021kfc,Fang:2022ttm,Hidaka:2022dmn,Fang:2023bbw}
and relativistic hydrodynamics with a quantum spin tensor \citep{Florkowski:2017ruc,Florkowski:2017dyn,Montenegro:2017rbu,Montenegro:2017lvf,Wang:2017jpl,Florkowski:2018myy,Hattori:2019lfp,Weickgenannt:2019dks,Fukushima:2020ucl,Li:2020eon,Bhadury:2020puc,Weickgenannt:2020aaf,Shi:2020htn,Speranza:2020ilk,Bhadury:2020cop,Singh:2020rht,Gallegos:2020otk,Garbiso:2020puw,Becattini:2020ngo,Becattini:2020sww,Gao:2020vbh,Liu:2020ymh,Gallegos:2021bzp,She:2021lhe,Hongo:2021ona,Peng:2021ago,Sheng:2021kfc,Weickgenannt:2022jes,Weickgenannt:2022qvh,Weickgenannt:2022zxs,Cao:2022aku,Biswas:2023qsw}.
At the same time, new calculations within the local equilibrium quantum-statistical
framework revealed the existence of an unexpected contribution from
the symmetric part of the gradient of the four-temperature, the so-called
thermal shear tensor \citep{Becattini:2021suc,Liu:2021uhn}, which
turned out to be as large as the original term from thermal vorticity.
The combination of these two terms seems to remove the discrepancy
between data and hydrodynamic model \citep{Becattini:2021iol,STAR:2023eck},
although recent analyses, with 3+1D hydrodynamic simulations, showed
that the magnitude of the effect depends on the initial conditions
\citep{Alzhrani:2022dpi,Wu:2022mkr}.


The mean spin polarization vector is sufficient to completely describe
the polarization state of a spin $1/2$ fermion, but it is not for
higher spin particle that requires more quantities. A vector meson,
with spin $1$, requires one more quantity besides the mean spin vector,
which is an Euclidean symmetric and traceless rank 2 tensor; therefore,
a vector meson can be polarized even though its mean spin vector vanishes.
Among the five independent components of this Euclidean tensor, an
easily accessible one in experiments is the so-called spin alignment,
which can be defined as the difference between the spin density matrix
element $\rho_{00}$ and its value in case of vanishing polarization,
that is $1/3$. Such quantity can be measured through the angular
distribution of the vector meson's decay product in a two-body decay
even if the decay is parity-conserving. It was indeed proposed in
2004 that a non-vanishin spin alignment could occur in peripheral
relativistic nuclear collisions \citep{Liang:2004xn}. The prediction
of a finite spin alignment was inspired by a naive quark combination
model: $\rho_{00}-1/3\sim P_{q}P_{\bar{q}}$, where $P_{q}$ and $P_{\bar{q}}$
are the polarizations of the quark and antiquark in the vector meson
respectively.

The first measurement of $\rho_{00}$ for $\phi$ and $K^{*0}$ vector
mesons in Au+Au collisions at 200 GeV was carried out in 2007-2008
by the STAR collaboration with results consistent with 1/3 within
errors due to limited statistics \citep{STAR:2008lcm}. Almost ten
years later, the STAR collaboration reported their preliminary data
for the $\phi$ meson's $\rho_{00}$ in Au+Au collisions at 11.5,
19.6, 27, 39 and 200 GeV \citep{Zhou:2017nwi} and a final measurement
in Ref.~\citep{STAR:2022fan}. The data shows that the alignment
$\rho_{00}-1/3\approx{\cal O}(10^{-2})$ at $\sqrt{s_{{\rm NN}}}=200$
GeV and growing at lower energy. The hydrodynamic-local equilibrium
model maintains that the spin alignment of vector mesons is quadratic
in thermal vorticity and the measured global polarization implies
a value which is roughly of the order of $10^{-4}$, which is consistently
smaller than the measured value. It was then realized that also thermal
shear can contribute to spin alignment of vector mesons at local equilibrium
but, unlike for spin $1/2$ particle mean spin vector, not at the
linear order. Indeed, the leading term for spin alignment can only
be quadratic in thermal shear or in second order derivatives, which,
most likely, make theoretical predictions of local equilibrium still
much lower than the measured value.

The measured values seemed to be incompatible with a quark combination
model as well. Indeed, from the observed global polarization of $\Lambda$
hyperons, one can estimate $\rho_{00}^{\phi}-1/3\sim P_{s}P_{\bar{s}}\sim\mathcal{O}(10^{-4})$.
However, it was later realized, after a thorough analysis of the quark
recombination model's arguments, that $\rho_{00}$ gives information
on $\langle P_{q}P_{\bar{q}}\rangle$, the correlation of $P_{q}$
and $P_{\bar{q}}$ inside the vector meson \citep{Sheng:2019kmk},
whereas $P_{\Lambda}$ and $P_{\bar{\Lambda}}$ can only give information
on the mean values $\langle P_{s}\rangle$ and $\langle P_{\bar{s}}\rangle$,
which cannot constrain $\rho_{00}$ for the $\phi$ meson. With this
idea, the non-relativistic quark coalescence model was upgraded. The
analysis shows that $\rho_{00}$ is determined by the local correlation
of quark's and antiquark's polarization functions $P_{q}(\mathbf{x}_{1},\mathbf{p}_{1})$
and $P_{\bar{q}}(\mathbf{x}_{2},\mathbf{p}_{2})$ inside the phase
space limited by the meson's wave function. There are many potential
sources for the polarization of the quarks and yet local equilibrium
at the quark level, even if including the electromagnetic field is
not sufficient to provide the observed large deviation of $\rho_{00}$
from 1/3 for the $\phi$ meson. It hase been then proposed that a
kind of vector field in strong interaction (called the $\phi$ field)
coupled to the strange and antistrange quark plays an important role
in the $\phi$ meson's $\rho_{00}$: the local correlation or fluctuation
of the $\phi$ field can give rise to the a large deviation of $\rho_{00}$
from 1/3. The non-relativistic quark coalescence model \citep{Sheng:2019kmk}
has been later extended to the relativistic one \citep{Sheng:2022wsy,Sheng:2022ffb,Sheng:2023urn}
in the framework of quantum transport theory. The prediction made
by the relativistic model provides a good description of STAR's data
on $\rho_{00}$ for the $\phi$ meson in many respects.

As it is apparent from the above account, a full understanding of
spin physics in relativistic nuclear collisions is far from being
achieved. Nevertheless, this endeavour is a highly rewarding one as
spin, being sensitive to gradients of the hydrodynamic-thermodynamic
fields - probes the hydrodynamic picture of the QGP to a much deeper
accuracy than traditional correlations in momentum space. Just to
mention some possible fruitful applications, polarization has been
proposed as an instrument to probe local parity violation \citep{Du:2008zzb,Becattini:2020xbh}
complementary to the Chiral Magnetic Effect; to investigate the energy
loss of highly energetic partons in the QGP \citep{Serenone:2021zef,Ribeiro:2023waz};
and even as a signature of the critical point \citep{Singh:2021yba}.

In this work, we are going to review the status of the field up to
2023 both from a theoretical and experimental standpoint. It should
be emphasized that this field is under a quick development, as has
been mentioned, and our experience taught us that some of the arguments
or even conclusions that we hereby present may become obsolete in
a couple of years.

\section{Theoretical models for global and local polarization in equilibrium}

\label{sec:polarization_theoretical_model}In this section, we are
going to summarize the main theoretical tools to calculate spin polarization
in a relativistic fluid and especially QGP. We will focus on spin
polarization of fermions, leaving vector mesons and spin alignment
to a dedicated section, Section~\ref{sec:spin-alignment} \footnote{ In this section natural units with $\hbar=c=K=1$ are used. Repeated
indices are assumed to be summed over, and sometimes contractions
of indices will be denoted with, e.g. $\beta_{\mu}p^{\mu}=\beta\cdot p$.
The Minkowskian metric tensor is ${\rm diag}(1,-1,-1,-1)$, and the
Levi-Civita symbol is chosen such that $\epsilon^{0123}=1$. Operators
in Hilbert space are denoted by a large upper hat ($\widehat{T}$)
while unit vectors with a small upper hat ($\hat{v}$); only the Dirac
field is expressed by $\Psi$ without an upper hat. The symbol ${\rm Tr}$
denotes the trace over all states in the Hilbert space, whereas the
symbol ${\rm tr}$ denotes the trace over polarization states or traces
of finite dimensional matrices.}.

\subsection{Quantum statistical field theory}

\label{subsec:polarization_qstat}

The quantum statistical field theory based on quantum field operators
and quantum density operators is the most fundamental approach to
deal with spin in relativistic fluids; we refer to Ref.~\citep{Becattini:2020sww}
for an extensive discussion. The essential and fundamental ingredient
to connect theory and measurements is the formula relating the mean
spin vector $S^{\mu}(p)$ with the covariant Wigner function of particle
$W_{+}(x,p)$: 
\begin{equation}
S^{\mu}(p)=\frac{1}{2}\frac{\int{\rm d}\Sigma\cdot p\;{\rm tr}_{4}\left[\gamma^{\mu}\gamma^{5}W_{+}(x,p)\right]}{\int{\rm d}\Sigma\cdot p\;{\rm tr}_{4}\left[W_{+}(x,p)\right]}.\label{meanspf0}
\end{equation}
The mean spin vector is directly what can be measured and the Wigner
function can be obtained with different methods as described below.
However, the most fundamental tool to describe a many body system
with spin degrees of freedom is the quantum statistical field theory;
any other method should reproduce the same results once the density
operator, that is the quantum state of the system, has been chosen
or determined. According to the successful hydrodynamic picture of
QGP, the system is close to local thermodynamic equilibrium until
hadronization, followed by a quick decoupling of the produced particles
which become free. Therefore, in principle, one can calculate the
leading term of the spin polarization at the decoupling by using the
density operator describing local thermodynamic equilibrium at a quantum
level. This is the final goal of this Section.




\subsubsection{The mean spin vector and the spin density matrix}

\label{subsubsec:spindensitymatrix}The spin vector of a single massive
particle in relativistic quantum mechanics is defined by means of
the Pauli-Lubanski operator: 
\begin{equation}
\widehat{S}^{\mu}=-\frac{1}{2m}\epsilon^{\mu\nu\rho\sigma}\widehat{J}_{\nu\rho}\widehat{P}_{\sigma},\label{luba}
\end{equation}
where $\widehat{J}_{\nu\rho}$ is the angular momentum-boost operator
and $\widehat{P}_{\sigma}$ the energy-momentum operator. This operator
is orthogonal to the energy-momentum, $\widehat{S}\cdot\widehat{P}=0$,
and satisfies the angular momentum algebra: 
\begin{equation}
[\widehat{S}_{\mu},\widehat{P}_{\nu}]=0,\quad[\widehat{S}_{\mu},\widehat{S}_{\nu}]=-i\epsilon_{\mu\nu\rho\sigma}\widehat{S}^{\rho}\widehat{P}^{\sigma}.
\end{equation}
It follows that the eigenstate $|p\rangle$ of $\widehat{P}$, which
is the momentum state measured by a detector, is also an eignevector
of $\widehat{S}^{\mu}$. For any momentum $p$, we denote with $\widehat{S}^{\mu}(p)$
the restriction of $\widehat{S}^{\mu}$ to the eigenspace spanned
by $|p\rangle$. Given that $\widehat{S}(p)\cdot p=0$, we have three
independent operators $\widehat{S}_{i}(p)$ that form a SU(2) algebra
and are the generators of the \emph{little group} of massive particles.
Therefore, we can choose one of these operators, for instance $\widehat{S}_{3}(p)$,
and create a set of mutually commuting operators $\{\widehat{P},\,\widehat{S}_{3}(p),\,\widehat{S}^{2}\}$.
A state $|p,s\rangle$ denotes the corresponding eignevector with
eigenvalues $\{p,\,S_{z},\,S(S+1)\}$ respectively, where $S_{z}$
is the spin quantum number along the z-direction or the eigenvalue
of $\widehat{S}_{3}(p)$, $S$ is referred as \emph{the} spin of the
particle. In order to simplify the notation, hereafter we denote $S_{z}$
as $r$, $s$ or $t$.


Consider now a single relativistic quantum particle in the statistical
ensemble described by the density operator $\widehat{\rho}$. The
average spin vector, denoted by $S^{\mu}$, is the mean value of the
operator (\ref{luba}): 
\begin{equation}
S^{\mu}={\rm Tr}(\,\widehat{\rho}\,\widehat{S}^{\mu}\,).\label{spinvector}
\end{equation}
Since particles are measured in a definite momentum state, we restrict
the operator (\ref{luba}) to the subspace of four-momentum $p$,
obtaining the mean value of the spin vector for a particle with momentum
$p$: 
\begin{equation}
S^{\mu}(p)={\rm Tr}[\widehat{\rho}\widehat{S}^{\mu}(p)]={\rm Tr}[\widehat{\Theta}(p)\widehat{S}^{\mu}(p)],\label{means}
\end{equation}
where we defined the more convenient \emph{spin density operator}
$\widehat{\Theta}(p)$. The matrix elements in the basis of the eigenvalues
of $\widehat{S}_{3}(p)$ gives the spin density matrix 
\begin{equation}
\Theta(p)_{rs}\equiv\langle p,r|\widehat{\rho}|p,s\rangle=\langle p,r|\widehat{\Theta}(p)|p,s\rangle.\label{sdm}
\end{equation}
The spin density matrix encodes all the information about the spin
state of the particle and it borrows form $\widehat{\rho}$ the properties
of being positive definite, Hermitian and of trace one. To express
the mean spin vector entirely in terms of spin matrix elements, the
first step is to expand the trace in~\eqref{means}: 
\begin{equation}
S^{\mu}(p)=\sum_{r}\langle p,r|\widehat{S}^{\mu}(p)\widehat{\Theta}(p)|p,r\rangle=\sum_{r,s}\langle p,r|\widehat{S}^{\mu}(p)|p,s\rangle\langle p,s|\widehat{\Theta}(p)|p,r\rangle.\label{meansp2}
\end{equation}
The matrix elements of the Pauli-Lubanski vector can be obtained by
taking advantage of the fact that $\widehat{S}^{\mu}(p)$ generates
the little group SU(2) algebra in the subspace spanned by $|p\rangle$
and the properties of the Pauli-Lubanski vector. One can show that
\begin{equation}
\langle p,r|\widehat{S}^{\mu}(p)|p,s\rangle=-\frac{1}{2m}\epsilon^{\mu\nu\rho\tau}p_{\tau}\left[D^{S}([p])^{-1}D^{S}(J_{\nu\rho})D^{S}([p])\right]_{rs},
\end{equation}
where the $D^{S}(J_{\lambda\nu})$ are matrices of the angular momentum
generators in the representation with spin $S$, $[p]$ is the \emph{standard
Lorentz transformation} bringing the timelike vector $p_{0}=(m,0,0,0)$
into the four-momentum $p$, and $D^{S}([p])$ its representation.
The mean vector is then obtained by 
\begin{equation}
S^{\mu}(p)=-\frac{1}{2m}\epsilon^{\mu\nu\rho\tau}p_{\tau}{\rm tr}\left(D^{S}([p])^{-1}D^{S}(J_{\nu\rho})D^{S}([p])\Theta(p)\right).\label{meansp3}
\end{equation}

In quantum field theory the spin operator (\ref{luba}) and the statistical
operator $\widehat{\rho}$ are still well defined concept, but the
concept of a single particle state with definite momentum must be
revised. A proper sound definition of a particle can only be given
in a free or weakly interacting theory, for instance in the perturbative
limit of QCD. In those cases, the one-particle states with momentum
$p$ and spin state $s$ are created by the action of the creation
operator $\widehat{a}_{s}^{\dagger}(p)$ to the vacuum, that is $|p,s\rangle=\widehat{a}_{s}^{\dagger}(p)|0\rangle$.
In quantum field theory, the mean spin vector for a particle with
momentum $p$ is given by Eq. (\ref{means}) but with the spin density
matrix defined as: 
\begin{equation}
\Theta(p)_{rs}=\frac{{\rm Tr}[\widehat{\rho}\,\widehat{a}_{s}^{\dagger}(p)\widehat{a}_{r}^{\dagger}(p)]}{\sum_{t}{\rm Tr}[\widehat{\rho}\widehat{a}_{t}^{\dagger}(p)\widehat{a}_{t}(p)]}.\label{spindens}
\end{equation}

It is instructive to derive the mean spin vector for the single quantum
relativistic particle when the system can be approximated as a collection
of distinguishable non-interacting particles such that the full density
operator is simply the tensor product of single-particle density operators:
$\widehat{\rho}=\otimes_{i}\widehat{\rho}_{i}$. To derive the mean
spin vector we evaluate the trace in Eq. (\ref{meansp3}). In order
to do that, we first need to evaluate the spin density matrix, which
contains the information about the physical state through the density
operator. For a relativistic system the most general form of the density
operator at global thermal equilibrium is \citep{Becattini:2012tc,Becattini:2014yxa}
\begin{equation}
\widehat{\rho}=\frac{1}{Z}\exp\left[-b\cdot\widehat{P}+\frac{1}{2}\varpi:\widehat{J}\right],\label{global}
\end{equation}
where $\widehat{P}$ and $\widehat{J}$ are operators for the total
energy-momentum and total angular momentum-boost, respectively. 
The four-vector $b$ is constant and time-like and $\varpi$ is a
constant anti-symmetric tensor. The inverse four-temperature of this
system is given by 
\begin{equation}
\beta^{\mu}=b^{\mu}+\varpi^{\mu\nu}x_{\nu}.\label{fourt}
\end{equation}
At global equilibrium it must fulfill the Killing equation, which
implies a constant thermal vorticity defined as 
\begin{equation}
\varpi_{\mu\nu}=-\frac{1}{2}\left(\partial_{\mu}\beta_{\nu}-\partial_{\nu}\beta_{\mu}\right).\label{thvort}
\end{equation}
Out of global equilibrium, the thermal vorticity does not have to
be constant. Taking advantage of the Poincare algebra, one can show\citep{Becattini:2020qol,Palermo:2021hlf}
that the density operator can be factorized as 
\begin{equation}
\widehat{\rho}=\frac{1}{Z}\exp[-b\cdot\widehat{P}+\varpi:\widehat{J}/2]=\frac{1}{Z}\exp[-\tilde{b}(\varpi)\cdot\widehat{P}]\exp[\varpi:\widehat{J}/2],\label{dopfact}
\end{equation}
with 
\begin{equation}
\tilde{b}(\varpi)=\sum_{k=0}^{\infty}\frac{i^{k}}{(k+1)!}\underbrace{\left(\varpi_{\mu\nu_{1}}\varpi^{\nu_{1}\nu_{2}}\ldots\varpi_{\nu_{k-1}\nu_{k}}\right)}_{\text{k times}}b^{\nu_{k}}\,.
\end{equation}
To calculate the spin density matrix in Eq. (\ref{meansp3}), we plug
Eq. (\ref{dopfact}) in Eq.~(\ref{sdm}). Since the factor $\exp(-\tilde{b}\cdot p)$
cancels out in the ratio, the spin density matrix is 
\begin{equation}
\Theta(p)_{rs}=\frac{\langle p,r|\exp[\varpi:\widehat{J}/2]|p,s\rangle}{\sum_{t=-S}^{S}\langle p,t|\exp[\varpi:\widehat{J}/2]|p,t\rangle}.
\end{equation}
The thermal vorticity, which is an adimensional quantity, is always
small $\varpi\ll1$, and the spin density matrix can be obtained at
first order: 
\begin{align}
\Theta(p)_{rs}\simeq & \frac{\delta_{r,s}}{2S+1}+\frac{\varpi_{\alpha\beta}}{2(2S+1)}\langle p,r|\widehat{J}^{\alpha\beta}|p,s\rangle\nonumber \\
= & \frac{\delta_{r,s}}{2S+1}+\frac{\varpi_{\alpha\beta}}{2(2S+1)}D^{S}(J^{\alpha\beta})_{rs}.\label{Thetafirstord}
\end{align}
Finally, by plugging \eqref{Thetafirstord} into \eqref{meansp3},
we get: 
\begin{align}
S^{\mu}(p)= & -\frac{1}{2m}\epsilon^{\mu\nu\rho\tau}p_{\tau}\frac{\varpi_{\alpha\beta}}{2(2S+1)}{\rm tr}\left[D^{S}([p])^{-1}D^{S}(J_{\nu\rho})D^{S}([p])D^{S}(J^{\alpha\beta})\right]\nonumber \\
= & -\frac{1}{2m}\epsilon^{\mu\nu\rho\tau}p_{\tau}\frac{\varpi_{\alpha\beta}}{2(2S+1)}\frac{S(S+1)(2S+1)}{3}g_{\,\nu}^{\alpha}g_{\,\rho}^{\beta}\nonumber \\
= & -\frac{1}{2m}\frac{S(S+1)}{3}\epsilon^{\mu\nu\rho\tau}p_{\tau}\varpi_{\nu\rho}.
\end{align}
Furthermore, we can extend this result to the case of local equilibrium
by allowing the thermal vorticity to be a function of the coordinates
and averaging over the 3D hypersurface $\Sigma$ where particles are
emitted weighting with the distribution function $f(x,p)$: 
\begin{equation}
S^{\mu}(p)=-\frac{1}{2m}\frac{S(S+1)}{3}\epsilon^{\mu\alpha\beta\nu}p_{\nu}\frac{\int_{\Sigma}d\Sigma_{\lambda}p^{\lambda}f(x,p)\varpi_{\alpha\beta}(x)}{\int_{\Sigma}d\Sigma_{\lambda}p^{\lambda}f(x,p)}.\label{spinvecSingleParticle}
\end{equation}
This extends the original result of spin polarization of spin $1/2$
particles\citep{Becattini:2013fla,Becattini:2016gvu} to any spin.


\subsubsection{Wigner functions}

\label{subsubsec:wignerf}

In a more general quantum field theory framework, the spin density
matrix is given by Eq. (\ref{spindens}). However, that form is not
well suited for making predictions and modeling. It is more convenient
to connect the spin density matrix with the Wigner function \citep{Heinz:1983nx,Vasak:1987um,Zhuang:1998bqx,Wang:2001dm,Blaizot:2001nr},
a well-known tool to describe quantum effects in relativistic fluids~\citep{DeGroot:1980dk,Hakim2011}.
Quantum kinetic or transport theory can be constructed in terms of
Wigner functions, which provides a microscopic description for spin
transport processes in heavy ion collisions \citep{Gao:2012ix,Chen:2012ca,Hidaka:2016yjf,Hidaka:2017auj,Gao:2019znl,Weickgenannt:2019dks,Weickgenannt:2020aaf,Hattori:2019ahi,Yang:2020hri,Liu:2020flb,Weickgenannt:2021cuo,Sheng:2021kfc,Wang:2019moi,Fang:2022ttm,Fang:2023bbw},
see, e.g., Refs. \citep{Gao:2020pfu,Hidaka:2022dmn} for recent reviews.


The covariant Wigner operator is defined as the Wigner transform of
the two-point function \citep{DeGroot:1980dk} 
\begin{eqnarray}
\widehat{W}_{\alpha\beta}(x,k) & = & -\int\frac{{\rm d}^{4}y}{(2\pi)^{4}}\;e^{-ik\cdot y}:\Psi_{\alpha}\left(x-\frac{y}{2}\right)\overline{\Psi}_{\beta}\left(x+\frac{y}{2}\right):\nonumber \\
 & = & \int\frac{{\rm d}^{4}y}{(2\pi)^{4}}\;e^{-ik\cdot y}:\overline{\Psi}_{\beta}\left(x+\frac{y}{2}\right)\Psi_{\alpha}\left(x-\frac{y}{2}\right):\,,\label{wigdirop}
\end{eqnarray}
where $\Psi$ is the Dirac field, and $:$ denotes the normal ordering
of creation and destruction operators. With $\Psi$ being a spinor,
the covariant Wigner operator is a $4\times4$ spinorial matrix. Thanks
to the integral transform, the non-locality of the two-point function
is mapped into a quasi-local operator depending on the coordinate
$x$ and the momentum $k$. For a system described by the density
operator $\wrho$, the Wigner function is the mean value of the Wigner
operator: 
\begin{equation}
W(x,p)={\rm Tr}\left[\,\widehat{\rho}\,\widehat{W}(x,p)\right].\label{wignerfunc}
\end{equation}
It follows that the Wigner function is real, but differently from
the classical distribution function it is not always positive definite.
If $\Psi$ is the free Dirac field (solving the free Dirac equation),
then the Wigner function satisfies the so-called Wigner equation 
\begin{equation}
\left(m-\slashed{k}-\frac{i}{2}\slashed{\partial}\right)\widehat{W}(x,k)=0.\label{wignereq}
\end{equation}

The Dirac field can be expanded in plane waves 
\begin{equation}
\Psi(x)=\sum_{r}\frac{1}{(2\pi)^{3/2}}\int\frac{{\rm d}^{3}{\rm p}}{2E_{p}}\left[\widehat{a}_{r}(p)u_{r}(p)e^{-ip\cdot x}+\widehat{b}_{r}^{\dagger}(p)v_{r}(p)e^{ip\cdot x}\right],\label{dirf}
\end{equation}
where $E_{p}=\sqrt{p^{2}+m^{2}}$, $u_{r}(p)$ and $v_{r}(p)$ are
the spinors of free particles and antiparticles in the polarization
state $r$ normalized as $\bar{u}_{r}u_{s}=2m\delta_{rs}$, $\bar{v}_{r}v_{s}=-2m\delta_{rs}$,
and the commutation relation of creation and destruction operators
are $\{\widehat{a}_{s}(p),\widehat{a}_{s^{\prime}}^{\dagger}(p^{\prime})\}=2E_{p}\delta_{s,s'}\delta^{3}({\bf p}-{\bf p}')$.
In terms of $\Psi(x)$ in Eq. (\ref{dirf}) the Wigner function can
be put into the form 
\begin{align}
\widehat{W}(x,k)= & \sum_{r,s}\frac{1}{(2\pi)^{3}}\int\frac{{\rm d}^{3}{\rm p}}{2E_{p}}\frac{{\rm d}^{3}{\rm p}^{\prime}}{2E_{p^{\prime}}}\nonumber \\
 & \times\left\{ e^{-i(p-p^{\prime})\cdot x}\left[\delta^{4}(k-(p+p^{\prime})/2)\widehat{a}_{s}^{\dagger}(p^{\prime})\widehat{a}_{r}(p)u_{r}(p)\bar{u}_{s}(p')\right.\right.\nonumber \\
 & -\left.\delta^{4}(k+(p+p^{\prime})/2)\widehat{b}_{r}^{\dagger}(p^{\prime})\widehat{b}_{s}(p)v_{r}(p')\bar{v}_{s}(p)\right]\nonumber \\
 & -\delta^{4}(k-(p-p^{\prime})/2)\left[e^{-i(p+p^{\prime})\cdot x}\widehat{a}_{r}(p)\widehat{b}_{s}(p^{\prime})u_{r}(p)\bar{v}_{s}(p^{\prime})\right.\nonumber \\
 & \left.\left.+e^{i(p+p^{\prime})\cdot x}\widehat{b}_{r}^{\dagger}(p^{\prime})\widehat{a}_{s}^{\dagger}(p)v_{r}(p^{\prime})\bar{u}_{s}(p)\right]\right\} .\label{wignerdop}
\end{align}
While the momenta $p$ and $p'$ are on-shell, from the delta functions
in the above expression it is clear that the momentum $k$ is not
on-shell. However, we see that each term in Eq. (\ref{wignerdop})
corresponds to the future time-like (particle), past time-like (antiparticle)
and space-like parts of the Wigner operator. Equation (\ref{wignerdop})
can then be written as 
\begin{align}
\widehat{W}(x,k)= & \widehat{W}(x,k)\theta(k^{2})\theta(k^{0})+\widehat{W}(x,k)\theta(k^{2})\theta(-k^{0})+\widehat{W}(x,k)\theta(-k^{2})\nonumber \\
\equiv & \widehat{W}_{+}(x,k)+\widehat{W}_{-}(x,k)+\widehat{W}_{S}(x,k).
\end{align}
For free fields, despite $k$ not being on-shell, when one integrates
the Wigner function over a 3D hypersurface, then $k$ becomes an on-shell
vector. This feature allows us to relate the spin density matrix with
the Wigner function in a linear way. To prove the previous statement,
we take the derivative $k\cdot\partial$ of Eq. (\ref{wignerdop})
and use $(p-p^{\prime})\cdot(p+p^{\prime})=0$, then we obtain $k\cdot\partial\widehat{W}_{\pm}(x,k)=k^{\cdot}\partial\widehat{W}_{S}(x,k)=0$.
This implies that with appropriate boundary conditions the integral
over a space-like 3D hypersurface: 
\begin{equation}
\int_{\Sigma}\text{{\rm d}}\Sigma_{\mu}k^{\mu}\widehat{W}(x,k)
\end{equation}
does not depend on the hypersurface $\Sigma$. Taking advantage of
this, without loosing generality we can choose $\Sigma$ at the hyperplane
$t=0$ and integrate the Eq. \eqref{wignerdop} explicitly: 
\begin{align}
\int_{t=0}\text{{\rm d}}\Sigma_{\mu}k^{\mu}\widehat{W}(x,k)= & k^{0}\int\text{{\rm d}}^{3}{\rm x}\;\widehat{W}(x,k)\nonumber \\
= & \sum_{r,s}\frac{1}{2}\delta(k^{2}-m^{2})\left[\theta(k^{0})\widehat{a}_{s}^{\dagger}(k)\widehat{a}_{r}(k)u_{r}(k)\bar{u}_{s}(k)\right.\nonumber \\
 & \left.+\theta(-k^{0})\widehat{b}_{r}^{\dagger}(-k)\widehat{b}_{s}(-k)v_{r}(-k)\bar{v}_{s}(-k)\right]\;,\label{wignerint}
\end{align}
where the mixed term vanished because of the factor $k^{0}\delta(k^{0})$.
We find that the integral of the Wigner operator over any space-like
hypersurface is proportional to $\delta(k^{2}-m^{2})$, proving that,
after integration, $k$ becomes an on-shell four-vector. Furthermore,
after integrating over $\Sigma$, the Wigner operator is split into
an on-shell particle and antiparticle parts $\widehat{w}_{\pm}$ from
Eq.~(\ref{wignerint}): 
\begin{eqnarray}
\frac{1}{2E_{k}}\delta(k^{0}-E_{k})\widehat{w}_{+}(k) & = & \int\text{{\rm d}}\Sigma_{\mu}k^{\mu}\widehat{W}_{+}(x,k),\nonumber \\
\frac{1}{2E_{k}}\delta(k^{0}+E_{k})\widehat{w}_{-}(k) & = & \int\text{{\rm d}}\Sigma_{\mu}k^{\mu}\widehat{W}_{-}(x,k),\label{reducedwig}
\end{eqnarray}
where $\widehat{w}_{+}(k)$ and $\widehat{w}_{-}(k)$ are defined
as 
\begin{eqnarray}
\widehat{w}_{+}(k) & = & \frac{1}{2}\sum_{r,s}\widehat{a}_{s}^{\dagger}(k)\widehat{a}_{r}(k)u_{r}(k)\bar{u}_{s}(k)\,,\nonumber \\
\widehat{w}_{-}(k) & = & \frac{1}{2}\sum_{r,s}\widehat{b}_{r}^{\dagger}(-k)\widehat{b}_{s}(-k)v_{r}(-k)\bar{v}_{s}(-k)\,.\label{wignerint2}
\end{eqnarray}
We are now in the position to prove the relation that connects the
Wigner operator of free fields with the spin density matrix \eqref{spindens}.
This can be achieved by expressing the product of the creation and
destruction operators in terms of the particle part of the Wigner
operator integrated over $\Sigma$. Simply multiplying the first line
of Eq. \eqref{wignerint2} by $u_{s}(k)$ to the right and by $\bar{u}_{r}(k)$
to the left, and keeping in mind the normalization of the spinors
$u$, we obtain: 
\begin{equation}
\bar{u}_{r}(k)\widehat{w}_{+}(k)u_{s}(k)=2m^{2}\widehat{a}_{s}^{\dagger}(k)\widehat{a}_{r}(k).\label{wignerint3}
\end{equation}
Plugging this expression into the definition of spin density matrix
(\ref{spindens}), we finally arrive at 
\begin{equation}
\Theta(p)_{rs}=\frac{\bar{u}_{r}(p)w_{+}(p)u_{s}(p)}{\sum_{t}\bar{u}_{t}(p)w_{+}(p)u_{t}(p)},\label{theta-spin-density}
\end{equation}
where $w_{+}$ is the mean value 
\begin{equation}
w_{+}(p)={\rm Tr}\left[\,\widehat{\rho}\,\widehat{w}_{+}\right].
\end{equation}
We can put Eq. (\ref{theta-spin-density}) into a more familiar form
using the definition of $\widehat{w}_{+}$ in Eq. (\ref{reducedwig})
\begin{equation}
\Theta(p)_{rs}=\frac{\int\text{{\rm d}}\Sigma_{\mu}p^{\mu}\bar{u}_{r}(p)W_{+}(x,p)u_{s}(p)}{\sum_{t}\int\text{{\rm d}}\Sigma_{\mu}p^{\mu}\bar{u}_{t}(p)W_{+}(x,p)u_{t}(p)},\label{spindensw}
\end{equation}
where the ratio of the delta functions has been simplified. In terms
of the $4\times2$ spinorial matrix $U$ (and correspondingly $2\times4$
matrix $\bar{U}$) such that $U_{\alpha,r}(p)=u_{r}^{\alpha}(p)$
the spin density matrix can be written as 
\begin{equation}
\Theta(p)=\frac{\int\text{{\rm d}}\Sigma_{\mu}p^{\mu}\bar{U}(p)W_{+}(x,p)U(p)}{{\rm tr}_{2}\int\text{{\rm d}}\Sigma_{\mu}p^{\mu}\bar{U}(p)W_{+}(x,p)U(p)},\label{spindensw2}
\end{equation}
where, from now on, we will distinguish between the trace over four
spinorial indices ${\rm tr}_{4}$ and the trace over two polarization
states ${\rm tr}_{2}$.


Before deriving the formula (\ref{meanspf0}), we need to present
other properties of the Wigner function which are also useful for
the kinetic theory. The definition (\ref{wigdirop}) of the Wigner
function is manifestly covariant, and indeed under a Lorentz transformation
with parameters $\omega$, the Wigner function becomes \citep{DeGroot:1980dk}
\begin{equation}
W'(x',k')=e^{i\omega:\Sigma/2}W(x,k)e^{-i\omega:\Sigma/2},\label{wignerfLorentz}
\end{equation}
where $\Sigma_{\alpha\beta}=(i/4)[\gamma_{\alpha},\,\gamma_{\beta}]$.
Instead of dealing with spinorial matrices, it is more practical to
use functions of scalars, vectors and tensors. The Wigner function
can be expanded on the basis of 16 independent generators of Clifford
algebra as 
\begin{eqnarray}
W(x,k) & = & \sum_{i}W_{i}\Gamma_{i}\nonumber \\
 & = & \frac{1}{4}\left[\mathcal{F}+i\gamma^{5}\mathcal{P}+\gamma^{\mu}\mathcal{V}_{\mu}+\gamma^{5}\gamma^{\mu}\mathcal{A}_{\mu}+\Sigma^{\mu\nu}\mathcal{S}_{\mu\nu}\right]\,,\label{WignerClifford}
\end{eqnarray}
where Dirac $\Gamma_{i}$ matrices are defined as 
\begin{equation}
\{\Gamma_{i}\}_{i=1,\,2,\,\dots,\,16}=\left\{ 1,\,i\gamma^{5},\,\gamma^{\mu},\,\gamma^{5}\gamma^{\mu},\,\Sigma^{\mu\nu}\right\} \,.
\end{equation}
Some component functions can be extracted as 
\begin{align}
\mathcal{F}(x,k)= & {\rm tr}_{4}\left[\,W(x,k)\right]\,,\nonumber \\
\mathcal{V}^{\mu}(x,k)= & {\rm tr}_{4}\left[\gamma^{\mu}W(x,k)\right]\,,\nonumber \\
\mathcal{A}^{\mu}(x,k)= & {\rm tr}_{4}\left[\gamma^{\mu}\gamma^{5}W(x,k)\right]\,,
\end{align}
and similarly for other components. From the cyclicity of the trace,
the Lorentz transformation of the gamma matrices and of the Wigner
function in Eq. (\ref{wignerfLorentz}), it follows that $\mathcal{F},\,\mathcal{P},\,\mathcal{V},\,\mathcal{A},\,\mathcal{S}$
transform respectively as a scalar, pseudo-scalar, four-vector, axial
four-vector and rank-2 tensor under Lorentz and charge, parity and
time reversal transformations. The mean values of the vector current
and energy-momentum tensor are given directly as momentum integrals
of these functions: 
\begin{align}
\langle\widehat{j}^{\mu}(x)\rangle= & \langle\bar{\psi}\gamma^{\mu}\psi\rangle=\int{\rm d}^{4}k\,\mathcal{V}^{\mu}(x,k)\,,\\
\langle\widehat{T}_{C}^{\mu\nu}(x)\rangle= & \int{\rm d}^{4}k\,\mathcal{V}^{\mu}(x,k)k^{\nu}\,,
\end{align}
where $\widehat{T}_{C}^{\mu\nu}$ is the canonical form of the energy-momentum
tensor for free Dirac fields. The Wigner equation (\ref{wignereq})
can also be written in terms of these quantities \citep{Hakim2011}.
For instance, the kinetic equations for the vector and axia-vector
components read 
\begin{equation}
k\cdot\mathcal{V}(x,k)=m\mathcal{F}(x,k),\quad k\cdot\mathcal{A}(x,k)=0.\label{wignereq2}
\end{equation}
This will be used below to work out Eq. (\ref{spindensw}).


\subsubsection{Polarization of fermions}

\label{subsubsec:fermionpolarization}

We can now calculate the mean spin vector from Eq. (\ref{meansp3})
using the spin density matrix expressed through the Wigner function
as given in Eq. (\ref{spindensw2}). To take advantage of the form
(\ref{meansp3}) that uses the representation of Lorentz's group in
the spin $S$, we use the Weyl's representation for the $U$ spinors~\citep{Weinberg:1964cn}:
\begin{equation}
U(p)=\sqrt{m}\left(\begin{array}{cc}
D^{S}([p])\\
D^{S}([p]^{\dagger-1})
\end{array}\right),\qquad V(p)=\sqrt{m}\left(\begin{array}{cc}
D^{S}([p]C^{-1})\\
D^{S}([p]^{\dagger-1}C)
\end{array}\right),\label{uvspin}
\end{equation}
where $C=i\sigma_{2}$ and $\sigma_{i}$ are the Pauli matrices. Taking
advantage of the Hermiticity of $\Theta(p)$, the cyclicity of the
trace and the fact that the mean spin vector is real, we obtain 
\begin{align*}
S^{\mu}(p)= & -\frac{1}{4m}\epsilon^{\mu\beta\gamma\delta}p_{\delta}\left\{ {\rm tr}_{2}\left[D^{S}([p]^{-1})D^{S}(J_{\beta\gamma})D^{S}([p])\Theta(p)\right]\right.\\
 & \left.+{\rm tr}_{2}\left[D^{S}([p]^{-1})D^{S}(J_{\beta\gamma})D^{S}([p])\Theta(p)\right]^{*}\right\} \\
= & -\frac{1}{4m}\epsilon^{\mu\beta\gamma\delta}p_{\delta}\left\{ {\rm tr}_{2}\left[D^{S}([p]^{-1})D^{S}(J_{\beta\gamma})D^{S}([p])\Theta(p)\right]\right.\\
 & \left.+{\rm tr}_{2}\left[D^{S}([p])^{\dagger}D^{S}(J_{\beta\gamma})^{\dagger}D^{S}([p]^{-1})^{\dagger}\Theta(p)\right]\right\} .
\end{align*}
Plugging Eq. (\ref{spindensw2}) with the representation in Eq. (\ref{uvspin}),
the mean spin vector is proportional to 
\begin{align}
{\rm tr}_{2} & \left[D^{S}([p]^{-1})D^{S}(J_{\beta\gamma})D^{S}([p])\bar{U}(p)W_{+}(x,p)U(p)\right]\nonumber \\
 & +{\rm tr}_{2}\left[D^{S}([p])^{\dagger}D^{S}(J_{\beta\gamma})^{\dagger}D^{S}([p]^{-1})^{\dagger}\bar{U}(p)W_{+}(x,p)U(p)\right].\label{numtrac}
\end{align}
Since the representation of the Lorentz transformations $D^{S}(J_{\beta\gamma})$
are related to the Dirac gamma matrices as follows: 
\begin{equation}
\Sigma_{\beta\gamma}=(i/4)[\gamma_{\beta},\gamma_{\gamma}]=\left(\begin{array}{cc}
D^{S}(J_{\beta\gamma})\; & \;0\\
0\; & \;D^{S}(J_{\beta\gamma})^{\dagger}
\end{array}\right),\label{Sigma}
\end{equation}
Eq. (\ref{numtrac}) is equivalent to: 
\begin{equation}
\frac{1}{m}{\rm tr}_{2}\left[\bar{U}(p)\Sigma_{\beta\gamma}U(p)\bar{U}(p)W_{+}(x,p)U(p)\right].\label{numtrac-1}
\end{equation}
Reminding that when $A$ is a $2\times4$ and $B$ is a $4\times2$
matrix, we simply have ${\rm tr}_{2}AB={\rm tr}_{4}BA$, Eq. (\ref{numtrac-1})
can also be evaluated as 
\begin{equation}
\frac{1}{m}{\rm tr}_{4}\left[U(p)\bar{U}(p)\Sigma_{\beta\gamma}U(p)\bar{U}(p)W_{+}(x,p)\right],
\end{equation}
which we can be further simplified as 
\begin{equation}
\frac{1}{m}{\rm tr}_{4}\left[(\slashed{p}+m)\Sigma_{\beta\gamma}(\slashed{p}+m)W_{+}(x,p)\right].
\end{equation}
Finally the mean spin vector becomes 
\begin{equation}
S^{\mu}(p)=\frac{N^{\mu}}{D},
\end{equation}
where 
\begin{equation}
N^{\mu}=\epsilon^{\mu\beta\gamma\delta}p_{\delta}\frac{1}{m}{\rm tr}_{4}\left[(\slashed{p}+m)\Sigma_{\beta\gamma}(\slashed{p}+m)W_{+}(x,p)\right],
\end{equation}
and 
\begin{equation}
D={\rm tr}_{2}\left[\bar{U}(p)W_{+}(x,p)U(p)\right]={\rm tr}_{4}\left[(\slashed{p}+m)W_{+}(x,p)\right].
\end{equation}
Using the decomposition in Eq. (\ref{WignerClifford}), but only for
the particle part of the Wigner function, the traces in $N^{\mu}$
and $D$ are obtained using the well-known traces of gamma matrices.
Taking those traces and using the relations $p\cdot\mathcal{A}_{+}(x,p)=0$
and $p\cdot\mathcal{V}_{+}(x,p)=m\mathcal{F}_{+}(x,p)$ in Eq. (\ref{wignereq2})
coming from the decomposition of the Wigner equation, we obtain 
\begin{equation}
N^{\mu}=m\mathcal{A}_{+}^{\mu}(x,p),\quad D=2m\mathcal{F}_{+}(x,p)\,,
\end{equation}
and finally 
\begin{equation}
S^{\mu}(p)=\frac{1}{2}\frac{\int\text{{\rm d}}\Sigma\cdot p\;\mathcal{A}_{+}^{\mu}(x,p)}{\int\text{{\rm d}}\Sigma\cdot p\;\mathcal{F}_{+}(x,p)}=\frac{1}{2}\frac{\int\text{{\rm d}}\Sigma\cdot p\;{\rm tr}_{4}\left[\gamma^{\mu}\gamma^{5}W_{+}(x,p)\right]}{\int\text{{\rm d}}\Sigma\cdot p\;{\rm tr}_{4}\left[W_{+}(x,p)\right]}.\label{meanspf2}
\end{equation}
This expression can also be derived using other methods~\citep{Becattini:2013fla,Fang:2016vpj,Weickgenannt:2019dks,Liu:2020flb}.
What has been presented here is the derivation from first principles.
However, it must be stressed that the relation (\ref{meanspf2}) is
strictly speaking valid only for the non-interacting case, as it was
derived using the relations in Eqs. (\ref{wignerint2}) and (\ref{wignereq2}),
which are consequences of the free Dirac equation. Corrections due
to weak interactions can be obtained following the steps described
above, once Eqs. (\ref{wignerint2}) and (\ref{wignereq2}) have been
modified to take interactions into account. Instead the relations
(\ref{means}), (\ref{meansp3}) and (\ref{spindens}) do not rely
on the non-interacting equations of motion, but as discussed above,
they can only be applied when the notion of particle is a sensible
concept. To evaluate the spin polarization of $\Lambda$ particles
in heavy-ion collisions, it is convenient to choose the freeze-out
hypersurface as the domain of integration in Eq. (\ref{meanspf2}),
as it is where they are formed and where the Wigner function of $\Lambda$
is well defined.

Using known identities for the gamma matrices and the Dirac equation,
one can also find equivalent forms of (\ref{meanspf2}), for instance
\begin{equation}
S^{\mu}(p)=-\frac{1}{2m}\epsilon^{\mu\beta\gamma\delta}p_{\delta}\frac{\int\text{{\rm d}}\Sigma_{\lambda}p^{\lambda}{\rm tr}_{4}(\Sigma_{\beta\gamma}W_{+}(x,p))}{\int\text{{\rm d}}\Sigma_{\lambda}p^{\lambda}{\rm tr}_{4}W_{+}(x,p)},\label{meanspffree}
\end{equation}
or 
\begin{equation}
S^{\mu}(p)=-\frac{1}{4}\epsilon^{\mu\beta\gamma\delta}p_{\delta}\frac{\int\text{{\rm d}}\Sigma_{\lambda}{\rm tr}_{4}\left[\{\gamma^{\lambda},\Sigma_{\beta\gamma}\}W_{+}(x,p)\right]}{\int\text{{\rm d}}\Sigma_{\lambda}p^{\lambda}{\rm tr}_{4}W_{+}(x,p)}.\label{meanspf3}
\end{equation}
While some of these have been previously obtained choosing a particular
form for the spin tensor, the derivation presented here shows that
these expressions for the mean spin vector do not require the introduction
of a spin tensor and are therefore independent thereof. Indeed the
Pauli-Lubanski operator defined in Eq. (\ref{luba}) is given in terms
of globally conserved operators that are independent of pseudo-gauge
transformations. The mean spin vector in an off-equilibrium system
has been indeed found to depend on the particular form of the spin
tensor~\citep{Buzzegoli:2021wlg}, however this dependence comes
from the density operator contained inside the Wigner function~\citep{Becattini:2018duy}.

\subsection{Local thermodynamic equilibrium}

\label{subsec:lte}In this section we calculate the mean spin vector
of a free Dirac field at local thermodynamic equilibrium (LTE) using
quantum statistical field theory at the first order in gradients of
$\beta^{\mu}$, the inverse four-temperature, and of $\zeta$, the
ratio of the chemical potential to temperature. So far we have reduced
the problem of evaluating the mean spin vector to that of evaluating
the axial and scalar parts of the covariant Wigner function. As the
Wigner function is the expectation value of the Wigner operator, as
shown in Eq. (\ref{wignerfunc}), the next step is to provide the
(covariant) density operator $\widehat{\rho}$ describing LTE.


The Zubarev's method of the stationary Non-Equilibrium Density Operator
(NEDO) \citep{Zubarev:1966,Zubarev:1979,vanWeert1982,Zubarev:1989su,Morzov:1998}
provides the framework to obtain the quantum density operator describing
a relativistic fluid which can be assumed to have achieved the condition
of local thermodynamic equilibrium at some stage; an updated and upgraded
version of this theory can be found in Ref.~\citep{Becattini:2019dxo}.
If the system achieves LTE at some initial 3D hypersurace $\Sigma_{0}$,
as it is supposedly the case for the QGP, the NEDO is obtained by
maximizing the entropy $S=-{\rm tr}(\widehat{\rho}\log\widehat{\rho})$
at fixed energy-momentum density \citep{Becattini:2014yxa}. This
procedure yields: 
\begin{equation}
\widehat{\rho}=\dfrac{1}{Z}\exp\left[-\int_{\Sigma_{0}}\text{{\rm d}}\Sigma\;n_{\mu}\left(\widehat{T}^{\mu\nu}(x)\beta_{\nu}(x)-\zeta(x)\widehat{j}^{\mu}(x)\right)\right],\label{densop}
\end{equation}
where $\widehat{T}^{\mu\nu}$ is the Belinfante symmetrized energy-momentum
tensor operator and $\widehat{j}^{\mu}$ a conserved current. As mentioned,
the form of the NEDO depends on the particular form used to describe
the energy-momentum tensor and the spin tensor operators \citep{Becattini:2018duy}.
The relation between different forms of the energy-momentum tensor
and the spin tensor operator that leaves the global charges unaffected,
i.e. the total four-momentum $\widehat{P}$ and the angular-boost
$\widehat{J}$ of the system, is called a pseudo-gauge transformation.
Here we adopted the Belinfante form. This dependence disappears at
global thermal equilibrium \citep{Becattini:2018duy} where the density
operator only depends on the pseudo-gauge invariant conserved operators
$\widehat{P}$, $\widehat{J}$ and $\widehat{Q}$.


The operators in Eq. (\ref{densop}) are evaluated at the initial
time of the QGP phase where the degrees of freedom are the quarks
and the gluons, but we are interested in the action of this operator
on the Wigner operator for hadronic fields in Eq. (\ref{wigdirop}).
Instead of solving the full dynamics of interacting quantum fields,
it is more convenient to write the density matrix in terms of operators
evaluated at later times, and more precisely at the time of freeze-out
as done for Eqs. (\ref{spinvecSingleParticle}) and (\ref{meanspf2}).
This is simply done by applying the Gauss theorem \citep{Becattini:2012tc,Becattini:2019dxo}
\begin{align}
\widehat{\rho} & =\frac{1}{Z}\exp\left[-\int_{\Sigma(\tau_{0})}\!\!\!\!\!\!{\rm d}\Sigma_{\mu}\;\left(\widehat{T}^{\mu\nu}\beta_{\nu}-\widehat{j}^{\mu}\zeta\right)\right]\nonumber \\
 & =\frac{1}{Z}\exp\left[-\int_{\Sigma({\rm FO})}\!\!\!\!\!\!{\rm d}\Sigma_{\mu}\;\left(\widehat{T}^{\mu\nu}\beta_{\nu}-\widehat{j}^{\mu}\zeta\right)\right.\nonumber \\
 & +\left.\int_{\Omega}{\rm d}\Omega\;\left(\widehat{T}^{\mu\nu}\nabla_{\mu}\beta_{\nu}-\widehat{j}^{\mu}\nabla_{\mu}\zeta\right)\right],
\end{align}
where $\Omega$ is the region of space-time between $\Sigma(\tau_{0})$
and $\Sigma({\rm FO})$. The values of $\beta^{\mu}$ and $\zeta$
at the freeze-out can be obtained starting from their values at initial
time $\tau_{0}$ using hydrodynamic equations. This procedure is much
simpler than solving the dynamics of strongly interacting quantum
field theory.


It can be shown \citep{Becattini:2019dxo} that the contributions
from the integral over the 3D hypersurface does not increase the rate
of entropy and dissipative effects are only contained in the integral
over the volume $\Omega$. The non-dissipative part of the density
operator is then 
\begin{equation}
\widehat{\rho}\simeq\widehat{\rho}_{\mathrm{LE}}=\frac{1}{Z}\exp\left[-\int_{\Sigma(\mathrm{FO})}\!\mathrm{d}\Sigma_{\mu}\left(\widehat{T}^{\mu\nu}\beta_{\nu}-\zeta\,\widehat{j}^{\mu}\right)\right],\label{eq:LEStatOper}
\end{equation}
and it is denoted as the local equilibrium (LE) density operator.
In order to obtain the non-dissipative contributions to the spin polarization,
we first have to calculate the non-dissipative part of the Wigner
function which is given by 
\begin{equation}
W_{{\rm LE}}(x,k)={\rm Tr}\left(\widehat{\rho}_{{\rm LE}}\widehat{W}(x,k)\right).\label{wignerle1}
\end{equation}
The exact calculation of Eq. (\ref{wignerle1}) is again a difficult
task. However, by taking advantage of the hydrodynamic regime of the
QGP and that correlation functions between operators evaluated at
different points go rapidly to zero with their distance, we can approximate
the density operator by expanding the thermodynamic quantities as
follows: 
\begin{equation}
\beta_{\nu}(y)\simeq\beta_{\nu}(x)+\partial_{\lambda}\beta_{\nu}(x)(y-x)^{\lambda},\quad\zeta(y)\simeq\zeta_{\nu}(x)+\partial_{\lambda}\zeta(x)(y-x)^{\lambda}\,,\label{eq:TaylorBeta}
\end{equation}
and we obtain 
\begin{eqnarray}
\widehat{\rho}_{{\rm LE}} & \simeq & \frac{1}{Z}\exp\left[-\beta_{\nu}(x)\widehat{P}^{\nu}+\zeta(x)\widehat{Q}+\frac{1}{2}\varpi_{\mu\nu}(x)\widehat{J}_{x}^{\mu\nu}-\frac{1}{2}\xi_{\mu\nu}(x)\widehat{Q}_{x}^{\mu\nu}\right.\nonumber \\
 &  & \left.+\partial_{\lambda}\zeta(x)\!\!\!\int_{\Sigma({\rm FO})}\!\!\!{\rm d}\Sigma_{\mu}(y-x)^{\lambda}\widehat{j}^{\mu}(y)+\cdots\right]\,,\label{eq:ApproxHydro}
\end{eqnarray}
where 
\begin{equation}
\widehat{J}_{x}^{\mu\nu}=\!\int\!{\rm d}\Sigma_{\lambda}\left[(y-x)^{\mu}\widehat{T}^{\lambda\nu}(y)-(y-x)^{\nu}\widehat{T}^{\lambda\mu}(y)\right]
\end{equation}
is the conserved angular momentum operator, and 
\begin{equation}
\widehat{Q}_{x}^{\mu\nu}=\!\int_{\Sigma_{FO}}\!{\rm d}\Sigma_{\lambda}\left[(y-x)^{\mu}\widehat{T}^{\lambda\nu}(y)+(y-x)^{\nu}\widehat{T}^{\lambda\mu}(y)\right]
\end{equation}
is a non-conserved symmetric quadrupole like operator, and we introduced
the thermal shear 
\begin{equation}
\xi_{\mu\nu}=\frac{1}{2}\left(\partial_{\mu}\beta_{\nu}+\partial_{\nu}\beta_{\mu}\right).
\end{equation}
Since the thermal vorticity couples to a conserved operator, a system
can reach global thermal equilibrium with non-vanishing thermal vorticity
and its thermal properties are different from the usual homogeneous
non vorticous equilibrium \citep{Buzzegoli:2017cqy,Becattini:2020qol,Palermo:2021hlf}.
The thermal shear in Eq. (\ref{eq:ApproxHydro}) is instead coupled
to a non-conserved operator and gives rise to off-equilibrium but
non-dissipative effects.


Since both thermal vorticity and thermal shear are usually small,
the non-dissipative expectation value of the Wigner operator in Eq.
(\ref{wignerle1}) can be calculated with the density operator (\ref{eq:ApproxHydro})
using linear response theory and standard techniques of thermal field
theory. For a free Dirac field, plugging the resulting Wigner function
in Eq. (\ref{meanspf2}), we obtain: 
\begin{eqnarray}
S^{\mu}(p) & = & -\frac{\epsilon^{\mu\rho\sigma\tau}p_{\tau}}{8m\int_{\Sigma}{\rm d}\Sigma\cdot p\;n_{F}}\int_{\Sigma}{\rm d}\Sigma\cdot p\nonumber \\
 &  & \times n_{F}(1-n_{F})\left[\varpi_{\rho\sigma}+2\hat{t}_{\rho}\frac{p^{\lambda}}{E_{p}}\xi_{\lambda\sigma}-\frac{\hat{t}_{\rho}\partial_{\sigma}\zeta}{2E_{p}}\right]\,,\label{eq:SpinPolFirstOrder}
\end{eqnarray}
where $\hat{t}$ is the time direction in the laboratory frame and
$n_{F}$ is the Fermi-Dirac phase-space distribution function: 
\begin{equation}
n_{F}=\frac{1}{\exp[\beta\cdot p-\mu q]+1},
\end{equation}
where $q$ is the charge of the particle and $\mu$ the corresponding
chemical potential. We derived the first order non-dissipative contributions
to spin polarization, the first term of Eq. (\ref{eq:SpinPolFirstOrder})
is the polarization induced by thermal vorticity which is the main
contribution for global spin polarization, the second term is the
shear induced polarization \citep{Becattini:2021suc,Becattini:2021iol,Liu:2021uhn,Liu:2021nyg,Yi:2021ryh},
and the last one is the contribution from the gradient of fugacity
\citep{Liu:2020dxg,Fu:2022myl,Ambrus:2020oiw,Ambrus:2019ayb,Buzzegoli:2022qrr},
also referred as the spin Hall effect. In addition to these terms
the other known non-dissipative contributions are coming from an imbalance
of chiral charges \citep{Becattini:2020xbh,Gao:2021rom}, and is obtained
including an axial charge in the density operator, and the spin polarization
induced by an external magnetic field \citep{Gao:2012ix,Becattini:2016gvu,Yi:2021ryh,Buzzegoli:2022qrr}
which can also be included in the Zubarev approach \citep{Buzzegoli:2020ycf,Buzzegoli:2022qrr}.



\subsection{Spin hydrodynamics}

\label{sbusec:polarization_spin_transport}

Spin hydrodynamics refers to the inclusion of the spin tensor in the
equations of relativistic hydrodynamics \footnote{In this subsection we define $\Delta_{\mu\nu}=g_{\mu\nu}-u_{\mu}u_{\nu}$
with $u^{\mu}$ being the fluid velocity. For a rank-$2$ tensor $A^{\mu\nu}$,
we introduce the short hand notation $A^{<\mu\nu>}\equiv(1/2)[\Delta^{\mu\alpha}\Delta^{\nu\beta}+\Delta^{\mu\beta}\Delta^{\nu\alpha}]A_{\alpha\beta}-(1/3)\Delta^{\mu\nu}(\Delta^{\alpha\beta}A_{\alpha\beta})$.}. To make sense of this simple sentence it is necessary to go through
some general arguments of symmetries in quantum field theory. In general,
in Minkowski space-time, the conserved currents from translation and
Lorentz invariance include the stress-energy tensor $\widehat{T}^{\mu\nu}$
and the angular momentum-boost current: 
\[
\widehat{{\cal J}}^{\mu,\lambda\nu}=x^{\lambda}\widehat{T}^{\mu\nu}-x^{\nu}\widehat{T}^{\mu\lambda}+\widehat{{\cal S}}^{\mu,\lambda\nu}.
\]
where $\widehat{{\cal S}}$ is the so-called \emph{spin tensor}, which
is anti-symmetric in the last two indices. Indeed, the stress-energy
and the spin tensor are not unique and can be changed by means of
a so-called pseudo-gauge transformation \citep{Hehl:1976vr,Leader:2013jra}:
\begin{eqnarray}
 &  & \widehat{T}^{\prime\mu\nu}=\widehat{T}^{\mu\nu}+\frac{1}{2}\nabla_{\lambda}\left(\widehat{\Phi}^{\lambda,\mu\nu}-\widehat{\Phi}^{\mu,\lambda\nu}-\widehat{\Phi}^{\nu,\lambda\mu}\right),\nonumber \\
 &  & \widehat{{\cal S}}^{\prime\lambda,\mu\nu}=\widehat{{\cal S}}^{\lambda,\mu\nu}-\widehat{{\cal S}}^{\lambda,\mu\nu},
\end{eqnarray}
where $\widehat{\Phi}$ is a rank-three tensor field antisymmetric
in the last two indices (often referred to as \emph{superpotential}).
In Minkowski space-time, the newly defined tensors preserve the total
energy, momentum, and angular momentum (herein expressed in Cartesian
coordinates): 
\begin{eqnarray}
\widehat{P}^{\nu}=\int_{\Sigma}\text{{\rm d}}\Sigma_{\mu}\widehat{T}^{\mu\nu},\qquad\widehat{J}^{\lambda\nu}=\int_{\Sigma}{\rm d}\Sigma_{\mu}\widehat{{\cal J}}^{\mu,\lambda\nu},\label{total}
\end{eqnarray}
as well as the conservation equations 
\begin{equation}
\nabla_{\mu}\widehat{T}^{\mu\nu}=0,\qquad\nabla_{\mu}\widehat{{\cal J}}_{C}^{\mu,\lambda\nu}=\widehat{T}^{\lambda\nu}-\widehat{T}^{\nu\lambda}+\nabla_{\mu}\widehat{{\cal S}}^{\mu,\lambda\nu}=0,\label{conserv}
\end{equation}
The pseudo-gauge freedom makes it possible to make the spin tensor
vanishing and, at the same time, to symmetrize the stress-energy tensor
operator; such choice is known as Belinfante pseudo-gauge or Belinfante
stress-energy tensor.

One of the main questions is whether the spin tensor has some physical
meaning in Minkowski space-time or, in other words, whether the pseudo-gauge
invariance can be broken by some measurement in flat space-time. This
question is similar to the better known gauge-independence in classical
electromagnetism, where only the fields have a physical meaning and
not the potentials. This issue has been the subject of investigations
\citep{Becattini:2018duy,Becattini:2020riu} and the conclusion was
that while operators cannot depend on the pseudo-gauge, quantum states
can and, particularly, the local equilibrium density operator \eqref{eq:LEStatOper}
is not pseudo-gauge invariant. Therefore, in principle, the mean spin
polarization vector or any other mean value is affected by the superpotential
and the particular spin tensor. For the spin polarization, the formula
including the spin potential associated to the spin tensor, for instance,
has been worked out in Ref.~\citep{Buzzegoli:2021wlg}.

These considerations have spurred the quest of extending conventional
hydrodynamics, where only the stress-energy tensor is conserved, to
include the conservation of angular momentum-boost current with a
generic spin tensor: 
\begin{eqnarray}
\partial_{\mu}\Theta^{\mu\nu} & = & 0,\nonumber \\
\partial_{\mu}J^{\mu\alpha\beta} & = & 0,\nonumber \\
\partial_{\mu}j^{\mu} & = & 0.\label{eq:total_conservation_01}
\end{eqnarray}
where: 
\begin{equation}
J^{\mu\alpha\beta}=x^{\alpha}\Theta^{\mu\beta}-x^{\beta}\Theta^{\mu\alpha}+S^{\mu\alpha\beta},\label{eq:decomposition_J}
\end{equation}
In the above equations the symbols are meant to be mean values of
quantum operators, e.g. $\Theta^{\mu\nu}\equiv{\rm Tr}(\widehat{\rho}\widehat{T}{}^{\mu\nu})$;
$\Theta^{\mu\nu}$ is the mean stress-energy tensor, $J^{\mu\alpha\beta}$
the mean total angular momentum current, $S^{\mu\alpha\beta}$ the
mean spin tensor and $j^{\mu}$ a mean abelian current such as e.g.
baryon number current.

Over the past few years, various approaches have been proposed to
construct the spin hydrodynamics, such as the entropy current analysis
\citep{Hattori:2019lfp,Fukushima:2020ucl,Li:2020eon,Gallegos:2021bzp,She:2021lhe,Hongo:2021ona,Cao:2022aku,Biswas:2023qsw},
the kinetic approach \citep{Florkowski:2017ruc,Florkowski:2017dyn,Florkowski:2018myy,Weickgenannt:2019dks,Bhadury:2020puc,Weickgenannt:2020aaf,Shi:2020htn,Speranza:2020ilk,Bhadury:2020cop,Singh:2020rht,Peng:2021ago,Sheng:2021kfc,Weickgenannt:2022zxs,Weickgenannt:2022jes,Weickgenannt:2022qvh},
the effective field theory \citep{Gallegos:2020otk,Garbiso:2020puw},
and the effective Lagrangian method \citep{Montenegro:2017rbu,Montenegro:2017lvf}.
For recent reviews, we refer the readers to Refs.~\citep{Wang:2017jpl,Becattini:2020ngo,Becattini:2022zvf,Gao:2020vbh,Liu:2020ymh}
and references therein.

As we have emphasized above, the spin tensor is not uniquely defined
and can be changed with pseudo-gauge transformations. Therefore, different
pseudo-gauge choices give rise to different forms of the spin hydrodynamics:
the canonical\citep{Hattori:2019lfp,Hongo:2021ona,Dey:2023hft}, Belinfante
\citep{Fukushima:2020ucl} , Hilgevoord-Wouthuysen (HW) \citep{Hilgevoord:1965,Weickgenannt:2022zxs}
, de Groot-van Leeuwen-van Weert (GLW) \citep{DeGroot:1980dk,Florkowski:2018fap}
forms. So far, no compelling theoretical argument has been found as
to which pseudo-gauge is physical and the debate is ongoing \citep{Becattini:2018duy,Speranza:2020ilk,Fukushima:2020ucl,Das:2021aar,Buzzegoli:2021wlg,Weickgenannt:2022jes,Daher:2022xon}.


Here, we discuss spin hydrodynamics in the canonical form, meaning
that we assume the spin tensor to be that obtained from the Lagrangian
by means of the Noether's theorem. The canonical energy-momentum tensor
can be further decomposed into a symmetric (s) and an anti-symmetric
(a) parts, 
\begin{equation}
\Theta^{\mu\nu}=\Theta_{(s)}^{\mu\nu}+\Theta_{(a)}^{\mu\nu}.\label{eq:decomposition_EMT_01}
\end{equation}
From Eqs. (\ref{eq:total_conservation_01}), (\ref{eq:decomposition_J})
and (\ref{eq:decomposition_EMT_01}) we obtain 
\begin{equation}
\partial_{\alpha}S^{\alpha\mu\nu}=-2\Theta_{(a)}^{\mu\nu}.
\end{equation}


Similar to the conventional hydrodynamics, the constitutive relations
for $\Theta^{\mu\nu}$ and $j^{\mu}$ can be written as: 
\begin{eqnarray}
\Theta^{\mu\nu} & = & (e+p)u^{\mu}u^{\nu}-pg^{\mu\nu}\nonumber \\
 &  & +h^{\mu}u^{\nu}+h^{\nu}u^{\mu}+\pi^{\mu\nu}\nonumber \\
 &  & +q^{\mu}u^{\nu}-q^{\nu}u^{\mu}+\phi^{\mu\nu},\nonumber \\
j^{\mu} & = & nu^{\mu}+\nu^{\mu},
\end{eqnarray}
where $e,p,n,u^{\mu}$ are the energy density, pressure, number density
and fluid velocity, respectively. The heat flow $h^{\mu}$, viscous
tensor $\pi^{\mu\nu}$ and diffusion current $\nu^{\mu}$ are conventional
dissipative quantities in the first order of the gradient expansion
which satisfy $u\cdot h=u\cdot\nu=u_{\mu}\pi^{\mu\nu}=0$, while the
first order terms $q^{\mu}$ and $\phi^{\mu\nu}$ are related to spin
degrees of freedom.

In general, not even the canonical spin tensor $S^{\lambda\mu\nu}$
is uniquely defined. For example, here are two equivalent Lagrangian
density for free Dirac fields, 
\begin{eqnarray}
\mathcal{L}_{1} & = & \overline{\psi}(i\gamma\cdot\partial-m)\psi,\nonumber \\
\mathcal{L}_{2} & = & \frac{1}{2}\overline{\psi}(i\gamma\cdot\overrightarrow{\partial}-m)\psi-\frac{1}{2}\overline{\psi}(i\gamma\cdot\overleftarrow{\partial}-m)\psi.
\end{eqnarray}
By applying Noether's theorem, the corresponding rank-$3$ spin tensor
operators are 
\begin{eqnarray}
\hat{S}_{1}^{\lambda\mu\nu} & = & \frac{1}{4}\overline{\psi}i\gamma^{\lambda}[\gamma^{\mu},\gamma^{\nu}]\psi,\\
\hat{S}_{2}^{\lambda\mu\nu} & = & \frac{1}{8}\overline{\psi}i\{\gamma^{\lambda},[\gamma^{\mu},\gamma^{\nu}]\}\psi.
\end{eqnarray}
where $\hat{S}_{1}^{\lambda\mu\nu}$ is antisymmetric only with respect
to $\mu$ and $\nu$, while $\hat{S}_{2}^{\lambda\mu\nu}$ is antisymmetric
for all three indices. In principle, one can derive the expectation
value $\hat{S}_{1,2}^{\lambda\mu\nu}$ in kinetic theory \citep{Florkowski:2018fap,Bhadury:2022ulr,Weickgenannt:2022zxs,Weickgenannt:2022jes,Weickgenannt:2022qvh}
or statistical field theory \citep{Becattini:2012tc,Becattini:2018duy,Becattini:2020sww}.
An alternative method to derive $S^{\lambda\mu\nu}$ is to map the
tensor structure of hydrodynamical variables to operators mentioned
above, as, e.g., in Refs.\citep{Hattori:2019lfp,Fukushima:2020ucl,Hongo:2021ona}.
Due to the symmetry, we have 
\begin{eqnarray}
S_{1}^{\alpha\mu\nu} & = & u^{\alpha}S^{\mu\nu}+\Sigma^{\alpha\mu\nu},\nonumber \\
S_{2}^{\alpha\mu\nu} & = & u^{\alpha}S^{\mu\nu}+u^{\mu}S^{\nu\alpha}+u^{\nu}S^{\alpha\mu}+\Sigma^{\alpha\mu\nu},
\end{eqnarray}
where the rank-$2$ tensor $S^{\mu\nu}$ denotes the spin density
and the tensor $\Sigma^{\alpha\mu\nu}$ satisfies $u_{\alpha}\Sigma^{\alpha\mu\nu}=0$.
In this review, we follow Refs. \citep{Hattori:2019lfp,Fukushima:2020ucl}
and adopt the spin tensor $S_{1}^{\lambda\mu\nu}$. For convenience,
we will omit the subscript "1" and set $S^{\lambda\mu\nu}=S_{1}^{\lambda\mu\nu}$.
For other choices, see Refs. \citep{Speranza:2020ilk,Hongo:2021ona,Daher:2022xon}
and references therein.


The spin density $S^{\mu\nu}$ is not a conserved quantity, but we
can assume that the decay of $S^{\mu\nu}$ is as slow as the characteristic
time scale of conventional hydrodynamics. The analytic solutions to
spin hydrodynamics in both Bjorken and Gubser flows are consistent
with this assumption. In this sense, the spin density $S^{\mu\nu}$
plays the same role as the number density $n$. Then the Gibbs-Duhem
relation in spin hydrodynamics can be generalized to 
\begin{equation}
e+p=Ts+\mu n+\omega_{\mu\nu}S^{\mu\nu},\label{eq:thermo_relation_01}
\end{equation}
where $T$, $s$ and $\mu$ are the temperature, entropy density and
chemical potential, respectively, and $\omega_{\mu\nu}$ is defined
as the spin potential. The corresponding thermodynamical relations
read \footnote{It should be pointed out that recently these relations have been studied
within a quantum statistical framework and found not to hold in general
\citep{Becattini:2023ouz}.}, 
\begin{eqnarray}
de & = & Tds+\mu dn+\omega_{\mu\nu}dS^{\mu\nu},\nonumber \\
dp & = & sdT+nd\mu+S^{\mu\nu}d\omega_{\mu\nu}.
\end{eqnarray}


Now let us briefly discuss the power counting scheme. The spin density
$S^{\mu\nu}$ can be in the same order as the number density $n$,
which corresponds to the case that a large part of particles are polarized
in the system. The correction $\omega_{\mu\nu}S^{\mu\nu}$ in Eq.
(\ref{eq:thermo_relation_01}) must be quantum and at the next-to-leading
order in the space-time gradient. Therefore we assume 
\begin{equation}
S^{\mu\nu}\sim O(1),\ \omega_{\mu\nu}\sim O(\partial),\ \Sigma^{\lambda\mu\nu}\sim O(\partial).\label{eq:PowerCounting-1}
\end{equation}
In contrast, a different power counting scheme is chosen in Refs.\citep{Hattori:2019lfp,Hongo:2021ona}:
$S^{\mu\nu}\sim O(\partial)$, $\omega_{\mu\nu}\sim O(\partial)$,
$\Sigma^{\lambda\mu\nu}\sim O(\partial^{2})$.


By taking $u_{\nu}\partial_{\mu}\Theta^{\mu\nu}=\mu\partial\cdot j$,
it is straightforward to get the entropy production rate \citep{Hattori:2019lfp,Fukushima:2020ucl},
\begin{eqnarray}
\partial_{\mu}\mathcal{S}_{\textrm{can}}^{\mu} & = & \left(h^{\mu}-\frac{e+p}{n}\nu^{\mu}\right)\left[\partial_{\mu}\frac{1}{T}+\frac{1}{T}(u\cdot\partial)u_{\mu}\right]\nonumber \\
 &  & +\frac{\pi^{\mu\nu}}{T}\partial_{\mu}u_{\nu}+\frac{1}{T}\phi^{\mu\nu}(\partial_{\mu}u_{\nu}+2\omega_{\mu\nu})\nonumber \\
 &  & +\frac{q^{\mu}}{T}\left[T\partial_{\mu}\frac{1}{T}-(u\cdot\partial)u_{\mu}+4\omega_{\mu\nu}u^{\nu}\right]+O(\partial^{3}),\label{eq:entropy_flow_01}
\end{eqnarray}
where $\mathcal{S}_{\textrm{can}}^{\mu}$ is the entropy current density.
The second law of thermodynamics $\partial_{\mu}\mathcal{S}_{\textrm{can}}^{\mu}\geq0$
leads to the first order constitutive relations \citep{Hattori:2019lfp,Fukushima:2020ucl},
\begin{eqnarray}
h^{\mu}-\frac{e+p}{n}\nu^{\mu} & = & \kappa\Delta^{\mu\nu}\left[\frac{1}{T}\partial_{\nu}T-(u\cdot\partial)u_{\nu}\right],\label{eq:SixHeatCurrent}\\
\pi^{\mu\nu} & = & 2\eta\partial^{<\mu}u^{\nu>}-\zeta\partial_{\mu}u^{\mu},\label{eq:SixTau}\\
q^{\mu} & = & \lambda\Delta^{\mu\nu}\left[\frac{1}{T}\partial_{\nu}T+(u\cdot\partial)u_{\nu}-4\omega_{\nu\alpha}u^{\alpha}\right],\label{eq:SixQ}\\
\phi^{\mu\nu} & = & 2\gamma_{s}T\Delta^{\mu\rho}\Delta^{\nu\sigma}\left[\partial_{\rho}\left(\frac{u_{\sigma}}{T}\right)-\partial_{\sigma}\left(\frac{u_{\rho}}{T}\right)+2\frac{\omega_{\rho\sigma}}{T}\right],\label{eq:SixPhi}
\end{eqnarray}
where the heat conductivity $\kappa$, shear viscosity $\eta$, and
bulk viscosity $\zeta$ also exist in conventional hydrodynamics,
but $\lambda$ and $\gamma_{s}$ are new coefficients corresponding
to the coupling of the spin and orbital angular momentum. All these
coefficients are positive quantities 
\begin{equation}
\kappa,\eta,\zeta,\lambda,\gamma_{s}>0.\label{eq:coefficients_01}
\end{equation}
As the system approaches a global equilibrium, we have $\partial_{\mu}\mathcal{S}_{\textrm{can}}^{\mu}=0$
in Eq.(\ref{eq:entropy_flow_01}), leading to the well-known killing
conditions \citep{Becattini:2012tc,Becattini:2018duy}, 
\begin{equation}
\partial_{\mu}\left(\frac{u^{\nu}}{T}\right)+\partial_{\nu}\left(\frac{u^{\mu}}{T}\right)=0,\label{eq:Killing_condition_01}
\end{equation}
and 
\begin{equation}
\omega_{\rho\sigma}=-\frac{1}{2}T\Delta^{\mu\rho}\Delta^{\nu\sigma}\left[\partial_{\rho}\left(\frac{u_{\sigma}}{T}\right)-\partial_{\sigma}\left(\frac{u_{\rho}}{T}\right)\right]\equiv-\frac{1}{2}T\Omega^{\mu\nu}.\label{eq:Killing_condition_02}
\end{equation}
The condition in Eq. (\ref{eq:Killing_condition_01}) agrees with
the one in conventional hydrodynamics, which gives the general solution
to the fluid velocity at global equilibrium, $u^{\mu}=T(b^{\mu}+a^{\mu\nu}x_{\nu})$,
where $b^{\mu}$ and the anti-symmetric tensor $a^{\mu\nu}$ are constants.
The condition in Eq. (\ref{eq:Killing_condition_02}) tells us that
the spin potential is proportional to the thermal vorticity tensor
$\Omega^{\mu\nu}$ in the global equilibrium, consistent with the
analysis from quantum statistic theory \citep{Becattini:2012tc,Becattini:2018duy}.
We also notice that in the global equilibrium we have $q^{\mu},\phi^{\mu\nu}=0$,
so $\Theta^{\mu\nu}$ is symmetric. In general, cross terms between
different dissipative currents may also exist due to Onsager's relation
\citep{Cao:2022aku,Hu:2022azy} which are neglected for simplicity.
One can use acceleration equations for the fluid velocity in the leading
order, $(u\cdot\partial)u^{\mu}=(1/T)\Delta^{\mu\nu}\partial_{\nu}T+O(\partial^{2})$,
to rewrite $q^{\mu}$ in Eq. (\ref{eq:SixQ}) as \citep{Hattori:2019lfp,Sarwar:2022yzs,Daher:2022wzf}
\begin{equation}
q^{\mu}=\lambda\left(\frac{2\Delta^{\mu\nu}\partial_{\nu}p}{e+p}-4\omega^{\mu\nu}u_{\nu}\right)+O(\partial^{2}).\label{eq:RewriteSixQ}
\end{equation}
The entropy production rate of spin hydrodynamics can also be derived
by quantum statistical theory \citep{Becattini:2023ouz}.


We can take the linear mode analysis \citep{Hiscock:1985zz,Hiscock:1987zz,Denicol:2008ha,Pu:2009fj}
for the first-order spin hydrodynamics with constitutive relations
(\ref{eq:SixHeatCurrent})-(\ref{eq:RewriteSixQ}). The linear mode
analysis can give causality conditions. It is found that the first-order
spin hydrodynamics is acausal and unstable \citep{Xie:2023gbo}. There
are two ways to construct causal hydrodynamics. One way is to add
the second order corrections to the dissipative terms, e.g. the Müller-Israel-Stewart
(MIS) theory \citep{Israel:1979,Israel:1979wp} or extended MIS theory
\citep{Baier:2007ix}. Recently, the second-order spin hydrodynamics
similar to MIS theory has been introduced \citep{Biswas:2023qsw}
by using the entropy principle. Another way is called Bemfica-Disconzi-Noronha-Kovtun
(BDNK) theory \citep{Bemfica:2017wps,Bemfica:2019knx,Kovtun:2019hdm,Hoult:2020eho,Bemfica:2020zjp,Hoult:2021gnb},
which is a first-order casual hydrodynamical theory in general (fluid)
frames. The analysis for the casual spin hydrodynamics in the first
order similar to BDNK theory can be found in Ref. \citep{Weickgenannt:2023btk}.
Here we concentrate on the minimal causal spin hydrodynamics proposed
in Refs. \citep{Liu:2020ymh,Xie:2023gbo} with $h^{\mu}=\nu^{\mu}=0$.
Then Eqs. (\ref{eq:SixQ}) and (\ref{eq:SixPhi}) can be extended
as, 
\begin{eqnarray}
\tau_{q}\Delta^{\mu\nu}\frac{d}{d\tau}q_{\nu}+q^{\mu} & = & \lambda[T^{-1}\Delta^{\mu\alpha}\partial_{\alpha}T\nonumber \\
 &  & +(u\cdot\partial)u^{\mu}-4\omega^{\mu\nu}u_{\nu}],\label{eq:Type1q}\\
\tau_{\phi}\Delta^{\mu\alpha}\Delta^{\nu\beta}\frac{d}{d\tau}\phi_{\alpha\beta}+\phi^{\mu\nu} & = & 2\gamma_{s}T\Delta^{\mu\rho}\Delta^{\nu\sigma}\nonumber \\
 &  & \times\left[\partial_{\rho}\left(\frac{u_{\sigma}}{T}\right)-\partial_{\sigma}\left(\frac{u_{\rho}}{T}\right)+2\frac{\omega_{\rho\sigma}}{T}\right]\,,\label{eq:Type1phi}
\end{eqnarray}
where $\tau_{q}$ and $\tau_{\phi}$ are positive relaxation times
for $q^{\mu}$ and $\phi^{\mu\nu}$ respectively. It is found that
stability conditions derived in the small and large wavelength limits
cannot guarantee the stability of the system for finite wavelength
\citep{Xie:2023gbo}. Therefore, the linear stability of the minimal
causal spin hydrodynamics remains uncertain. The analysis beyond linear
modes may provide an answer to the stability problem in general cases.


Let us briefly discuss analytic solutions to the spin hydrodynamics.
In conventional relativistic hydrodynamics, the Bjorken's \citep{Bjorken:1982qr}
and Gubser's solutions\citep{Gubser:2010ui,Gubser:2010ze} are widely
used. In the Bjorken's flow, the fluid velocity is given by $u_{\textrm{Bjorken}}^{\mu}=\left(t/\tau,0,0,z/\tau\right)$
with $\tau=\sqrt{t^{2}-z^{2}}$ being the proper time. The system
in the Bjokren's flow is assumed to be homogeneous in transverse plane,
and therefore, all macroscopic quantities depend on the proper time
$\tau$ only. For the spin hydrodynamics, we consider the simplest
equations of states for the relativistic fluid, $e=3p$ and $S^{\mu\nu}=a_{1}T^{2}\omega^{\mu\nu}$
with $a_{1}$ being a constant. The former is the equation of state
for the ideal gas, while the latter is in analogy with the equation
of state for the number density as the function of the chemical potential
in the high temperature limit. As the initial time, we assume that
the fluid velocity is in the Bjorken form $u_{\textrm{Bjorken}}^{\mu}$
and all thermodynamic quantities are functions of $\tau$ only. Then
we search for those configurations of $\omega^{\mu\nu}$ that can
keep the initial Bjorken velocity $u_{\textrm{Bjorken}}^{\mu}$ unchanged.
To keep the whole system boost invariant in later time, only $\omega^{xy}$
is allowed to be nonzero at the initial time. Eventually, the analytic
solution for the spin density is \citep{Wang:2021ngp} 
\begin{equation}
S^{xy}(\tau)=a_{1}\omega_{0}^{xy}T_{0}^{2}\left(\frac{\tau_{0}}{\tau}\right)\exp\left[-\frac{2\gamma\tau_{0}}{a_{1}T_{0}^{3}}\left(\frac{\tau^{2}}{\tau_{0}^{2}}-1\right)\right]+...,\label{eq:S_xy_01}
\end{equation}
where the subscript ``0" stands for the quantities at the initial
time $\tau_{0}$, and ``..." stands for corrections from viscous
tensors and other second order terms. We notice that the typical time
behavior for $S^{xy}$ is $\sim\tau^{-1}$ similar to that for the
number density in the Bjorken flow. Therefore, the assumption that
the spin density can be approximated as a hydrodynamical variable
holds. The discussion for the analytic solutions to the spin hydrodynamics
in the Gubser's flow can be found in Ref. \citep{Wang:2021wqq}. We
also refer the readers to the studies of the spin hydrodynamics in
expanding backgrounds with the Bjorken's \citep{Florkowski:2019qdp,Singh:2021man}
and Gubser's \citep{Singh:2020rht} flows.


We now briefly discuss the spin hydrodynamic in Belinfante's form.
As we mentioned, the choice of the angular momentum and energy momentum
tensor is not unique and subject to the pseudo-gauge transformation
characterized by an arbitrary tensor $K^{\lambda\mu\nu}=-K^{\mu\lambda\nu}$,
\begin{eqnarray}
J^{\mu\nu\alpha} & = & J^{\mu\nu\alpha}+\partial_{\rho}(x^{\nu}K^{\rho\mu\alpha}-x^{\alpha}K^{\rho\mu\nu})\,,\nonumber \\
T^{\mu\nu} & = & \Theta^{\mu\nu}+\partial_{\lambda}K^{\lambda\mu\nu}\,.\label{eq:EM_Belinfant}
\end{eqnarray}
In Belinfante's form, $K^{\lambda\mu\nu}$ is chosen as 
\begin{equation}
K_{\textrm{Bel}}^{\lambda\mu\nu}=\frac{1}{2}(S^{\lambda\mu\nu}-S^{\mu\lambda\nu}+S^{\nu\mu\lambda})\,,\label{eq:K_01}
\end{equation}
which leads to 
\begin{equation}
J_{\textrm{Bel}}^{\mu\nu\alpha}=x^{\nu}T_{\textrm{Bel}}^{\mu\alpha}-x^{\alpha}T_{\textrm{Bel}}^{\mu\nu}\,.\label{eq:J_Bel}
\end{equation}
We see that the angular momentum tensor in Eq. (\ref{eq:J_Bel}) does
not contain the spin tensor. Inserting Eq. (\ref{eq:K_01}) into Eq.
(\ref{eq:EM_Belinfant}), we obtain \citep{Fukushima:2020ucl}, 
\begin{eqnarray}
T_{\textrm{Bel}}^{\mu\nu} & = & (e+p+\delta e)u^{\mu}u^{\nu}-pg^{\mu\nu}\nonumber \\
 &  & +(h^{\mu}+\delta h^{\mu})u^{\nu}+(h^{\nu}+\delta h^{\nu})u^{\mu}+\pi^{\mu\nu}+\delta\pi^{\mu\nu}\,,\label{em-tensor-belinfangte}
\end{eqnarray}
where 
\begin{eqnarray}
\delta e & = & u_{\rho}\partial_{\sigma}S^{\rho\sigma}\,,\nonumber \\
\delta h^{\mu} & = & \frac{1}{2}\Delta_{\sigma}^{\mu}\partial_{\lambda}S^{\sigma\lambda}+\frac{1}{2}u_{\rho}S^{\rho\lambda}\partial_{\lambda}u^{\mu}\,,\nonumber \\
\delta\pi^{\mu\nu} & = & \partial_{\lambda}(u^{<\mu}S^{\nu>\lambda})-\frac{1}{3}\partial_{\lambda}(u^{\sigma}S^{\rho\lambda})\Delta_{\rho\sigma}\,,
\end{eqnarray}
are the corrections from the spin part to conventional hydrodynamical
variables. We see in Eq. (\ref{em-tensor-belinfangte}) that $T^{\mu\nu}$
becomes symmetric in Belinfante's form.





\section{Spin alignment of vector mesons}

\label{sec:spin-alignment}

The spin alignment of vector mesons (SAV) in the direction of the
global orbital angular momentum in heavy-ion collisions was first
proposed by Liang and Wang in 2005 \citep{Liang:2004xn}. The SAV
is characterized by a deviation of $\rho_{00}$, the 00-component
of the spin density matrix, from 1/3. The first attempt to measure
the global SAV was made by the STAR collaboration in 2008 but failed
to find non-vanishing signals within statistical errors \citep{STAR:2008lcm}.
The ALICE collaboration at the Large Hadron Collider measured the
global SAV of $K^{*0}$ and $\phi$ vector mesons at the collision
energy 2.76 TeV \citep{ALICE:2019aid}. It was found that $\rho_{00}<1/3$
for $K^{*0}$ and $\phi$ at low transverse momenta at the level of
$3\sigma$ and $2\sigma$ respectively. Recently the STAR collaboration
finally measured non-vanishing global SAV of $\phi$ mesons at collision
energies from 11.5 GeV to 62.4 GeV with significant deviations of
$\rho_{00}$ from 1/3 \citep{STAR:2022fan}. In contrast, the values
of $\rho_{00}$ for $K^{*0}$ are consistent with 1/3.

In this section, we will give an overview on theoretical models for
the SAV and how to understand experimental measurements by these models.
These models are based on quantum kinetic or transport theory for
spin transport processes \citep{Gao:2019znl,Weickgenannt:2019dks,Weickgenannt:2020aaf,Hattori:2019ahi,Yang:2020hri,Liu:2020flb,Weickgenannt:2021cuo,Sheng:2021kfc,Wang:2019moi,Fang:2022ttm,Fang:2023bbw}
in terms of Wigner functions. For recent reviews on quantum kinetic
or transport theory, see, e.g., Refs. \citep{Gao:2020pfu,Hidaka:2022dmn}
, for recent reviews on SAV, see, e.g., Refs. \citep{Chen:2023hnb,Wang:2023fvy,Sheng:2023chinphyb}.

The definition of two-point Green's functions $G$ and $\Sigma$ in
this section differs by a factor $i=\sqrt{-1}$ from the usual one
in quantum field theory, which are related by $G=i\widetilde{G}$
and $\Sigma=i\widetilde{\Sigma}$.

\subsection{Spin density matrix and angular distribution of decay daughters}

The spin ensemble of a particle system can be described by the spin
density matrix 
\begin{equation}
\rho=\sum_{i}\mathcal{P}_{i}\left|\psi_{i}\right\rangle \left\langle \psi_{i}\right|,
\end{equation}
where $\left|\psi_{i}\right\rangle $ is the normalized spin state
of the particle and $\mathcal{P}_{i}$ is the probability on the spin
state satisfying $\sum_{i}\mathcal{P}_{i}=1$. The spin density matrix
for the spin-$S$ particle is a $(2S+1)\times(2S+1)$ Hermitian matrix
with positive eigenvalues and unity trace. From these conditions the
number of independent real variables is $4S(S+1)$. For examples,
the numbers of independent real variables for spin-1/2 and spin-1
particles are 3 and 8 respectively.

The spin density matrix for the spin-1/2 particle can be put into
the form 
\begin{equation}
\rho=\frac{1}{2}\left(1+\boldsymbol{P}\cdot\boldsymbol{\sigma}\right),
\end{equation}
where $\boldsymbol{\sigma}=(\sigma_{x},\sigma_{y},\sigma_{z})$ are
Pauli matrices and $\boldsymbol{P}=(P_{x},P_{y},P_{z})$ is the spin
polarization vector. So we confirm that the number of independent
real variables for the spin-1/2 particle is 3 represented by three
components of $\boldsymbol{P}$. The spin density matrix for the spin-1
particle is 
\begin{equation}
\rho=\left(\begin{array}{ccc}
\rho_{11} & \rho_{10} & \rho_{1,-1}\\
\rho_{01} & \rho_{00} & \rho_{0,-1}\\
\rho_{-1,1} & \rho_{-1,0} & \rho_{-1,-1}
\end{array}\right)=\left(\begin{array}{ccc}
\rho_{11} & \rho_{01}^{*} & \rho_{-1,1}^{*}\\
\rho_{01} & \rho_{00} & \rho_{-1,0}^{*}\\
\rho_{-1,1} & \rho_{-1,0} & \rho_{-1,-1}
\end{array}\right),\label{eq:sdm-vector-m-1}
\end{equation}
which satisfies $\ensuremath{\rho=\rho^{\dagger}}$ and $\ensuremath{\mathrm{Tr}\rho=1}$.
From these conditions one can immediately see that the diagonal elements
$\rho_{11}$, $\rho_{00}$ and $\rho_{-1,-1}$ are real parameters.
The spin density matrix (\ref{eq:sdm-vector-m-1}) can be decomposed
into a vector and a tensor part as 
\begin{equation}
\rho=\frac{1}{3}\left(1+\frac{3}{2}P_{i}\Sigma_{i}+3T_{ij}\Sigma_{ij}\right),\label{eq:spin-density}
\end{equation}
where $\Sigma_{i}$ ($i=1,2,3$ or $x,y,z$) are three spin matrices
with their corresponding coefficients $P_{i}$ representing three
components of the spin polarization vector $\boldsymbol{P}$, $\Sigma_{ij}=\Sigma_{ji}$
are five traceless matrices, and their coefficients $T_{ij}=T_{ji}$
form a traceless rank-2 tensor \citep{Sheng:2023chinphyb}. The matrices
$\Sigma_{i}$ are defined as 
\begin{equation}
\Sigma_{1}=\frac{1}{\sqrt{2}}\left(\begin{array}{ccc}
0 & 1 & 0\\
1 & 0 & 1\\
0 & 1 & 0
\end{array}\right),\ \ \Sigma_{2}=\frac{1}{\sqrt{2}}\left(\begin{array}{ccc}
0 & -i & 0\\
i & 0 & -i\\
0 & i & 0
\end{array}\right),\ \ \Sigma_{3}=\left(\begin{array}{ccc}
1 & 0 & 0\\
0 & 0 & 0\\
0 & 0 & -1
\end{array}\right).
\end{equation}
One can verify that the commutators of $\Sigma_{i}$ follow those
of angular momenta, $\left[\Sigma_{i},\Sigma_{j}\right]=i\epsilon_{ijk}\Sigma_{k}$.
The matrices $\Sigma_{ij}$ can be expressed by $\Sigma_{i}$ as 
\begin{equation}
\Sigma_{ij}=\frac{1}{2}(\Sigma_{i}\Sigma_{j}+\Sigma_{j}\Sigma_{i})-\frac{2}{3}\mathbf{1}\delta_{ij}.
\end{equation}
The coefficients $P_{i}$ can be extracted by $P_{i}=\mathrm{Tr}(\rho\boldsymbol{\Sigma}_{i})$
as 
\begin{align}
P_{x}= & \sqrt{2}\;\mathrm{Re}(\rho_{-1,0}+\rho_{01}),\nonumber \\
P_{y}= & \sqrt{2}\;\mathrm{Im}(\rho_{-1,0}+\rho_{01}),\nonumber \\
P_{z}= & \rho_{11}-\rho_{-1,-1},
\end{align}
using the property $\ensuremath{\mathrm{Tr}(\boldsymbol{\Sigma}_{i}\boldsymbol{\Sigma}_{jk})=0}$.
The coefficients $T_{ij}$ with $i\neq j$ can also be extracted in
the same way, $T_{ij}=\mathrm{Tr}(\rho\boldsymbol{\Sigma}_{ij})$,
\begin{align}
T_{12}= & \mathrm{Im}\rho_{-1,1},\nonumber \\
T_{23}= & \frac{1}{\sqrt{2}}\;\mathrm{Im}(\rho_{01}-\rho_{-1,0}),\nonumber \\
T_{31}= & \frac{1}{\sqrt{2}}\;\mathrm{Re}(\rho_{01}-\rho_{-1,0}).
\end{align}
But it is not possible to extract $T_{11}$, $T_{22}$ and $T_{33}$
in the same way since $\ensuremath{\mathrm{Tr}(\boldsymbol{\Sigma}_{ii}\boldsymbol{\Sigma}_{jj})\neq0}$.
They can only be determined directly from $\rho_{-1,1}$, $\rho_{11}+\rho_{-1,-1}$
and the traceless condition for $T_{ij}$. The results read 
\begin{align}
T_{11}= & \frac{1}{2}\left(\rho_{00}-\frac{1}{3}\right)+\mathrm{Re}\rho_{-1,1},\nonumber \\
T_{22}= & \frac{1}{2}\left(\rho_{00}-\frac{1}{3}\right)-\mathrm{Re}\rho_{-1,1},\nonumber \\
T_{33}= & -\left(\rho_{00}-\frac{1}{3}\right).
\end{align}

It is known that $K^{*0}$ and $\phi$ vector mesons decay mainly
into pseudoscalar mesons through strong interaction which respect
parity symmetry 
\begin{align}
K^{*0}\rightarrow & K^{+}+\pi^{-},\ \ (\sim100\%),\nonumber \\
\phi\rightarrow & K^{+}+K^{-},\ \ (\sim49\%),\label{eq:strong-decays}
\end{align}
where the percentages inside brackets are branching ratios. The lifetime
of $K^{*0}$ mesons is about 4 fm/c while that of $\phi$ mesons is
about 45 fm/c, so most of $K^{*0}$ mesons decay inside the fireball
and suffer from in-medium effects such as rescattering and regeneration
\citep{Kumar:2015uxe}. In contrast, $\phi$ mesons are expected to
freeze-out early and may not suffer much from in-medium effects. In
decay channels (\ref{eq:strong-decays}), decay mothers are spin-1
particles while decay daughters are scalar particles, these decays
are in P-wave with the orbital angular momentum $L=1$. The decay
amplitude of the $\phi$ meson, for example, can be put into the form
\begin{equation}
\left\langle K^{+},K^{-}\right|\mathcal{M}\left|\phi;S_{z}\right\rangle =Y_{1,S_{z}}(\theta,\varphi),
\end{equation}
where $S_{z}=0,\pm1$ is the spin quantum number in the spin quantization
direction $z$, $Y_{1,S_{z}}(\theta,\varphi)$ is the spherical harmonic
function $Y_{LM}$ with $L=1$ and $M=S_{z}$, and $(\theta,\varphi)$
denotes the polar and azimuthal angle of one decay daughter $K^{+}$
or $K^{-}$ in the rest frame of the $\phi$ meson. Suppose that the
$\phi$ meson is in the spin state $S_{z}$ with the probability $\mathcal{P}_{S_{z}}$,
the angular distribution of the daughter particle can be written as
\citep{Yang:2017sdk}
\begin{align}
\frac{dN}{d\Omega}= & \sum_{S_{z}}\mathcal{P}_{S_{z}}\left|\left\langle K^{+},K^{-}\right|\mathcal{M}\left|\phi;S_{z}\right\rangle \right|^{2}\nonumber \\
\rightarrow & \sum_{S_{z1},S_{z2}}\left\langle K^{+},K^{-}\right|\mathcal{M}\left|\phi;S_{z1}\right\rangle \left\langle \phi;S_{z1}\right|\rho\left|\phi;S_{z2}\right\rangle \left\langle \phi;S_{z2}\right|\mathcal{M}^{\dagger}\left|K^{+},K^{-}\right\rangle \nonumber \\
= & \sum_{S_{z1},S_{z2}}\rho_{S_{z1},S_{z2}}Y_{1,S_{z1}}(\theta,\varphi)Y_{1,S_{z2}}^{*}(\theta,\varphi),\label{eq:dn-d-omega}
\end{align}
where $\rho_{S_{z1},S_{z2}}=\left\langle \phi;S_{z1}\right|\rho\left|\phi;S_{z2}\right\rangle $
is the spin density matrix. Here we have generalized the expression
with the probability $\mathcal{P}_{S_{z}}$ to that with the spin
density matrix $\rho_{S_{z1},S_{z2}}$. With $Y_{1,\pm1}=\mp\sqrt{3/8\pi}\sin\theta e^{\pm i\varphi}$
and $Y_{10}=\sqrt{3/4\pi}\cos\theta$, we obtain the explicit form
of Eq. (\ref{eq:dn-d-omega}) 
\begin{align}
\frac{dN}{d\Omega}= & \frac{3}{8\pi}\left[(1-\rho_{00})+(3\rho_{00}-1)\cos^{2}\theta\right.\nonumber \\
 & -(T_{11}-T_{22})\sin^{2}\theta\cos(2\varphi)-2T_{12}\sin^{2}\theta\sin(2\varphi)\nonumber \\
 & \left.-2T_{31}\sin(2\theta)\cos\varphi-2T_{23}\sin(2\theta)\sin\varphi\right],
\end{align}
which is a normalized distribution satisfying $\int d\Omega(dN/d\Omega)=1$.
We see that the angular distribution only depends on the tensor part
of the spin density matrix, which is the consequence of the parity
symmetry in strong interaction. So five parameters of $T_{ij}$ can
be measured via the angular distribution of the daughter particle.
Due to statistics in experiments, it is impossible to measure all
five parameters. Instead, by integrating over the azimuthal angle,
one is left with the polar angle distribution 
\begin{equation}
\left.\frac{dN}{d\cos\theta}\right|_{V\rightarrow\mathrm{scalars}}=\frac{3}{4}\left[(1-\rho_{00})+(3\rho_{00}-1)\cos^{2}\theta\right],\label{eq:scalar-decay}
\end{equation}
which depends on $\rho_{00}$ only. If $\rho_{00}=1/3$, the polar
angle distribution is constant, implying that three spin states are
equally populated. If $\rho_{00}\neq1/3$, the polar angle distribution
has a $\cos^{2}\theta$ dependence, implying that three spin states
are not equally populated and there is a spin alignment along the
spin quantization direction. For $\rho_{00}\gtrless1/3$, the $S_{z}=0$
state is occupied with more/less probability and the polar angle distribution
is donut-like/peanut-like. Therefore the spin alignment of the vector
meson can be measured via the polar angle distribution of its daughter
particle in its rest frame.

The vector meson can also have di-lepton decays to $e^{+}e^{-}$ or
$\mu^{+}\mu^{-}$ through electromagnetic interaction. Unlike its
decay to pseudoscalar mesons, the daughters in di-lepton decays are
spin-1/2 particles. The helicity amplitude is normally used to express
the angular distribution of the decay daughter with spin in a general
two-body decay $A\rightarrow1+2$. The decay amplitude can be written
as 
\begin{equation}
\mathcal{M}(S_{z},\lambda_{1},\lambda_{2})=\left\langle \mathbf{p},\lambda_{1},\lambda_{2}\right|\mathcal{M}\left|S,S_{z}\right\rangle .\label{eq:amp}
\end{equation}
Here $\left|S,S_{z}\right\rangle $ is the spin state of $A$ with
$S$ and $S_{z}$ being the spin quantum number and that in the spin
quantization direction $z$, and $\left|\mathbf{p},\lambda_{1},\lambda_{2}\right\rangle $
is the helicity state of daughter particles, where $\lambda_{1}$
and $\lambda_{2}$ are the helicity of particle 1 and 2 respectively,
$\mathbf{p}=|\mathbf{p}|\widehat{\mathbf{p}}$ is the momentum of
particle 1 in the rest frame of the mother particle, and $\widehat{\mathbf{p}}$
is the unit vector of the direction $(\theta,\varphi)$. Then the
angular distribution of the daughter particle can be put into the
form 
\begin{align}
\frac{dN}{d\Omega}\propto & \sum_{\lambda_{1},\lambda_{2}}\sum_{S_{z1},S_{z2}}\rho_{S_{z1},S_{z2}}\mathcal{M}(S_{z1},\lambda_{1},\lambda_{2})\mathcal{M}^{*}(S_{z2},\lambda_{1},\lambda_{2})\nonumber \\
= & \frac{2S+1}{4\pi}\sum_{\lambda_{1},\lambda_{2}}\left|H(S;\lambda_{1},\lambda_{2})\right|^{2}\nonumber \\
 & \times\sum_{S_{z1},S_{z2}}\rho_{S_{z1},S_{z2}}D_{S_{z1},\lambda_{1}-\lambda_{2}}^{(S)*}(R)D_{S_{z2},\lambda_{1}-\lambda_{2}}^{(S)}(R),\label{eq:angular-dist-1}
\end{align}
where $D_{S_{z1},S_{z2}}^{(S)}(R)$ is the Wigner rotation matrix
with $S_{z}=-S,-S+1,\cdots,S$, $R$ represents the rotation from
the spin quantization direction $z$ to the direction $\widehat{\mathbf{p}}$
as a function of Euler angles $R(\alpha,\beta,\gamma)=R(\varphi,\theta,-\varphi)$,
and $H(S;\lambda_{1},\lambda_{2})$ denotes the helicity amplitude
of the decay.

We can apply Eq. (\ref{eq:angular-dist-1}) to $J/\psi\rightarrow l^{+}l^{-}$
with $l$ being $e$ or $\mu$. In the massless limit of the lepton,
we have $\lambda_{1}=-\lambda_{2}=\pm1/2$ and $\lambda_{1}-\lambda_{2}=\pm1$.
So Eq. (\ref{eq:angular-dist-1}) can be written as 
\begin{align}
\frac{dN}{d\Omega}\propto & \frac{3}{4\pi}\left|H(1;1/2,-1/2)\right|^{2}\sum_{S_{z1},S_{z2}}\rho_{S_{z1},S_{z2}}D_{S_{z1},1}^{(1)*}(R)D_{S_{z2},1}^{(1)}(R)\nonumber \\
 & +\frac{3}{4\pi}\left|H(1;-1/2,1/2)\right|^{2}\sum_{S_{z1},S_{z2}}\rho_{S_{z1},S_{z2}}D_{S_{z1},-1}^{(1)*}(R)D_{S_{z2},-1}^{(1)}(R)\nonumber \\
= & \frac{3}{4\pi}\left|H(S_{A};1/2,-1/2)\right|^{2}\sum_{S_{z1},S_{z2}}\rho_{S_{z1},S_{z2}}e^{i(S_{z1}-S_{z2})\varphi}\nonumber \\
 & \times\left[d_{S_{z1},1}^{(1)}(\theta)d_{S_{z2},1}^{(1)}(\theta)+d_{S_{z1},-1}^{(1)}(\theta)d_{S_{z2},-1}^{(1)}(\theta)\right],
\end{align}
where we have used one property of helicity amplitude $\left|H(S;\lambda_{1},\lambda_{2})\right|^{2}=\left|H(S;-\lambda_{1},-\lambda_{2})\right|^{2}$,
and the Wigner rotation matrix for $S=1$, 
\begin{equation}
D_{S_{z1}S_{z2}}^{(1)}(R)=e^{-i(S_{z1}-S_{z2})\varphi}d_{S_{z1}S_{z2}}^{(1)}(\theta),
\end{equation}
with $d_{S_{z1}S_{z2}}^{(1)}(\theta)$ being given by 
\begin{equation}
d_{S_{z1}S_{z2}}^{(1)}(\theta)=\left(\begin{array}{ccc}
\frac{1+\cos\theta}{2} & -\frac{\sin\theta}{\sqrt{2}} & \frac{1-\cos\theta}{2}\\
\frac{\sin\theta}{\sqrt{2}} & \cos\theta & -\frac{\sin\theta}{\sqrt{2}}\\
\frac{1-\cos\theta}{2} & \frac{\sin\theta}{\sqrt{2}} & \frac{1+\cos\theta}{2}
\end{array}\right),
\end{equation}
where the order of $S_{z1}$ is $(1,0,-1)$ fro the first row/column
to the third row/column. Then the polar angle distribution has the
explicit form 
\begin{equation}
\left.\frac{dN}{d\cos\theta}\right|_{V\rightarrow\mathrm{dilepton}}=\frac{3}{8}\left[(1+\rho_{00})+(1-3\rho_{00})\cos^{2}\theta\right].\label{eq:dilepton}
\end{equation}
One can compare the above distribution for the di-lepton decay with
that for the pseudoscalar meson decay in Eq. (\ref{eq:scalar-decay}):
the main difference is that the coefficient of $\cos^{2}\theta$ has
an opposite sign.

\subsection{Green's functions for vector mesons in CTP formalism}

\label{sec:green-functions}The Lagrangian density for unflavored
vector mesons with spin-1 and mass $m_{V}$ reads 
\begin{eqnarray}
\mathcal{L} & = & -\frac{1}{4}F_{\mu\nu}F^{\mu\nu}+\frac{m_{V}^{2}}{2}A_{\mu}A^{\mu}-A_{\mu}j^{\mu}.\label{eq:Lagrangian density}
\end{eqnarray}
where $A^{\mu}(x)$ is the real vector field for the meson, $F_{\mu\nu}=\partial_{\mu}A_{\nu}-\partial_{\nu}A_{\mu}$
is the field strength tensor, and $j^{\mu}$ is the source coupled
to $A^{\mu}(x)$. The vector field satisfies the Proca equation 
\begin{equation}
\left(\partial^{2}+m_{V}^{2}\right)A^{\mu}(x)-\partial^{\mu}\partial_{\nu}A^{\nu}(x)=j^{\mu}(x),\label{eq:proca}
\end{equation}
following the Euler-Lagrange equation.


The quantized form of the vector field is 
\begin{eqnarray}
A^{\mu}(x) & = & \sum_{\lambda=0,\pm1}\int\frac{d^{3}p}{(2\pi)^{3}}\frac{1}{2E_{p}^{V}}\nonumber \\
 &  & \times\left[\epsilon^{\mu}(\lambda,{\bf p})a_{V}(\lambda,{\bf p})e^{-ip\cdot x}+\epsilon^{\mu\ast}(\lambda,{\bf p})a_{V}^{\dagger}(\lambda,{\bf p})e^{ip\cdot x}\right],\label{eq:a-quantization}
\end{eqnarray}
where $p^{\mu}=(E_{p}^{V},\mathbf{p})$ is the on-shell momentum of
the vector meson, $E_{p}^{V}=\sqrt{|{\bf p}|^{2}+m_{V}^{2}}$ is the
vector meson's energy, $\lambda$ denotes the spin state, $a(\lambda,{\bf p})$
and $a^{\dagger}(\lambda,{\bf p})$ are annihilation and creation
operators respectively, and $\epsilon^{\mu}(\lambda,{\bf p})$ represents
the polarization vector obeying the following relations 
\begin{eqnarray}
p^{\mu}\epsilon_{\mu}(\lambda,{\bf p}) & = & 0,\nonumber \\
\epsilon(\lambda,{\bf p})\cdot\epsilon^{*}(\lambda^{\prime},{\bf p}) & = & -\delta_{\lambda\lambda^{\prime}},\nonumber \\
\Sigma_{\lambda}\epsilon^{\mu}(\lambda,{\bf p})\epsilon^{\nu,*}(\lambda,{\bf p}) & = & -\left(g^{\mu\nu}-\frac{p^{\mu}p^{\nu}}{m_{V}^{2}}\right),
\end{eqnarray}
where $p^{\mu}$ is on-shell. One can check that the quantum field
$A^{\mu}$ defined in Eq. (\ref{eq:a-quantization}) is Hermitian,
$A^{\mu}=A^{\mu\dagger}$. The annihilation and creation operators
satisfy the commutator 
\begin{equation}
\left[a_{V}(\lambda,{\bf p}),a_{V}^{\dagger}(\lambda^{\prime},{\bf p}^{\prime})\right]=\delta_{\lambda\lambda^{\prime}}2E_{p}^{V}(2\pi)^{3}\delta^{(3)}({\bf p}-{\bf p}^{\prime}).
\end{equation}


The non-equilbrium phenomena are known as the initial value problems:
given the system's state at the initial time and find how the system
evolves in later time. The elegant tool to solve these problems is
the closed-time-path (CTP) formalism invented by Schwinger and later
developed by Mahanthappa \citep{Mahanthappa:1962ex,Bakshi:1962dv}
and Keldysh \citep{Keldysh:1964ud}, see e.g. Refs.\citep{Chou:1984es,Blaizot:2001nr,Berges:2004yj,Cassing:2008nn,Crossley:2015evo}
for reviews on the CTP formalism.


The two-point Green's function on the CTP is defined as 
\begin{eqnarray}
G_{\mathrm{CTP}}^{\mu\nu}(x_{1},x_{2}) & = & \left\langle T_{C}A^{\mu}(x_{1})A^{\nu\dagger}(x_{2})\right\rangle ,\label{eq:green-fun-ctp}
\end{eqnarray}
where $\left\langle \cdots\right\rangle $ denotes the ensemble average
and $T_{C}$ denotes time order operator on the CTP contour. We can
put $G_{\mathrm{CTP}}^{\mu\nu}(x_{1},x_{2})$ in a matrix form as
\begin{equation}
G_{\mathrm{CTP}}^{\mu\nu}(x_{1},x_{2})=\left(\begin{array}{cc}
G_{F}^{\mu\nu}(x_{1},x_{2}) & G_{<}^{\mu\nu}(x_{1},x_{2})\\
G_{>}^{\mu\nu}(x_{1},x_{2}) & G_{\overline{F}}^{\mu\nu}(x_{1},x_{2})
\end{array}\right),\label{eq:green-func}
\end{equation}
depending on whether the field $A^{\mu}$ lives on the positive or
negative time branch. The four elements of $G_{\mathrm{CTP}}^{\mu\nu}$
are 
\begin{eqnarray}
G_{F}^{\mu\nu}(x_{1},x_{2}) & \equiv & G_{++}^{\mu\nu}(x_{1},x_{2})\nonumber \\
 & = & \theta(t_{1}-t_{2})\left\langle A^{\mu}(x_{1})A^{\nu}(x_{2})\right\rangle \nonumber \\
 &  & +\theta(t_{2}-t_{1})\left\langle A^{\nu}(x_{2})A^{\mu}(x_{1})\right\rangle ,\nonumber \\
G_{<}^{\mu\nu}(x_{1},x_{2}) & = & G_{+-}^{\mu\nu}(x_{1},x_{2})=\left\langle A^{\nu}(x_{2})A^{\mu}(x_{1})\right\rangle ,\nonumber \\
G_{>}^{\mu\nu}(x_{1},x_{2}) & = & G_{-+}^{\mu\nu}(x_{1},x_{2})=\left\langle A^{\mu}(x_{1})A^{\nu}(x_{2})\right\rangle ,\nonumber \\
G_{\overline{F}}^{\mu\nu}(x_{1},x_{2}) & \equiv & G_{--}^{\mu\nu}(x_{1},x_{2})\nonumber \\
 & = & \theta(t_{2}-t_{1})\left\langle A^{\mu}(x_{1})A^{\nu}(x_{2})\right\rangle \nonumber \\
 &  & +\theta(t_{1}-t_{2})\left\langle A^{\nu}(x_{2})A^{\mu}(x_{1})\right\rangle .\label{eq:def_4}
\end{eqnarray}
From the constraint $G_{F}^{\mu\nu}+G_{\overline{F}}^{\mu\nu}=G_{<}^{\mu\nu}+G_{>}^{\mu\nu}$,
only three of them are independent. In the so-called physical representation
\citep{Chou:1984es,kadanoff1962quantum,fetter2003quantum}, three
independent two-point Green's functions are 
\begin{eqnarray}
G_{R}^{\mu\nu}(x_{1},x_{2}) & = & (G_{F}^{\mu\nu}-G_{<}^{\mu\nu})(x_{1},x_{2}),\nonumber \\
G_{A}^{\mu\nu}(x_{1},x_{2}) & = & (G_{F}^{\mu\nu}-G_{>}^{\mu\nu})(x_{1},x_{2}),\nonumber \\
G_{C}^{\mu\nu}(x_{1},x_{2}) & = & G_{>}^{\mu\nu}(x_{1},x_{2})+G_{<}^{\mu\nu}(x_{1},x_{2}),\label{eq:def_7}
\end{eqnarray}
where the subscripts $"A"$ and $"R"$ denote the advanced and retarded
Green's function respectively. The two-point Green's functions in
Eqs. (\ref{eq:def_4}-\ref{eq:def_7}) can be used to express any
two-point functions defined on the CTP contour. Note that the definition
of two-point Green's functions in Eqs. (\ref{eq:green-fun-ctp}) and
(\ref{eq:def_4}) differs from the usual one in quantum field theory
by an $i=\sqrt{-1}$ factor.


Normally one can make Fourier transform with respect to the relative
position $y=x_{1}-x_{2}$ for the two-point function $G(x_{1},x_{2})$
to obtain the corresponding Wigner function 
\begin{equation}
G(x,p)\equiv\int d^{4}y\,e^{ip\cdot y}G(x_{1},x_{2}),
\end{equation}
where $G(x_{1},x_{2})$ can be any two-point function in Eqs. (\ref{eq:def_4})
and (\ref{eq:def_7}). One immediately derive from Eq. (\ref{eq:def_7})
\begin{equation}
G_{R}^{\mu\nu}(x,p)-G_{A}^{\mu\nu}(x,p)=G_{>}^{\mu\nu}(x,p)-G_{<}^{\mu\nu}(x,p).\label{eq:gra-g-less-large}
\end{equation}
Using the property $\left[\widetilde{G}_{R}^{\mu\nu}(x,p)\right]^{*}=\widetilde{G}_{A}^{\nu\mu}(x,p)$
and assuming $G_{R,A}^{\nu\mu}=G_{R,A}^{\mu\nu}$, one can define
the spectra function 
\begin{equation}
\rho^{\mu\nu}(x,p)=-\frac{1}{\pi}\mathrm{Im}\widetilde{G}_{R}^{\mu\nu}(x,p)=\frac{1}{\pi}\mathrm{Im}\widetilde{G}_{A}^{\mu\nu}(x,p),\label{eq:spectral-density}
\end{equation}
through which $G_{>}^{\mu\nu}$ and $G_{<}^{\mu\nu}$ can be expressed
as 
\begin{align}
G_{>}^{\mu\nu}(x,p)= & 2\pi\rho^{\mu\nu}(x,p)\left[1+n_{B}(p_{0})\right],\nonumber \\
G_{<}^{\mu\nu}(x,p)= & 2\pi\rho^{\mu\nu}(x,p)n_{B}(p_{0}),\nonumber \\
\frac{G_{>}^{\mu\nu}(x,p)}{G_{<}^{\mu\nu}(x,p)}= & \frac{1+n_{B}(p_{0})}{n_{B}(p_{0})}=e^{\beta p_{0}}.\label{eq:spec-density-rep-1}
\end{align}
The above relations are general and valid for free and interacting
fields in local equilibrium. In terms of the spectral density (\ref{eq:spectral-density}),
$\widetilde{G}_{R,A}^{\mu\nu}$ can be expressed as 
\begin{align}
G_{R/A}^{\mu\nu}(x,p)= & -i\int d\omega\frac{\rho^{\mu\nu}(x,\omega,\mathbf{p})}{\omega-p_{0}\mp i\epsilon}\nonumber \\
= & \int\frac{d\omega}{2\pi i}\frac{1}{\omega-p_{0}\mp i\epsilon}\left[G_{>}^{\mu\nu}(x,\omega,\mathbf{p})-G_{<}^{\mu\nu}(x,\omega,\mathbf{p})\right].\label{eq:gra-spec}
\end{align}


Let us use the quantized form of the vector field (\ref{eq:a-quantization})
to express $G_{<}^{\mu\nu}(x,p)$ in terms of the matrix valued spin
dependent distribution (MVSD) \citep{Sheng:2022ffb} 
\begin{align}
G_{<}^{\mu\nu}(x,p)= & \int d^{4}y\,e^{ip\cdot y/\hbar}\left\langle A^{\nu}(x_{2})A^{\mu}(x_{1})\right\rangle \nonumber \\
= & 2\pi\sum_{\lambda_{1},\lambda_{2}}\delta\left(p^{2}-m_{V}^{2}\right)\left\{ \theta(p^{0})\epsilon^{\mu}\left(\lambda_{1},{\bf p}\right)\epsilon^{\nu\ast}\left(\lambda_{2},{\bf p}\right)f_{\lambda_{1}\lambda_{2}}^{V}(x,p)\right.\nonumber \\
 & +\left.\theta(-p^{0})\epsilon^{\mu\ast}\left(\lambda_{1},-{\bf p}\right)\epsilon^{\nu}\left(\lambda_{2},-{\bf p}\right)\left[\delta_{\lambda_{2}\lambda_{1}}+f_{\lambda_{2}\lambda_{1}}^{V}(x,\overline{p})\right]\right\} \nonumber \\
 & +\mathcal{O}(\partial f_{V}),\label{eq:g-less-mvsd}
\end{align}
where $p^{\mu}=(E_{p},\mathbf{p})$ and $\overline{p}^{\mu}=(E_{p},-{\bf p})$
are on-shell momenta, and $f_{\lambda_{2}\lambda_{1}}^{V}$ is the
MVSD defined as 
\begin{align}
f_{\lambda_{1}\lambda_{2}}^{V}(x,p) & \equiv\int\frac{d^{4}u}{2(2\pi)^{3}}\delta(p\cdot u)e^{-iu\cdot x}\nonumber \\
 & \times\left\langle a_{V}^{\dagger}\left(\lambda_{2},{\bf p}-\frac{{\bf u}}{2}\right)a_{V}\left(\lambda_{1},{\bf p}+\frac{{\bf u}}{2}\right)\right\rangle .
\end{align}
One can check that $f_{\lambda_{1}\lambda_{2}}^{V}(x,p)$ is Hermitian:
$\left[f_{\lambda_{1}\lambda_{2}}^{V}(x,p)\right]^{*}=f_{\lambda_{2}\lambda_{1}}^{V}(x,p)$.
We have neglected in Eq. (\ref{eq:g-less-mvsd}) gradient terms of
the MVSD including off-shell terms. In unpolarized system, $f_{\lambda_{1}\lambda_{2}}^{V}(x,p)$
is diagonal, $f_{\lambda_{1}\lambda_{2}}^{V}(x,p)=f_{V}\delta_{\lambda_{1}\lambda_{2}}$,
so $G_{<}^{\mu\nu}(x,p)$ in Eq. (\ref{eq:g-less-mvsd}) becomes 
\begin{align}
G_{<}^{\mu\nu}(x,p)= & -2\pi\sum_{\lambda_{1},\lambda_{2}}\delta\left(p^{2}-m_{V}^{2}\right)\left(g^{\mu\nu}-\frac{p^{\mu}p^{\nu}}{m_{V}^{2}}\right)\nonumber \\
 & \times\left\{ \theta(p^{0})f_{V}(x,p)+\theta(-p^{0})\left[1+f_{V}(x,-p)\right]\right\} +\mathcal{O}(\partial f),\label{eq:g-less-mvsd-1}
\end{align}
where $p^{\mu}=(p_{0},\mathbf{p})$.


We can extract the particle's contribution from $G_{<}^{\mu\nu}(x,p)$
in Eq. (\ref{eq:g-less-mvsd}) by 
\begin{align}
W^{\mu\nu}(x,p)= & \frac{E_{p}}{\pi\hbar}\int_{0}^{\infty}dp_{0}G_{<}^{\mu\nu}(x,p)\nonumber \\
= & \sum_{\lambda_{1},\lambda_{2}}\epsilon^{\mu}\left(\lambda_{1},{\bf p}\right)\epsilon^{\nu\ast}\left(\lambda_{2},{\bf p}\right)f_{\lambda_{1}\lambda_{2}}^{V}(x,{\bf p}).\label{eq:decomp}
\end{align}
Obviously we have $p_{\mu}W^{\mu\nu}=0$ with $p^{\mu}=(E_{p},\mathbf{p})$.
The Wigner function $W^{\mu\nu}(x,p)$ can be decomposed into the
scalar, polarization and tensor components as \citep{Li:2022vmb}
\begin{equation}
W^{\mu\nu}=-\frac{1}{3}\Delta^{\mu\nu}\mathcal{S}+W^{[\mu\nu]}+\mathcal{T}^{\mu\nu},
\end{equation}
where $\Delta^{\mu\nu}$ is the projector, and $W^{[\mu\nu]}$ and
$\mathcal{T}^{\mu\nu}$ are the polarization and tensor components
respectively 
\begin{align}
\Delta^{\mu\nu}= & g^{\mu\nu}-\frac{p^{\mu}p^{\nu}}{p^{2}},\nonumber \\
\mathcal{T}^{\mu\nu}= & \frac{1}{2}\left(W^{\mu\nu}+W^{\nu\mu}\right)+\frac{1}{3}\Delta^{\mu\nu}\mathcal{S},\nonumber \\
W^{[\mu\nu]}= & \frac{1}{2}\left(W^{\mu\nu}-W^{\nu\mu}\right).
\end{align}
Note that $W^{[\mu\nu]}$ is anti-symmetric and $\mathcal{T}^{\mu\nu}$
is symmetric for $\mu\leftrightarrow\nu$. Using $f_{\lambda_{1}\lambda_{2}}^{V}=\rho_{\lambda_{1}\lambda_{2}}\mathrm{Tr}(f_{V})$
and the decomposition in Eq. (\ref{eq:spin-density}), we can identify
\begin{align}
\mathcal{S}= & \mathrm{Tr}f,\nonumber \\
W^{[\mu\nu]}= & \frac{1}{2}\mathrm{Tr}(f)\sum_{\lambda_{1},\lambda_{2}}\epsilon^{\mu}\left(\lambda_{1},{\bf p}\right)\epsilon^{\nu\ast}\left(\lambda_{2},{\bf p}\right)P_{i}\Sigma_{\lambda_{1}\lambda_{2}}^{i},\nonumber \\
\mathcal{T}^{\mu\nu}= & \mathrm{Tr}(f)\sum_{\lambda_{1},\lambda_{2}}\epsilon^{\mu}\left(\lambda_{1},{\bf p}\right)\epsilon^{\nu\ast}\left(\lambda_{2},{\bf p}\right)T_{ij}\Sigma_{\lambda_{1}\lambda_{2}}^{ij}.
\end{align}
We can verify that $g_{\mu\nu}W^{[\mu\nu]}$ and $g_{\mu\nu}\mathcal{T}^{\mu\nu}$
are all vanishing due to $\mathrm{Tr}\Sigma_{i}=\mathrm{Tr}\Sigma_{ij}=0$.
We can extract $\rho_{00}-1/3$ which is called the spin alignment
by the contraction of $W^{\mu\nu}$ with 
\begin{equation}
L^{\mu\nu}(p)=\epsilon^{\mu,*}\left(0,{\bf p}\right)\epsilon^{\nu}\left(0,{\bf p}\right)+\frac{1}{3}\Delta^{\mu\nu}.
\end{equation}
The result is \citep{Dong:2023cng}
\begin{equation}
\frac{L_{\mu\nu}(p)W^{\mu\nu}}{-\Delta_{\mu\nu}W^{\mu\nu}(x,p)}=\frac{f_{00}^{V}(x,{\bf p})}{\mathrm{Tr}f_{V}}-\frac{1}{3}=\rho_{00}-\frac{1}{3}.
\end{equation}
We see that the spin density matrix $\rho_{\lambda_{1}\lambda_{2}}$
can be extracted from $W^{\mu\nu}$ using (\ref{eq:decomp}).


\subsection{Dyson-Schwinger equations in CTP formalism}

The Dyson-Schwinger equation on the CTP for the vectcor meson reads
\begin{eqnarray}
G^{\mu\nu}(x_{1},x_{2}) & = & G_{\text{free}}^{\mu\nu}(x_{1},x_{2})\nonumber \\
 &  & +\int_{C}d^{4}x_{3}d^{4}x_{4}\,G_{\text{free}}^{\mu\alpha}(x_{1},x_{3})\nonumber \\
 &  & \times\Sigma_{\alpha\beta}(x_{3},x_{4})G^{\beta\nu}(x_{4},x_{2}),\nonumber \\
G^{\mu\nu}(x_{1},x_{2}) & = & G_{\text{free}}^{\mu\nu}(x_{1},x_{2})\nonumber \\
 &  & +\int_{C}d^{4}x_{3}d^{4}x_{4}\,G^{\mu\alpha}(x_{1},x_{3})\nonumber \\
 &  & \times\Sigma_{\alpha\beta}(x_{3},x_{4})G_{\text{free}}^{\beta\nu}(x_{4},x_{2}),\label{eq:dse-ctp}
\end{eqnarray}
where $G^{\mu\nu}(x_{1},x_{2})$ is the full Green's function and
$G_{\text{free}}^{\mu\nu}(x_{1},x_{2})$ the free one, $\Sigma_{\alpha\beta}(x_{3},x_{4})$
is the self-energy which will be given later. All two-point functions
are defined on the CTP but we suppress the 'CTP' index in all of them
for notational simplicity. The integrals over $x_{3}$ and $x_{4}$
are taken on the CTP contour, which can be expressed as 
\begin{equation}
\int_{C}d^{4}x=\begin{cases}
\int_{t_{0}}^{\infty}d^{4}x, & x_{0}\in[t_{0}^{+},+\infty]\\
-\int_{t_{0}}^{\infty}d^{4}x, & x_{0}\in[t_{0}^{-},+\infty]
\end{cases}
\end{equation}
where $\int_{t_{0}}^{\infty}d^{4}x$ is the ordinary space-time integral.
The free Green's function satisfies 
\begin{equation}
H_{\mu}^{\lambda}(x_{1})G_{\text{free}}^{\mu\nu}(x_{1},x_{2})=H_{\mu}^{\lambda}(x_{2})G_{\text{free}}^{\nu\mu}(x_{1},x_{2})=i\delta_{\text{CTP}}^{(4)}(x_{1}-x_{2})g^{\lambda\nu},\label{eq:free-two-point}
\end{equation}
where $H_{\mu}^{\lambda}(x_{1})$ is the differential operator acting
on $x_{1}$ 
\begin{equation}
H_{\mu}^{\lambda}(x_{1})=(\partial_{x_{1},\rho}\partial_{x_{1}}^{\rho}+m_{V}^{2})g_{\mu}^{\lambda}-\partial_{x_{1}}^{\lambda}\partial_{\mu}^{x_{1}}.
\end{equation}
The delta-function on the CTP is defined as 
\begin{align}
\delta_{\text{CTP}}^{(4)}(x_{1}-x_{2})= & \delta_{\text{CTP}}(x_{1}^{0}-x_{2}^{0})\delta^{(3)}({\bf x}_{1}-{\bf x}_{2})\nonumber \\
= & \left\{ \begin{array}{cc}
\delta(x_{1}^{0}-x_{2}^{0})\delta^{(3)}({\bf x}_{1}-{\bf x}_{2}), & x_{1}^{0},x_{2}^{0}\in[t_{0}^{+},+\infty]\\
-\delta(x_{1}^{0}-x_{2}^{0})\delta^{(3)}({\bf x}_{1}-{\bf x}_{2}), & x_{1}^{0},x_{2}^{0}\in[t_{0}^{-},+\infty]\\
0, & x_{1}^{0},x_{2}^{0}\;\mathrm{on\;}\\
 & \mathrm{different\;branches}
\end{array}\right.
\end{align}
The minus sign in $\delta_{\text{CTP}}(x_{1}^{0}-x_{2}^{0})$ when
$x_{1}^{0}$ and $x_{2}^{0}$ are on the negative time branch comes
from the integral 
\begin{align}
\int_{C}dx_{1}^{0}\delta_{\text{CTP}}(x_{1}^{0}-x_{2}^{0})= & \int_{t_{0}^{+}}^{\infty}dx_{1}^{0}\delta_{\text{CTP}}(x_{1}^{0}-x_{2}^{0})-\int_{t_{0}^{-}}^{\infty}dx_{1}^{0}\delta_{\text{CTP}}(x_{1}^{0}-x_{2}^{0})\nonumber \\
= & -\int_{t_{0}^{-}}^{\infty}dx_{1}^{0}\delta_{\text{CTP}}(x_{1}^{0}-x_{2}^{0})=1.
\end{align}
Applying $H_{\mu}^{\lambda}(x_{1})$ and $H_{\mu}^{\lambda}(x_{2})$
to Eq. (\ref{eq:dse-ctp}) and using Eq. (\ref{eq:free-two-point}),
we obtain 
\begin{align}
H_{\mu}^{\lambda}(x_{1})G^{\mu\nu}(x_{1},x_{2}) & =i\delta_{\text{CTP}}^{(4)}(x_{1}-x_{2})g^{\lambda\nu}\nonumber \\
 & +i\int_{C}d^{4}x^{\prime}\,\Sigma_{\ \alpha}^{\lambda}(x_{1},x^{\prime})G^{\alpha\nu}(x^{\prime},x_{2}),\nonumber \\
H_{\nu}^{\lambda}(x_{2})G^{\mu\nu}(x_{1},x_{2}) & =i\delta_{\text{CTP}}^{(4)}(x_{1}-x_{2})g^{\mu\lambda}\nonumber \\
 & +i\int_{C}d^{4}x^{\prime}\,G^{\mu\alpha}(x_{1},x^{\prime})\Sigma_{\alpha}^{\;\;\lambda}(x^{\prime},x_{2}).
\end{align}
The above equations are in the CTP form: all functions are defined
on the CTP. We can rewrite them into normal forms for four different
cases that $(x_{1}^{0},x_{2}^{0})$ are on different time-branches.
This results in matrix form equations 
\begin{eqnarray}
 &  & H_{\mu}^{\lambda}(x_{1})\left(\begin{array}{cc}
G_{F}^{\mu\nu}(x_{1},x_{2}) & G_{<}^{\mu\nu}(x_{1},x_{2})\\
G_{>}^{\mu\nu}(x_{1},x_{2}) & G_{\overline{F}}^{\mu\nu}(x_{1},x_{2})
\end{array}\right)\nonumber \\
 & = & i\left(\begin{array}{cc}
g^{\lambda\nu}\delta^{(4)}(x_{1}-x_{2}) & 0\\
0 & -g^{\lambda\nu}\delta^{(4)}(x_{1}-x_{2})
\end{array}\right)\nonumber \\
 &  & +i\int d^{4}x^{\prime}\left(\begin{array}{cc}
\Sigma_{F,\alpha}^{\lambda}(x_{1},x^{\prime}) & -\Sigma_{<,\alpha}^{\lambda}(x_{1},x^{\prime})\\
\Sigma_{>,\alpha}^{\lambda}(x_{1},x^{\prime}) & -\Sigma_{\overline{F},\alpha}^{\lambda}(x_{1},x^{\prime})
\end{array}\right)\nonumber \\
 &  & \times\left(\begin{array}{cc}
G_{F}^{\alpha\nu}(x^{\prime},x_{2}) & G_{<}^{\alpha\nu}(x^{\prime},x_{2})\\
G_{>}^{\alpha\nu}(x^{\prime},x_{2}) & G_{\overline{F}}^{\alpha\nu}(x^{\prime},x_{2})
\end{array}\right),\label{eq:dse-matrix-1}
\end{eqnarray}
and 
\begin{eqnarray}
 &  & H_{\nu}^{\lambda}(x_{2})\left(\begin{array}{cc}
G_{F}^{\mu\nu}(x_{1},x_{2}) & G_{<}^{\mu\nu}(x_{1},x_{2})\\
G_{>}^{\mu\nu}(x_{1},x_{2}) & G_{\overline{F}}^{\mu\nu}(x_{1},x_{2})
\end{array}\right)\nonumber \\
 & = & i\left(\begin{array}{cc}
g^{\mu\lambda}\delta^{(4)}(x_{1}-x_{2}) & 0\\
0 & -g^{\mu\lambda}\delta^{(4)}(x_{1}-x_{2})
\end{array}\right)\nonumber \\
 &  & +i\int d^{4}x^{\prime}\left(\begin{array}{cc}
G_{F}^{\mu\alpha}(x_{1},x^{\prime}) & -G_{<}^{\mu\alpha}(x_{1},x^{\prime})\\
G_{>}^{\mu\alpha}(x_{1},x^{\prime}) & -G_{\overline{F}}^{\mu\alpha}(x_{1},x^{\prime})
\end{array}\right)\nonumber \\
 &  & \times\left(\begin{array}{cc}
\Sigma_{\alpha}^{F,\lambda}(x^{\prime},x_{2}) & \Sigma_{\alpha}^{<,\lambda}(x^{\prime},x_{2})\\
\Sigma_{\alpha}^{>,\lambda}(x^{\prime},x_{2}) & \Sigma_{\alpha}^{\overline{F},\lambda}(x^{\prime},x_{2})
\end{array}\right),\label{eq:dse-matrix-2}
\end{eqnarray}
where all integrals and functions are normal ones. We multiply the
unitary transformation matrix $U$ from the left and $U^{-1}$ from
the right to Eqs. (\ref{eq:dse-matrix-1}) and (\ref{eq:dse-matrix-2}),
where $U$ and $U^{-1}$ are defined as 
\begin{equation}
U=\frac{1}{\sqrt{2}}\left(\begin{array}{cc}
1 & -1\\
1 & 1
\end{array}\right),\;\;U^{-1}=\frac{1}{\sqrt{2}}\left(\begin{array}{cc}
1 & 1\\
-1 & 1
\end{array}\right)=U^{T}.
\end{equation}
The resulting equations are 
\begin{alignat}{1}
 & -iH_{\rho}^{\mu}(x_{1})\left(\begin{array}{cc}
0 & G_{A}^{\rho\nu}\\
G_{R}^{\rho\nu} & G_{C}^{\rho\nu}
\end{array}\right)(x_{1},x_{2})\nonumber \\
 & =\left(\begin{array}{cc}
0 & 1\\
1 & 0
\end{array}\right)g^{\mu\nu}\delta^{(4)}(x_{1}-x_{2})\nonumber \\
 & +\int dx^{\prime}\left(\begin{array}{ll}
0 & \;\;\Sigma_{A,\rho}^{\mu}\star G_{A}^{\rho\nu}\\
\Sigma_{R,\rho}^{\mu}\star G_{R}^{\rho\nu} & \;\;\Sigma_{C,\rho}^{\mu}\star G_{A}^{\rho\nu}+\Sigma_{R,\rho}^{\mu}\star G_{C}^{\rho\nu}
\end{array}\right)(x_{1},x_{2}),\label{eq:DSE_phy}
\end{alignat}
\begin{alignat}{1}
 & -iH_{\rho}^{\nu}(x_{2})\left(\begin{array}{cc}
0 & G_{A}^{\mu\rho}\\
G_{R}^{\mu\rho} & G_{C}^{\mu\rho}
\end{array}\right)(x_{1},x_{2})\nonumber \\
 & =\left(\begin{array}{cc}
0 & 1\\
1 & 0
\end{array}\right)g^{\mu\nu}\delta^{(4)}(x_{1}-x_{2})\nonumber \\
 & +\int dx^{\prime}\left(\begin{array}{ll}
0 & \;\;G_{A,\rho}^{\mu}\star\Sigma_{A}^{\rho\nu}\\
G_{R,\rho}^{\mu}\star\Sigma_{R}^{\rho\nu} & \;\;G_{C,\rho}^{\mu}\star\Sigma_{A}^{\rho\nu}+G_{R,\rho}^{\mu}\star\Sigma_{C}^{\rho\nu}
\end{array}\right)(x_{1},x_{2}),\label{eq:DSE_phy-1}
\end{alignat}
where we used the shorthand notation $O_{1}\star O_{2}(x_{1},x_{2})\equiv O_{1}(x_{1},x^{\prime})O_{2}(x^{\prime},x_{2})$
for $O=G,\Sigma$ and the formula 
\begin{eqnarray}
\left(\begin{array}{cc}
0 & O_{A}\\
O_{R} & O_{C}
\end{array}\right) & = & U\left(\begin{array}{cc}
O_{F} & O_{<}\\
O_{>} & O_{\overline{F}}
\end{array}\right)U^{-1},\nonumber \\
\left(\begin{array}{cc}
O_{A} & 0\\
O_{C} & O_{R}
\end{array}\right) & = & U\left(\begin{array}{cc}
O_{F} & -O_{<}\\
O_{>} & -O_{\overline{F}}
\end{array}\right)U^{-1}.\label{eq:physics-basis}
\end{eqnarray}

The off-diagonal elements of Eqs. (\ref{eq:DSE_phy}) and (\ref{eq:DSE_phy-1})
give Dyson-Schwinger equations for retarded and advanced two-point
Green's functions 
\begin{align}
-iH_{\;\;\rho}^{\mu}(x_{1})G_{R/A}^{\rho\nu}(x_{1},x_{2})= & g^{\mu\nu}\delta^{(4)}(x_{1}-x_{2})+\int dx^{\prime}\Sigma_{R/A,\rho}^{\mu}\star G_{R/A}^{\rho\nu}(x_{1},x_{2}),\nonumber \\
-iH_{\;\;\rho}^{\nu}(x_{2})G_{R/A}^{\mu\rho}(x_{1},x_{2})= & g^{\mu\nu}\delta^{(4)}(x_{1}-x_{2})+\int dx^{\prime}G_{R/A,\rho}^{\mu}\star\Sigma_{R/A}^{\rho\nu}(x_{1},x_{2}).\label{dse-gra}
\end{align}
We see that retarded or advanced two-point functions are always together:
there is no mixing between retarded and advanced two-point functions.
The $g^{\mu\nu}$ term indicates that retarded or advanced two-point
functions are off-shell functions in principle. For free particles
without interaction or in a homogeneous system, two-point Green's
functions only depend on the distance between two points. In this
case the retarded and advanced two-point Green's functions in momentum
space read \citep{Dong:2023cng}
\begin{align}
G_{R/A,\mathrm{free}}^{\rho\nu}(p)= & -i\frac{1}{p^{2}-m_{V}^{2}\pm i\mathrm{sgn}(p_{0})\varepsilon}\left(g^{\rho\nu}-\frac{p^{\rho}p^{\nu}}{m_{V}^{2}}\right),\nonumber \\
G_{R/A}^{\rho\nu}(p)= & \frac{-i\Delta_{T}^{\mu\nu}}{p^{2}-m_{V}^{2}-\widetilde{\Sigma}_{R/A}^{T}(p)\pm i\mathrm{sgn}(p_{0})\varepsilon}\nonumber \\
 & +\frac{-i\Delta_{L}^{\mu\nu}}{p^{2}-m_{V}^{2}-\widetilde{\Sigma}_{R/A}^{L}(p)\pm i\mathrm{sgn}(p_{0})\varepsilon}+i\frac{p^{\rho}p^{\nu}}{m_{V}^{2}p^{2}},\label{eq:g-ra-full}
\end{align}
where $\varepsilon$ is a small positive number, $\widetilde{\Sigma}_{R/A}^{T,L}(p)\equiv-i\Sigma_{R/A}^{T,L}(p)$
are retarded and advanced self-energies in momentum space for transverse
and longitudinal modes, and $\Delta_{T,L}^{\mu\nu}$ are projectors
for transverse and longitudinal modes defined as 
\begin{align}
\Delta_{T}^{\mu\nu}= & -g^{\mu0}g^{\nu0}+g^{\mu\nu}+\frac{\mathbf{p}^{\mu}\mathbf{p}^{\nu}}{|\mathbf{p}|^{2}},\nonumber \\
\Delta_{L}^{\mu\nu}= & \Delta^{\mu\nu}-\Delta_{T}^{\mu\nu},\label{eq:projectors}
\end{align}
where $p^{\mu}=(p_{0},\mathbf{p})$ is off-shell and $\mathbf{p}^{\mu}=(0,\mathbf{p})$.
One can verify that $p_{\mu}\Delta_{T}^{\mu\nu}=p_{\mu}\Delta_{L}^{\mu\nu}=0$.

\subsection{Kinetic equations}

The "12" element of the matrix equation (\ref{eq:dse-matrix-1})
gives the equation for $G_{<}^{\mu\nu}(x_{1},x_{2})$ as 
\begin{eqnarray}
 &  & H_{\;\;\rho}^{\mu}(x_{1})G^{<,\rho\nu}(x_{1},x_{2})\nonumber \\
 & = & i\int d^{4}x^{\prime}\left[\Sigma_{F,\rho}^{\mu}(x_{1},x^{\prime})G_{<}^{\rho\nu}(x^{\prime},x_{2})-\Sigma_{<,\rho}^{\mu}(x_{1},x^{\prime})G_{\overline{F}}^{\rho\nu}(x^{\prime},x_{2})\right]\nonumber \\
 & = & i\int d^{4}x^{\prime}\left[\Sigma_{R,\rho}^{\mu}(x_{1},x^{\prime})G^{<,\rho\nu}(x^{\prime},x_{2})+\Sigma_{<,\rho}^{\mu}(x_{1},x^{\prime})G_{A}^{\rho\nu}(x^{\prime},x_{2})\right].\label{eq:kinetic-g-less}
\end{eqnarray}
The companion equation with $H_{\nu}^{\lambda}(x_{2})G^{<,\mu\nu}(x_{1},x_{2})$
can be read out from the 12 element of Eq. (\ref{eq:dse-matrix-2}),
which can also be obtained from Eq. (\ref{eq:kinetic-g-less}) by
the replacement in the right-hand-side: $\Sigma\leftrightarrow G$.
The integrals in Eq. (\ref{eq:kinetic-g-less}) are all of the type
\begin{align}
I(x_{1},x_{2})= & \int d^{4}x^{\prime}O_{1}(x_{1},x^{\prime})O_{2}(x^{\prime},x_{2})\nonumber \\
= & \int_{-\infty}^{\infty}d^{4}y^{\prime}O_{1}\left(y-y^{\prime},X+\frac{1}{2}y^{\prime}\right)O_{2}\left(y^{\prime},X-\frac{1}{2}(y-y^{\prime})\right),\label{eq:convolution-1}
\end{align}
where we have expressed a two-point function in terms of the distance
and center position of the two space-time coordinates, and we have
used following variables for distances and center positions for $I(x_{1},x_{2})$,
$O_{1}(x_{1},x^{\prime})$ and $O_{2}(x^{\prime},x_{2})$, 
\begin{align}
y= & x_{1}-x_{2},\;\;X=\frac{1}{2}(x_{1}+x_{2}),\nonumber \\
y^{\prime}= & x^{\prime}-x_{2},\;\;X_{2}=\frac{1}{2}(x^{\prime}+x_{2})=X-\frac{1}{2}(y-y^{\prime}),\nonumber \\
y-y^{\prime}= & x_{1}-x^{\prime},\;\;X_{1}=\frac{1}{2}(x_{1}+x^{\prime})=X+\frac{1}{2}y^{\prime}.\label{eq:grad-xy}
\end{align}
Suppose $y,y^{\prime}\ll X$, we can expand the integrand in Eq. (\ref{eq:convolution-1})
in $y^{\prime}$ and $y-y^{\prime}$ relative to $X$, then we obtain
\begin{align}
I(x_{1},x_{2})= & \int_{-\infty}^{\infty}d^{4}y^{\prime}O_{1}\left(y-y^{\prime},X+\frac{1}{2}y^{\prime}\right)O_{2}\left(y^{\prime},X-\frac{1}{2}(y-y^{\prime})\right)\nonumber \\
= & \int_{-\infty}^{\infty}d^{4}y^{\prime}O_{1}\left(y-y^{\prime},X\right)O_{2}\left(y^{\prime},X\right)\nonumber \\
 & +\frac{1}{2}\int_{-\infty}^{\infty}d^{4}y^{\prime}y_{\mu}^{\prime}\partial_{X}^{\mu}O_{1}\left(y-y^{\prime},X\right)O_{2}\left(y^{\prime},X\right)\nonumber \\
 & -\frac{1}{2}\int_{-\infty}^{\infty}d^{4}y^{\prime}O_{1}\left(y-y^{\prime},X\right)(y_{\mu}-y_{\mu}^{\prime})\partial_{X}^{\mu}O_{2}\left(y^{\prime},X\right).\label{eq:ix1x2}
\end{align}
The Fourier transform of $I(x_{1},x_{2})$ with respect to $y$ gives
its form in Wigner functions 
\begin{align}
I(X,p)= & \int d^{4}ye^{ip\cdot y}I(y,X)\nonumber \\
= & O_{1}\left(X,p\right)O_{2}\left(X,p\right)-\frac{1}{2}i\partial_{X}^{\mu}O_{1}\left(X,p\right)\partial_{\mu}^{p}O_{2}\left(X,p\right)\nonumber \\
 & +\frac{1}{2}i\partial_{\mu}^{p}O_{1}\left(X,p\right)\partial_{X}^{\mu}O_{2}\left(X,p\right)\nonumber \\
= & O_{1}\left(X,p\right)O_{2}\left(X,p\right)-i\frac{1}{2}\left\{ O_{1}\left(X,p\right),O_{2}\left(X,p\right)\right\} _{\mathrm{PB}},\label{eq:fourier-tr}
\end{align}
where there is a correspondence from Eq. (\ref{eq:ix1x2}) to (\ref{eq:fourier-tr})
$y^{\mu}\rightarrow-i\partial_{p}^{\mu}$, and we have used the Poisson
brackets defined as 
\begin{equation}
\left\{ O_{1}(x,p),O_{2}(x,p)\right\} _{\mathrm{PB}}\equiv\partial_{x}^{\mu}O_{1}\partial_{\mu}^{p}O_{2}-\partial_{\mu}^{p}O_{1}\partial_{x}^{\mu}O_{2}.
\end{equation}
We perform the Fourier transform of Eq. (\ref{eq:kinetic-g-less})
with respect to $y$ by using Eq. (\ref{eq:fourier-tr}), the result
is 
\begin{align}
 & \left\{ g_{\rho}^{\mu}\left[-\left(p^{2}-m_{V}^{2}-\frac{1}{4}\partial_{x}^{2}\right)-ip\cdot\partial_{x}\right]\right.\nonumber \\
 & \left.-\frac{1}{4}\partial_{x}^{\mu}\partial_{\rho}^{x}+p^{\mu}p_{\rho}+\frac{1}{2}i\left(p_{\rho}\partial_{x}^{\mu}+p^{\mu}\partial_{\rho}^{x}\right)\right\} G_{<}^{\rho\nu}(x,p)\nonumber \\
= & i\Sigma_{R,\rho}^{\mu}(x,p)G_{<}^{\rho\nu}(x,p)+i\Sigma_{<,\rho}^{\mu}(x,p)G_{A}^{\rho\nu}(x,p)\nonumber \\
 & +\frac{1}{2}\left\{ \Sigma_{R,\rho}^{\mu}(x,p),G_{<}^{\rho\nu}(x,p)\right\} _{\mathrm{PB}}+\frac{1}{2}\left\{ \Sigma_{<,\rho}^{\mu}(x,p),G_{A}^{\rho\nu}(x,p)\right\} _{\mathrm{PB}},\label{eq:dse-sig-g}
\end{align}
where we have replaced $X$ by $x$ without ambiguity. In the same
way, we can also derive the companion equation from the '12' element
of Eq. (\ref{eq:dse-matrix-2}) in terms of Wigner functions, which
we can also obtain by flipping the sign of the $\partial_{x}$ term
in the left-hand-side and making the replacement $\Sigma\leftrightarrow G$
in the right-hand-side of the equation. Taking the difference between
Eq. (\ref{eq:dse-sig-g}) and its companion equation, we obtain the
kinetic equation 
\begin{align}
 & p\cdot\partial_{x}G_{<}^{\mu\nu}(x,p)-\frac{1}{4}\left[p^{\mu}\partial_{\rho}^{x}G_{<}^{\rho\nu}(x,p)+p^{\nu}\partial_{\rho}^{x}G_{<}^{\mu\rho}(x,p)\right]\nonumber \\
= & \frac{1}{2}\left[G_{R,\rho}^{\mu}(x,p),\Sigma_{<}^{\rho\nu}(x,p)\right]_{\star}+\frac{1}{2}\left[G_{<,\rho}^{\mu}(x,p),\Sigma_{A}^{\rho\nu}(x,p)\right]_{\star},\label{eq:kinetic-g<}
\end{align}
where we have defined a special commutator 
\begin{equation}
\left[G_{R,\rho}^{\mu}(x,p),\Sigma_{<}^{\rho\nu}(x,p)\right]_{\star}\equiv G_{R,\rho}^{\mu}(x,p)\Sigma_{<}^{\rho\nu}(x,p)-\Sigma_{R,\rho}^{\mu}(x,p)G_{<}^{\rho\nu}(x,p).
\end{equation}
In deriving Eq. (\ref{eq:kinetic-g<}), we have neglected terms with
Poisson brackets and the terms $p_{\rho}G_{<}^{\rho\nu}(x,p)$ and
$p_{\rho}G_{<}^{\mu\rho}(x,p)$ in the left-hand-side which are vanishing
in the approximation that we make in this part of the review.

From Eq. (\ref{eq:def_4}) one can choose $G_{R}^{\mu\nu}$, $G_{>}^{\mu\nu}$
and $G_{<}^{\mu\nu}$ as three independent variables. In local equilibrium,
we have Eqs. (\ref{eq:spectral-density}) and (\ref{eq:spec-density-rep-1}),
which relate $G_{>}^{\mu\nu}$ and $G_{<}^{\mu\nu}$ to $G_{R}^{\mu\nu}$.
We also see from (\ref{eq:g-ra-full}) that the dressed $G_{R}^{\mu\nu}$
depends on $\Sigma_{R}^{\mu\nu}$ which depends on $G_{R}^{\mu\nu}$,
$G_{>}^{\mu\nu}$ and $G_{<}^{\mu\nu}$ and finally on $G_{R}^{\mu\nu}$
in a self-consistent way. Similarly $\Sigma_{A}^{\mu\nu}$ also depends
on $G_{R}^{\mu\nu}$, $G_{>}^{\mu\nu}$ and $G_{<}^{\mu\nu}$ and
finally on $G_{R}^{\mu\nu}$. Therefore Eq. (\ref{eq:kinetic-g<})
can be reduced to the kinetic equation for $G_{R}^{\mu\nu}$ aided
by on-shell equation (\ref{dse-gra}) for $G_{R}^{\mu\nu}$. Once
we have $G_{R}^{\mu\nu}$ we have everything including the spin density
matrix for the vector meson.

\subsection{Spin Boltzmann equation in on-shell approximation}

One approximation that we can make in solving the kinetic equation
(\ref{eq:kinetic-g<}) is to neglect the real parts $\widetilde{G}_{R/A}^{\mu\nu}(x,p)$
by assuming 
\begin{equation}
G_{R/A}^{\mu\nu}(x,p)\approx\pm\frac{1}{2}\left(G_{>}^{\mu\nu}-G_{<}^{\mu\nu}\right)(x,p).\label{eq:g-on-shell}
\end{equation}
This means that we only consider the imaginary part of $1/(\omega-p_{0}\mp i\epsilon)$
in the integral in Eq. (\ref{eq:gra-spec}), i.e. the imaginary parts
of $\widetilde{G}_{R/A}^{\mu\nu}(x,p)$. We also assume the same relation
for $\Sigma_{R/A}^{\mu\nu}(x,p)$ as (\ref{eq:g-on-shell}). Applying
Eq. (\ref{eq:g-on-shell}) and the similar equation for $\Sigma_{R/A}^{\mu\nu}(x,p)$,
Eq. (\ref{eq:kinetic-g<}) can be simplified as 
\begin{align}
 & p\cdot\partial_{x}G_{<}^{\mu\nu}(x,p)-\frac{1}{4}\left[p^{\mu}\partial_{\rho}^{x}G_{<}^{\rho\nu}(x,p)+p^{\nu}\partial_{\rho}^{x}G_{<}^{\mu\rho}(x,p)\right]\nonumber \\
= & \frac{1}{4}\left[G_{>,\rho}^{\mu}(x,p),\Sigma_{<}^{\rho\nu}(x,p)\right]_{\star}-\frac{1}{4}\left[G_{<,\rho}^{\mu}(x,p),\Sigma_{>}^{\rho\nu}(x,p)\right]_{\star}.\label{eq:kinetic-g<-1}
\end{align}
We assume that $G_{\lessgtr}^{\mu\nu}(x,p)$ are on-shell and can
be expressed in terms of MVSD as in Eq. (\ref{eq:g-less-mvsd}).

Now we look at $\Sigma_{\lessgtr}^{\mu\nu}(x,p)$, the self-energies
of the vector meson. In a hardon gas, the vector meson's self-energy
depends on hadron's ``$<$'' and ``$>$'' propagators and hadron-vector-meson
vertices. In a quark matter with the quark-vector-meson interaction,
the vector meson's self-energy depends on quark's ``$<$'' and ``$>$''
propagators and quark-vector-meson vertices. All these ``$<$''
and ``$>$'' propagators are assumed to be on-shell and can be expressed
in terms of MVSDs. The latter case in the quark matter was considered
in the quark coalescence model. The resulting spin Boltzmann equation
for the unflavored vector meson formed by a quark and its antiquark
are about MVSDs of the vector meson, the quark and antiquark \citep{Sheng:2022ffb}.

The quark's and its antiquark's MVSDs \citep{Becattini:2013fla,Weickgenannt:2020aaf,Sheng:2021kfc,Weickgenannt:2021cuo}
can be parameterized as 
\begin{eqnarray}
f_{rs}^{(+)}(x,\mathbf{p}) & = & \frac{1}{2}f_{q}(x,\mathbf{p})\left[\delta_{rs}-P_{\mu}^{q}(x,\mathbf{p})n_{j}^{\mu}(\mathbf{p})\tau_{rs}^{j}\right],\nonumber \\
f_{rs}^{(-)}(x,-\mathbf{p}) & = & \frac{1}{2}f_{\overline{q}}(x,-\mathbf{p})\left[\delta_{rs}-P_{\mu}^{\overline{q}}(x,-\mathbf{p})n_{j}^{\mu}(-\mathbf{p})\tau_{rs}^{j}\right],\label{eq:f-rs-pol}
\end{eqnarray}
where $\tau^{j}$ ($j=1,2,3$) are Pauli matrices in spin space, $f_{q}(x,\mathbf{p})$
and $f_{\overline{q}}(x,-\mathbf{p})$ are unpolarized distributions
for the quark and its antiquark respectively, and $P_{q}^{\mu}(x,\mathbf{p})$
and $P_{\overline{q}}^{\mu}(x,-\mathbf{p})$ are polarization four-vectors
for the quark and its antiquark respectively. The four-vectors for
three basis directions are given by 
\begin{equation}
n_{j}^{\mu}(\mathbf{p})\equiv n^{\mu}(\mathbf{n}_{j},\mathbf{p})=\left(\frac{\mathbf{n}_{j}\cdot{\bf p}}{m_{q}},\mathbf{n}_{j}+\frac{(\mathbf{n}_{j}\cdot{\bf p}){\bf p}}{m_{q}(E_{{\bf p}}^{q}+m_{q})}\right),
\end{equation}
where $\mathbf{n}_{j}$ for $j=1,2,3$ are three basis unit vectors
that form a Cartesian coordinate system in the particle's rest frame
with $\mathbf{n}_{3}$ being the spin quantization direction. The
four-vectors $n_{j}^{\mu}(\mathbf{p})$ are the Lorentz transformed
four-vectors of $\mathbf{n}_{j}$ and obey the sum rule 
\begin{equation}
n_{j}^{\mu}(\mathbf{p})n_{j}^{\nu}(\mathbf{p})=-\Delta^{\mu\nu},
\end{equation}
where $p^{\mu}=(E_{p}^{q},\mathbf{p})$. There are many sources to
the spin polarization of the quark and antiquark described by $P_{q}^{\mu}(x,\mathbf{p})$
and $P_{\overline{q}}^{\mu}(x,-\mathbf{p})$: vorticity, magnetic
field, shear tensor, etc.. All these sources are thought to be not
enough to account for the observed large spin alignment (a large deviation
of $\rho_{00}$ from 1/3) for the $\phi$ meson \citep{Yang:2017sdk,Xia:2020tyd,Gao:2021rom,Muller:2021hpe}.
It was proposed that such a large spin alignment may possibly arise
from the $\phi$ vector field, a strong force field in connection
with the current of pseudo-Goldstone bosons \citep{Manohar:1983md}.

In the on-shell approximation, the spin Boltzman equation for $f_{\lambda_{1}\lambda_{2}}^{V}$,
the MVSD for the unflavored vector meson, has been derived from Eq.
(\ref{eq:kinetic-g<-1}) for the quark coalescence and dissociation
process $V\leftrightarrow q\overline{q}$, giving the gain term and
loss term respectively in the right-hand-side (collision terms). The
collisions depend on the $\phi$ vector field's strength tensor through
$f_{rs}^{(\pm)}(x,\mathbf{p})$ for the quark and antiquark. In heavy
ion collisions, the phase space distribution functions are normally
much less than 1, $f_{\lambda_{1}\lambda_{2}}(x,\mathbf{p})\sim f_{rs}^{(+)}\sim f_{rs}^{(-)}\ll1$,
so the spin Boltzmann equation can be approximated as 
\begin{equation}
\frac{p}{E_{p}^{V}}\cdot\partial_{x}f_{\lambda_{1}\lambda_{2}}(x,\mathbf{p})\approx R_{\lambda_{1}\lambda_{2}}^{\mathrm{coal}}(\mathbf{p})-R^{\mathrm{diss}}(\mathbf{p})f_{\lambda_{1}\lambda_{2}}(x,\mathbf{p}),\label{eq:boltzmann-v}
\end{equation}
where $R_{\lambda_{1}\lambda_{2}}^{\mathrm{coal}}$ and $R^{\mathrm{diss}}$
denote the coalescence and dissociation rates for the vector meson,
respectively. Note that $R^{\mathrm{diss}}$ is independent of spin
indices $\lambda_{1}\lambda_{2}$, therefore the spin structure of
$f_{\lambda_{1}\lambda_{2}}$ is controlled by $R_{\lambda_{1}\lambda_{2}}^{\mathrm{coal}}$.

\subsection{Spin alignment of the $\phi$ meson}

From the solution to the spin Boltzmann equation (\ref{eq:boltzmann-v}),
the space-time and momentum averaged $\rho_{00}$ for the $\phi$
meson can be expressed in terms of the field strength tensor of the
$\phi$ field \citep{Sheng:2022wsy,Sheng:2022ffb}, 
\begin{eqnarray}
\left\langle \rho_{00}\right\rangle  & \equiv & \left\langle \rho_{00}(x,{\bf p})\right\rangle _{x,\mathbf{p}}\approx\frac{1}{3}-\frac{4}{3}\sum_{i=1,2,3}\left\langle I_{B,i}({\bf p})\right\rangle _{\mathbf{p}}\frac{1}{m_{\phi}^{4}}\left\langle g_{\phi}^{2}(\mathbf{B}_{i}^{\phi})^{2}/T_{\mathrm{h}}^{2}\right\rangle _{x}\nonumber \\
 &  & -\frac{4}{3}\sum_{i=1,2,3}\left\langle I_{E,i}({\bf p})\right\rangle _{\mathbf{p}}\frac{1}{m_{\phi}^{4}}\left\langle g_{\phi}^{2}(\mathbf{E}_{i}^{\phi})^{2}/T_{\mathrm{h}}^{2}\right\rangle _{x},\label{eq:rho00-xp}
\end{eqnarray}
where $m_{\phi}$ is the $\phi$ meson's mass, $T_{\text{h}}$ is
the local temperature at the hadronization time, $I_{B,i}({\bf p})$
and $I_{E,i}({\bf p})$ are momentum functions given in Ref. \citep{Sheng:2022ffb},
${\bf E}_{\phi}$ and ${\bf B}_{\phi}$ are electric and magnetic
parts of the $\phi$ field in the lab frame as functions of spacetime,
$\left\langle O(x)\right\rangle _{x}$ denotes the space-time average,
and $\left\langle O(\mathbf{p})\right\rangle _{\mathbf{p}}$ denotes
the momentum average defined as 
\begin{equation}
\left\langle O(\mathbf{p})\right\rangle _{\mathbf{p}}=\frac{\int d^{3}\mathbf{p}\left(E_{p}^{\phi}\right)^{-1}O(\mathbf{p})f_{\phi}(\mathbf{p})}{\int d^{3}\mathbf{p}\left(E_{p}^{\phi}\right)^{-1}f_{\phi}(\mathbf{p})}.\label{eq:momentum-average}
\end{equation}
Here $E_{p}$ is the $\phi$ meson's energy, $f_{\phi}(\mathbf{p})$
is its momentum distribution which may contain information about collective
flows such as $v_{1}$ and $v_{2}$, etc., and $d^{3}\mathbf{p}\left(E_{p}^{\phi}\right)^{-1}=p_{T}dp_{T}d\varphi dY$
where $p_{T}$, $\varphi$ and $Y$ are the transverse momentum, azimuthal
angle and rapidity respectively. In Eq. (\ref{eq:rho00-xp}) $\left\langle (\mathbf{B}_{i}^{\phi})^{2}\right\rangle _{x}$
and $\left\langle (\mathbf{E}_{i}^{\phi})^{2}\right\rangle _{x}$
reflect the average fluctuations of $\phi$ fields which can be regarded
as parameters to be determined by fitting the data.

If we want to obtain the $p_{T}$ spectra of $\left\langle \rho_{00}\right\rangle $,
we can integrate over $\varphi$ and $Y$. If we want to obtain the
$Y$ spectra of $\left\langle \rho_{00}\right\rangle $, we can integrate
over $p_{T}$ and $\varphi$. The theoretical results for $\left\langle \rho_{00}\right\rangle $
as functions of transverse momenta, collision energies and centralities
are presented in Ref. \citep{Sheng:2022wsy}, which are in a good
agreement with recent STAR data \citep{STAR:2022fan}. The rapidity
dependence of $\left\langle \rho_{00}^{y}\right\rangle $ (spin quantization
in the $y$ direction) using fluctuation parameters that are extracted
from STAR data on momentum-integrated $\ensuremath{\rho_{00}^{y}}$
\citep{STAR:2022fan} was predicted, the results show that $\ensuremath{\rho_{00}^{y}}$
has a negative deviation from $1/3$ at mid-rapidity $Y=0$ and a
positive deviation at slightly forward rapidity $Y=1$ \citep{Sheng:2023urn}.
The trend agrees with the preliminary data of STAR.



\section{Global spin polarization and alignment: overview on experimental
results}

\label{sec:polarization_overview} 



\subsection{Hyperon global polarization}

When two nuclei collide with finite impact parameter, i.e., non head-on
collisions, the system carries a large orbital angular momentum ($L\sim10^{5}\!-\!10^{7}\hbar$
at RHIC top energy or LHC energies), which is partially kept by the
created medium. One cannot directly detect such a rotation with a
few femtometer size and $\sim\!10$ fm/$c$ time scale but instead
one can measure particle polarization. Particles produced in the collisions
are globally polarized on average along the direction of the orbital
angular momentum via spin-orbit coupling~\citep{Liang:2004ph,Voloshin:2004ha,Becattini:2007sr},
referred to as global polarization. In a non-relativistic limit, the
polarization of particles ${\bm{P}}$ can be related to the vorticity
$\omega$ assuming a local thermal equilibrium: 
\begin{eqnarray}
{\bm{P}}=\frac{(S+1)({\bm{\omega}}+\mu_{B}{\bm{B}}/S)}{3T},\label{eq:Pol}
\end{eqnarray}
where $S$ is spin quantum number and $\mu_{B}$ is the magnetic moment
of the particle, $T$ is the temperature, and ${\bm{B}}$ is the magnetic
field.

The natural way to measure such particle polarization is to utilize
hyperon weak decays. Because of parity-violation in the weak decay,
the momentum direction of the daughter product in the hyperon rest
frame is correlated with the hyperon polarization: 
\begin{equation}
\frac{dN}{d\Omega^{\ast}}=\frac{1}{4\pi}(1+\alpha_{H}\bm{P}_{H}^{\ast}\cdot\hat{\bm{p}}_{B}^{\ast}),
\end{equation}
where $\alpha_{H}$ is the decay parameter of hyperons, $\bm{P}_{H}$
is the hyperon polarization, $\hat{\bm{p}}_{B}$ is the direction
of the daughter baryon's momentum, and the asterisk denotes the rest
frame of the parent hyperon. In case for the global polarization,
one needs to calculate the projection of the polarization vector into
the angular momentum direction of the system, which is perpendicular
to the reaction plane~\citep{STAR:2007ccu}. 
\begin{equation}
P_{H}=\frac{8}{\pi\alpha_{H}A_{0}}\frac{\langle\sin(\Psi_{1}-\phi_{B}^{\ast})\rangle}{{\rm Res(\Psi_{1})}},
\end{equation}
where $\phi_{B}^{\ast}$ is the azimuthal angle of the daughter baryon
in the hyperon's rest frame and $\Psi_{1}$ is the first-order event
plane being an experimental proxy for azimuthal angle of the reaction
plane. The ${\rm Res(\Psi_{1})}$ represents the experimental resolution
of the $\Psi_{1}$ angle and $A_{0}$ is an acceptance correction
factor usually close to be unity. Note that the $\Psi_{1}$ angle
is experimentally determined by measuring spectator deflection using
forward/backward detectors as the spectators are known to deflect
outward in high-energy nuclear collisions~\citep{Voloshin:2016ppr}.

The first attempt to measure the global polarization was made using
Au+Au collisions at $\sqrt{s_{NN}}=200$ GeV by STAR experiment in
2007~\citep{STAR:2007ccu}, where the results reported were consistent
with zero having large uncertainties, giving un upper limit of $|P_{H}|<2\%$.
Ten years later the positive signal of the global polarization, on
the order of a few percent implyng its energy dependence, was first
observed in $\Lambda$ hyperons in lower collision energies ($\sqrt{s_{NN}}=7.7\text{\textendash}39$
GeV) from the beam energy scan (BES-I) program at RHIC by STAR~\citep{STAR:2017ckg}.
Higher statistics data at 200 GeV~\citep{STAR:2018gyt} confirmed
the positive signal of the order of a few tenth of a percentage as
well as the energy dependence of the global polarization, allowing
us to study the polarization more differentially as discussed later.
The results were further improved with recent data from the second
phase of BES (BES-II)~\citep{STAR:2021beb,STAR:2023nvo} and HADES
experiment~\citep{HADES:2022enx} which provide more precise results.

\begin{figure}[ht]
\centering{}\includegraphics[width=0.6\linewidth]{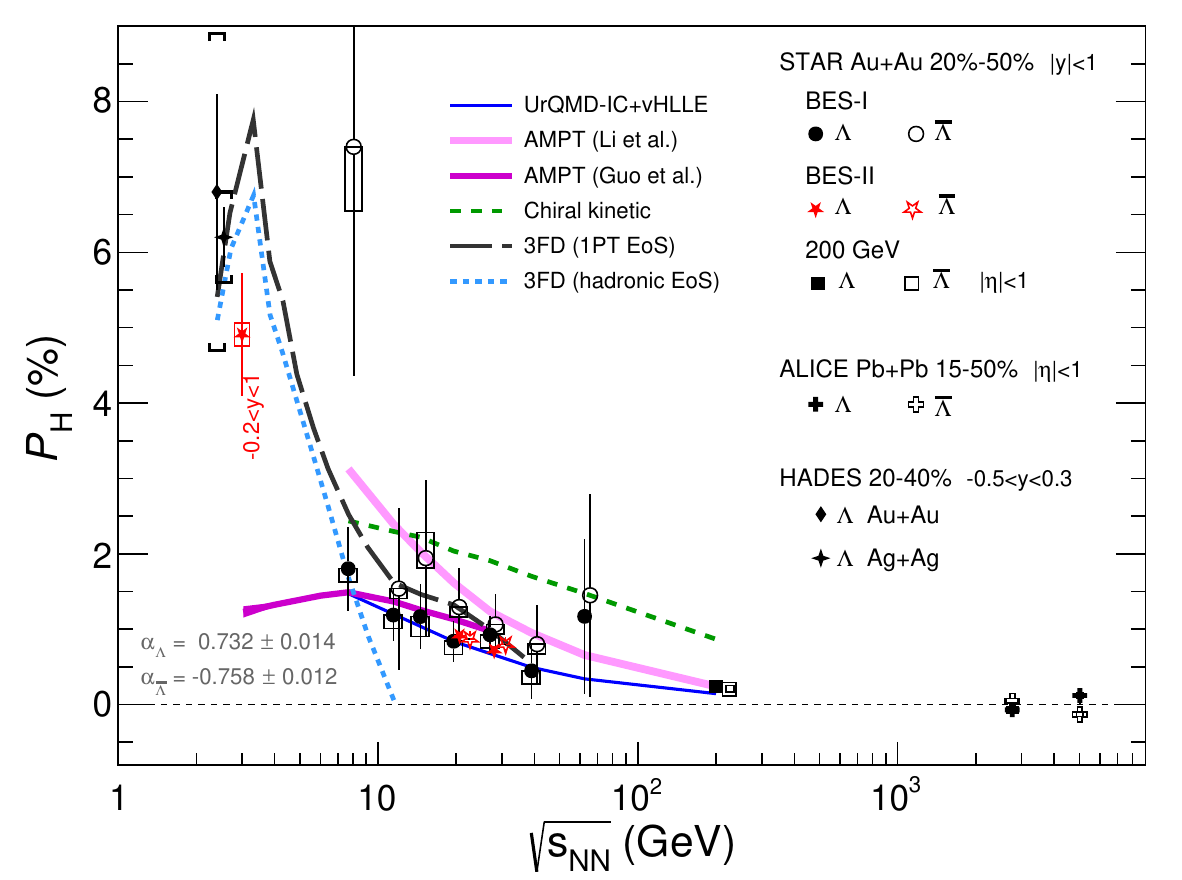}
\caption{Collision energy dependence of $\mbox{\ensuremath{\Lambda}}$ and
$\bar{\Lambda}$ global polarization from STAR~\citep{STAR:2017ckg,STAR:2018gyt,STAR:2021beb,STAR:2023nvo},
ALICE~\citep{ALICE:2019onw}, and HADES~\citep{HADES:2022enx} experiments.
Theoretical calculations such as viscous hydrodynamics~\citep{Karpenko:2016jyx},
a multiphase transport (AMPT) models~\citep{Li:2017slc,Guo:2021udq},
a three-fluid dynamics with two different equation-of-state~\citep{Ivanov:2020udj},
and chiral kinetic approach~\citep{Sun:2017xhx} are shown for comparison.}
\label{fig:PHvsRootS} 
\end{figure}

Figure~\ref{fig:PHvsRootS} shows a compilation of published experimental
results on $\Lambda$ and $\bar{\Lambda}$ global polarization vs.
collision energy. The results show a strong energy dependence, i.e.,
it increases with decreasing collision energy, which is described
well by various theoretical calculations~\citep{Karpenko:2016jyx,Li:2017slc,Guo:2021udq,Sun:2017xhx}.
Most of the models are based on the local vorticity of the fluid integrated
over freeze-out hypersurface as in Eq. (\ref{eq:SpinPolFirstOrder})
obtained assuming the local thermal equilibrium of the spin degrees
of freedom~\citep{Becattini:2007sr}. 
The total angular momentum of the system increases in higher energies~\citep{Jiang:2016woz}
but what is measured is just the polarization in the central rapidity
region where the vorticity field becomes smaller at higher energies
because of less baryon stopping and approximately longitudinal boost
invariance~\citep{Becattini:2015ska,Karpenko:2016jyx,Ivanov:2019ern}.
The dilution effect of the vorticity in a longer lifetime of the system
at higher energies would also contribute to the observed energy dependence~\citep{Karpenko:2016jyx}.
Following Eq.~\eqref{eq:Pol}, the fluid vorticity can be estimated
and is found to be the fastest vorticity ever observed~\citep{STAR:2017ckg},
$\omega\sim(9\pm1)\times10^{21}~\mathrm{s}^{-1}$.

\begin{figure}[ht]

\centering{}\includegraphics[width=0.6\linewidth]{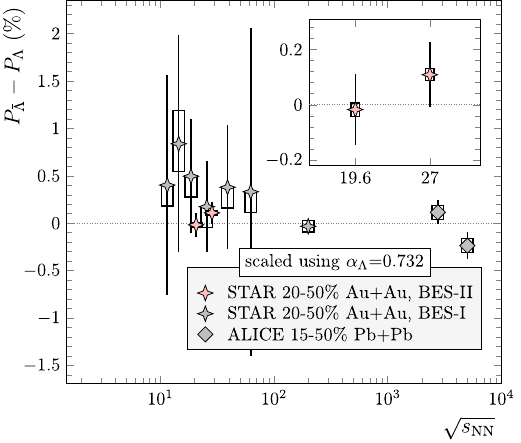}
\caption{Global polarization difference between $\mbox{\ensuremath{\Lambda}}$
and $\bar{\Lambda}$ hyperons as a function of the collision energy.
This figure is taken from Ref.~\citep{STAR:2023nvo}.}
\label{fig:deltaPH} 
\end{figure}

In the initial state of the collisions, a strong magnetic field would
be created by electric charges of protons that move to the opposite
direction in the speed of light. The magnitude of the field is expected
to be of the order of\citep{Kharzeev:2007jp,Skokov:2009qp,Voronyuk:2011jd,Ou:2011fm,Deng:2012pc,Tuchin:2013apa}
$B\sim10^{13}\!-\!10^{15}$~T, and the direction of the field coincides
with the initial angular momentum. 
Therefore the particles can also be polarized by the magnetic field
as indicated in Eq.~\eqref{eq:Pol}. Because the sign of the magnetic
moment is opposite for particles and antiparticles, one would expect
the difference in the global polarization between particles and anti-particles
if the effect is significant. Figure~\ref{fig:deltaPH} shows the
difference of the global polarization, $P_{\bar{\Lambda}}-P_{\Lambda}$,
as a function of the collision energy. There is no significant difference
in the $\mbox{\ensuremath{\Lambda}}$ and $\bar{\Lambda}$ global
polarization, which can be understandable because the lifetime of
the initial magnetic field, which depends on the electric conductivity
of the medium~\citep{McLerran:2013hla,Tuchin:2013apa,Zakharov:2014dia},
is expected to be very short ($\lesssim0.5$ fm/$c$). One can still
estimate the upper limit of the magnetic field based on Eq.~\eqref{eq:Pol}
as $|B|=T|P_{\bar{\Lambda}}-P_{\Lambda}|/(2|\mu_{\Lambda}|)$ ~\citep{Becattini:2016gvu,Muller:2018ibh},
where $\mu_{\Lambda}=-\mu_{\bar{\Lambda}}=-0.614\mu_{N}$ is the magnetic
moment of $\Lambda$ with $\mu_{N}$ being the nuclear magneton, and
$T$ is the temperature when $\mbox{\ensuremath{\Lambda}}$ and $\bar{\Lambda}$
are emitted. Assuming the temperature $T=150$~MeV, one obtain the
upper limit of the late-stage magnetic field to be $|B|<10^{12-13}$~T
~\citep{Muller:2018ibh,STAR:2023nvo}.


\begin{figure}[ht]

\centering{}\includegraphics[width=0.6\linewidth]{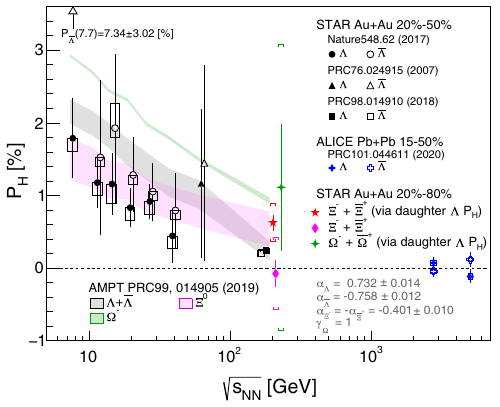}
\caption{Global polarization of $\Xi$ and $\Omega$ hyperons as well as that
of \lam hyperons as a function of the collision energy. This figure
is taken from Ref.~\citep{STAR:2020xbm}.}
\label{fig:PHxi} 
\end{figure}

The global vorticity picture has been confirmed by the measurement
of other hyperons such as $\Xi$ and $\Omega$. It would be of particular
interest to study spin and/or magnetic moment dependence of the polarization
with $\Xi$ and $\Omega$. The $\Xi$ hyperon has two-step decay:
$\Xi\rightarrow\Lambda\pi$ and $\Lambda\rightarrow p\pi$. In a similar
way to $\Lambda$'s case, one can measure $\Xi$ polarization by analyzing
its daughter $\Lambda$ distribution. Another independent way is to
measure the polarization of the daughter $\Lambda$ through the granddaughter
proton's distribution and to convert it to the parent $\Xi$ polarization
by utilizing the following relation~\citep{Lee:1957qs,Becattini:2016gvu}:
\begin{equation}
{\bm{P}}_{{\rm D}}^{\ast}=C_{{\rm PD}}{\bm{P}}_{{\rm P}}^{\ast},\label{eq:Pol_transfer}
\end{equation}
where $C_{{\rm PD}}$ is the polarization transfer coefficient in
the decay from parent particle P to daughter particle D and the asterisk
denotes the parent rest frame. For $\Xi^{-}$ decay, the transfer
coefficient is known as $C_{\Xi\Lambda}=0.944$. For the decay of
$\Omega^{-}\rightarrow\Lambda K^{-}$, the polarization transfer $C_{\Omega\Lambda}$
depends on the unmeasured decay parameter $\gamma_{\Omega}$ which
is expected to be $\gamma_{\Omega}\approx\pm1$: 
${\bm{P}}_{{\rm \Lambda}}^{\ast}=C_{{\rm \Omega\Lambda}}{\bm{P}}_{{\rm \Omega}}^{\ast}=\frac{1}{5}(1+4\gamma_{\Omega}){\bm{P}}_{\Omega}^{\ast}$.
The ambiguity on the sign of $\gamma_{\Omega}$ can be elucidated
by measuring global polarization of $\Omega$ hyperons based on the
global vorticity picture. Figure~\ref{fig:PHxi} presents $\Xi$
and $\Omega$ global polarization at $\sqrt{s_{NN}}=200$ GeV~\citep{STAR:2020xbm},
showing a hint of hierarchy: $P_{\Omega}>P_{\Xi}>P_{\Lambda}$. Such
a relation could be understood by the spin dependence of the polarization
as indicated in Eq.~(\ref{eq:Pol}) as well as the effect of feed-down
contributions~\citep{Li:2021zwq}, although the current uncertainties
are too large to show the particle species dependence. Note that the
data presented in this section contain contributions from the feed-down
and model studies show that the measured polarization is smeared by
$\sim$15\% for $\Lambda$ ~\citep{Becattini:2016gvu,Karpenko:2016jyx,Li:2017slc,Xia:2019fjf}
and is enhanced by $\sim$25\% for $\Xi$~\citep{Li:2021zwq}.


Since the orbital angular momentum carried by the medium depends on
the impact parameter~\citep{Becattini:2015ska,Jiang:2016woz}, the
observed global polarization is expected to depend on the impact parameter
as well. Such a trend was confirmed in the study of collision centrality
dependence. The global polarization is found to increase with going
from central to peripheral collisions for $\Lambda$~\citep{STAR:2018gyt,STAR:2021beb,STAR:2023nvo}
and $\Xi$~\citep{STAR:2020xbm} hyperons as expected 
from model calculations~\citep{Xie:2017upb,Xie:2019jun,Fu:2020oxj,Ryu:2021lnx}.

\subsection{Local polarization}

The global polarization refers to the polarization along the initial
angular momentum direction averaged over all particles and the phase
space, experimentally at mid-rapidity covered by the detector acceptance.
The magnitude of the same polarization component could depend on kinematics
such as momentum, rapidity, and azimuthal angle (``local polarization")
because of the complex structure of the local vorticity in a dynamically
expanding system. Recent high statistics data allow us to study the
polarization differentially as in Refs.~\citep{STAR:2018gyt,STAR:2021beb,STAR:2023nvo,HADES:2022enx},
where no significant dependence on the transverse momentum and rapidity
was observed. It is worth to mention that theoretical models~\citep{Deng:2016gyh,Wei:2018zfb,Wu:2019eyi,Xie:2019jun,Liang:2019pst}
predict the rapidity dependence differently at forward/backward regions
and the current data do not show such significant dependence with
large uncertainty, which should be explored in the future studies.

\begin{figure}[htb]

\centering{}\includegraphics[width=0.4\linewidth]{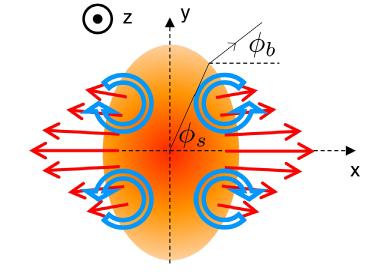}
\caption{A sketch of vorticities along the beam direction (open arrows) induced
by anisotropic flow (depicted by solid arrows) in the transverse plane
of a heavy-ion collision. This cartoon is taken from Ref.~\citep{STAR:2019erd}.}
\label{fig:wz} 
\end{figure}

Various complex vortical structures have been predicted to appear
in heavy-ion collisions due to the collective expansion of the system~\citep{Pang:2016igs,Becattini:2017gcx,Voloshin:2017kqp,Xia:2018tes}
and jet-medium interaction~\citep{Betz:2007kg,Tachibana:2012sa,Serenone:2021zef}.
Refs.~\citep{Becattini:2017gcx,Voloshin:2017kqp} suggest that the
vorticity, consequently particle polarization, can be induced by anisotropic
flow where the rotational axis is along the beam direction as shown
in Fig.~\ref{fig:wz}. The STAR Collaboration observed $\Lambda$
($\bar{\Lambda}$) polarization along the beam direction $P_{z}$
as expected~\citep{STAR:2019erd}, and later the ALICE Collaboration
confirmed it at the LHC energy~\citep{ALICE:2021pzu}.

\begin{figure}[ht]

\centering{}\includegraphics[width=0.49\linewidth]{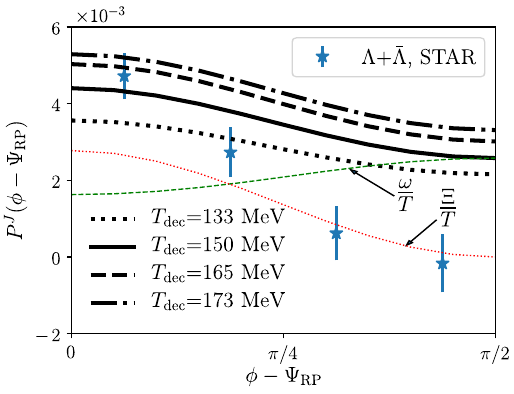}
\hspace{0.5pt} 
\includegraphics[width=0.49\linewidth]{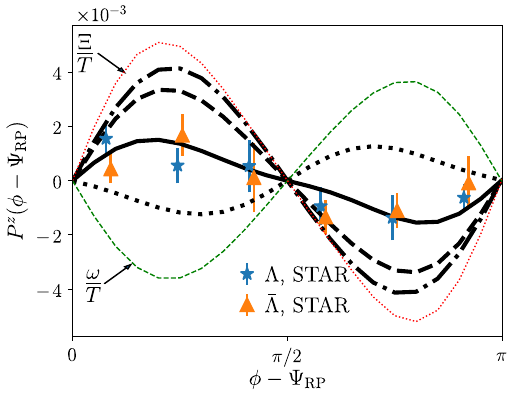} \caption{Azimuthal angle dependence of $\Lambda$ ($\bar{\Lambda}$) polarization
(left) along the initial angular momentum direction (STAR preliminary
result)~\citep{Niida:2018hfw} and (right) along the beam direction~\citep{STAR:2019erd}.
Also shown the contributions from the kinematic vorticity $\omega$
and kinematic shear $\Xi$ as well as their sum (thick lines) in the
hydrodynamic model under the hypothesis of isothermal hadronization~\citep{Becattini:2021iol}.
Taken from Ref.~\citep{Becattini:2021iol}.}
\label{fig:AzPol} 
\end{figure}

Figure~\ref{fig:AzPol} from ref.~\citep{Becattini:2021iol} shows
the azimuthal angle dependence of $\Lambda$ ($\bar{\Lambda}$) polarization
along the initial angular momentum (left) and the beam direction (right).
The data show cosine or sine patterns of the polarization, however
as indicated by thin lines the contribution from vorticity alone (marked
as $\omega/T$) cannot explain the phase or sign of the ``local"
polarization despite its reasonable description of the ``global"
polarization as in Fig.~\ref{fig:PHvsRootS}. This situation has
been called ``spin sign puzzle" in heavy-ion collisions. Recent
theoretical studies~\citep{Becattini:2021suc,Liu:2021uhn} show that
the thermal shear contribution to the polarization play an important
role and is necessary to understand the experimental data as shown
in Fig.~\ref{fig:AzPol}.

However, because of large cancellation of the thermal and shear contributions,
the sign of the polarization depends on the detailed implementation
of the shear contribution, which is still under intense discussion~\citep{Fu:2020oxj,Yi:2021ryh,Florkowski:2021xvy,Sun:2021nsg,Alzhrani:2022dpi},
and is found to be sensitive to initial conditions, shear viscosity,
and freeze-out temperature. It is worth to mention that a simple hydrodynamics-inspired
blast-wave model can also explain the sign and magnitude of the polarization
along the beam direction using the freeze-out parameters constrained
by other observables such as particle spectra and elliptic flow~\citep{Voloshin:2017kqp,STAR:2019erd}.

\begin{figure}[ht]
\includegraphics[scale=1.6]{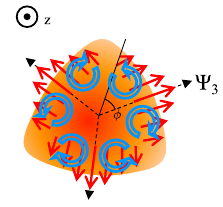}$\qquad\qquad$\includegraphics{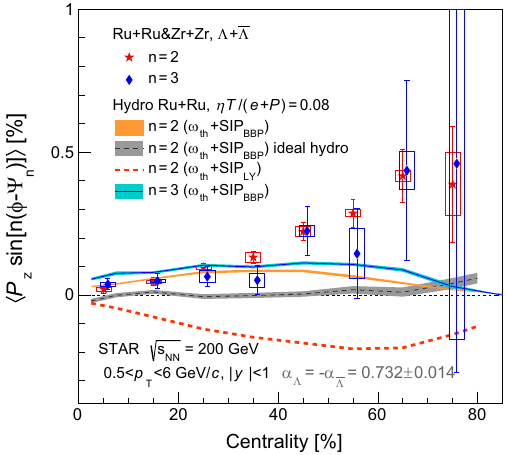}\caption{(Left)A sketch of triangular-flow-driven vorticities along the beam
direction (open arrows) with the triangular-shaped initial condition
due to event-by-event density fluctuations. (Right)Fourier sine coefficients
of $\Lambda$+$\bar{\Lambda}$ polarization along the beam direction
in Ru+Ru and Zr+Zr collisions at $\sqrt{s_{NN}}=200$ GeV~\citep{STAR:2023eck},
comparing to viscous hydrodynamic model calculations with two different
implementations of shear-induced polarization (SIP)~\citep{Becattini:2021suc,Liu:2021uhn}.
These figures are taken from Ref.~\citep{STAR:2023eck}.}
\label{fig:PzIsobar} 
\end{figure}

Such an anisotropic-flow-driven polarization is expected to be induced
even by higher harmonic flow~\citep{Voloshin:2017kqp}. Recent measurement
by STAR~\citep{STAR:2023eck} indeed shows a triangular-flow-driven
polarization along the beam direction as shown in Fig.~\ref{fig:PzIsobar}
and the results can be qualitatively explained except peripheral collisions
by hydrodynamics model with one of the implementations of the shear-induced
polarization (SIP). The contribution from the SIP in the higher harmonic
flow could be different from the case for the elliptic-flow-driven
polarization, therefore the result could provide us additional information
to constrain the shear contribution. 

\subsection{Global spin alignment of vector mesons}

\begin{figure}[ht]
\centering 
\includegraphics[width=0.7\linewidth]{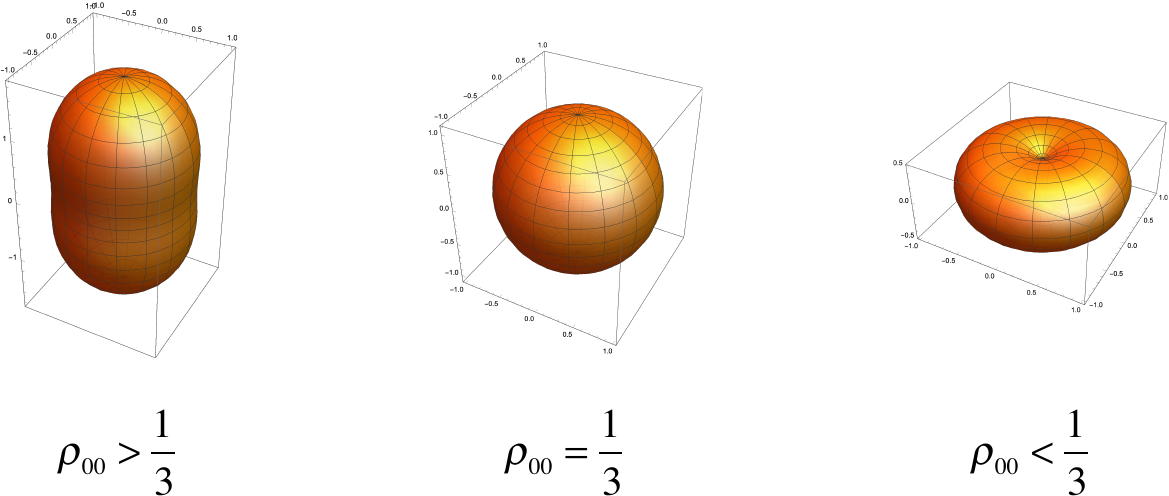} \caption{The daughter's distribution in vector meson's rest frame, corresponding
to three distinct $\rho_{00}$ values. The vertical axis serves as
the spin quantization axis.}
\label{fig-rho00_shapes} 
\end{figure}

In addition to hyperons, global quark polarization can also influence
vector mesons. Unlike $\Lambda(\bar{\Lambda})$ hyperons, which can
undergo weak decay with parity violation, the polarization of vector
mesons cannot be directly measured because they predominantly decay
through strong interactions, which conserve parity. However, vector
mesons like $\phi(1020)$ and $K^{*0}(892)$ can be characterized
by a 3 $\times$ 3 spin density matrix with a unit trace~\citep{Schilling:1969um}
as in Eq. (\ref{eq:sdm-vector-m-1}). The diagonal elements of this
matrix ($\rho_{-1,-1}$, $\rho_{0,0}$ and $\rho_{1,1}$) represent
the probabilities of finding a vector meson in spin states of $-1$,
0, and 1, respectively. If there is no spin alignment, $\rho_{00}$
is equal to 1/3, otherwise, $\rho_{00}$ deviates from this value.
One can extract $\rho_{00}$ from $dN/d(\mathrm{cos}\theta)$ measured
in experiments via Eq. (\ref{eq:scalar-decay}) 
where $\theta$ denotes the polar angle between the spin quantization
axis and the momentum direction of one daughter particle in the vector
meson's rest frame. 
Fig.~\ref{fig-rho00_shapes} provides a visual representation of
the daughter's distribution in the parent's rest frame, illustrating
three different scenarios for $\rho_{00}$.

\begin{figure}[ht]
\centering \includegraphics[clip,width=15cm]{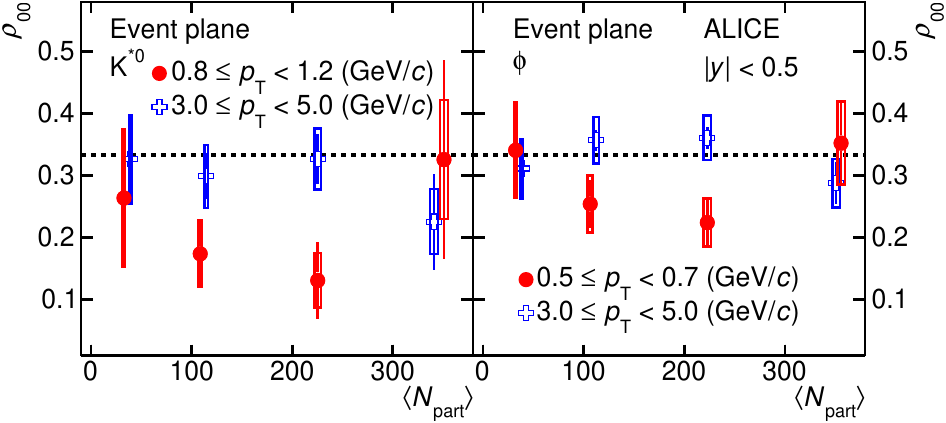} \caption{Measurements of $\rho_{00}$ as a function of $\langle N_{part}\rangle$
for $\phi$ and $K^{*0}$ mesons at low and high $p_{T}$ in Pb+Pb
collisions. The result is taken from ALICE publication~\citep{ALICE:2019aid}.}
\label{fig-rho00_alice} 
\end{figure}

In 2008, the STAR collaboration initiated the first effort~\citep{STAR:2008lcm}
to measure the global spin alignment of $\phi$ and $K^{*0}$ mesons
in Au+Au collisions at $\sqrt{s_{NN}}=200\mathrm{\,GeV}$. 
Subsequently, the ALICE collaboration conducted measurements~\citep{ALICE:2019aid}
(as shown in Fig.~\ref{fig-rho00_alice}) on the global spin alignment
of $\phi$ and $K^{*0}$ mesons in Pb+Pb collisions at $\sqrt{s_{NN}}=2.76\mathrm{,TeV}$.
For $p_{T}<2\mathrm{\,GeV}/c$, $\rho_{00}$ values were found to
be less than 1/3, with significance levels of 2$\sigma$ and 3$\sigma$
for $\phi$ and $K^{*0}$, respectively. Previous attempts, while
yielding limited significant results, provided some initial evidence
for the spin alignment along the event plane. Despite the challenges
faced in earlier studies, these results contribute to our evolving
comprehension of this intricate phenomenon.

\begin{figure}[ht]
\centering \includegraphics[clip,width=9.5cm]{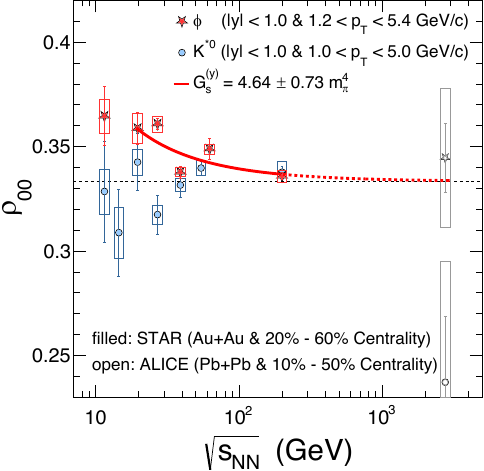}
\caption{The measured $\rho_{00}$ is plotted against the beam energy for $\phi$
and $K^{*0}$ vector mesons within specified windows of centrality,
transverse momentum ($p_{T}$), and rapidity ($y$). Open symbols
denote ALICE results \citep{ALICE:2019aid} for Pb+Pb collisions at
2.76 TeV. The red solid curve represents a fit to data across the
$\sqrt{s_{NN}}$ range of 19.6 to 200 GeV, based on a theoretical
calculation incorporating a $\phi$-meson field~\citep{Sheng:2019kmk}.
The red dashed line extends the solid curve with the fitted parameter
$G_{s}^{(y)}$. The black dashed line represents $\rho_{00}=1/3$.
This figure is sourced from the publication ~\citep{STAR:2022fan}
.}
\label{fig-rho00_phiKstar} 
\end{figure}

In a recent publication~\citep{STAR:2022fan} with data gathered
during the initial Beam Energy Scan program at RHIC, the STAR collaboration
disclosed a noteworthy global spin alignment for the $\phi$ meson.
Figure~\ref{fig-rho00_phiKstar} illustrates the measured global
spin alignment for both $\phi$ and $K^{*0}$ mesons in Au+Au collisions
varying with collision energies. The results reveal that while $\rho_{00}$
values for the $K^{*0}$ meson consistently hover around 1/3 within
the margin of error, the values of $\rho_{00}$ for the $\phi$ meson
exceed 1/3 significantly at collision energies lower than 62 GeV,
indicating tangible global spin alignment. There are many possible
contributions to the global spin alignment of the $\phi$ meson, such
as vorticities or electric and magnetic fields~\citep{Liang:2004xn,Becattini:2013vja,Yang:2017sdk,Xia:2020tyd}.
However, these contributions are insufficient to account for the observed
data~\citep{Sheng:2019kmk}. Furthermore, additional factors such
as local spin alignment~\citep{Xia:2020tyd,Gao:2021rom} and turbulent
color fields~\citep{Muller:2021hpe} negatively impacted $\rho_{00}$.

It was proposed that the strange and antistrange quarks can be polarized
by a kind of vector field, the $\phi$ field, induced by the current
of pseudosclar bosons~\citep{Sheng:2019kmk} when they form the $\phi$
meson. The local correlation or fluctuation of the $\phi$ field can
have significant contribution to the observed large deviation of the
$\phi$ meson's $\rho_{00}$ from 1/3~\citep{Sheng:2019kmk,Sheng:2022wsy,Sheng:2022ffb}.
It was also proposed that the local fluctuation in the glasma field~\citep{Kumar:2023ghs}
can also have a significant contribution to $\rho_{00}$ . 
The model with the $\phi$ field can qualitatively explain the collision
energy, transverse momentum and rapidity dependence of the observed
$\rho_{00}$~\citep{Sheng:2022wsy,Sheng:2023urn}. 
This observation underscores the pivotal role of the local correlation
or fluctuation in the strong force field in $\rho_{00}$ for the $\phi$
meson, in contrast to the mean value of the field that plays the role
in hyperon's polarization.

\begin{figure}
\centering \includegraphics[width=9.5cm]{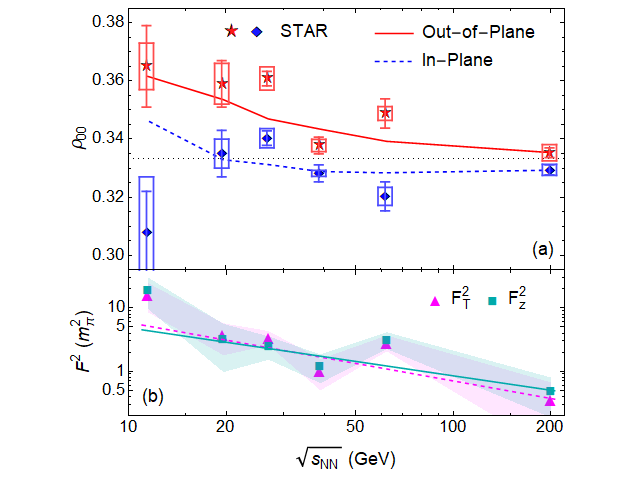} \caption{\label{fig:rho00-eng}(a) The STAR's data \citep{STAR:2022fan} on
$\phi$ meson's $\rho_{00}^{y}$ (out-of-plane, red stars) and $\rho_{00}^{x}$
(in-plane, blue diamonds) in 0-80\% Au+Au collisions as functions
of collision energies. The red-solid line (out-of-plane) and blue-dashed
line (in-plane) are calculated with values of $F_{T}^{2}$ and $F_{z}^{2}$
from fitted curves in (b). (b) Values of $F_{T}^{2}$ (magenta triangles)
and $F_{z}^{2}$ (cyan squares) with shaded error bands extracted
from the STAR's data on the $\phi$ meson's $\rho_{00}^{y}$ and $\rho_{00}^{x}$
in (a). The magenta-dashed line (cyan-solid line) is a fit to the
extracted $F_{T}^{2}$ ($F_{z}^{2}$) as a function of $\sqrt{s_{\mathrm{NN}}}$.
The definitions of transverse field squared $F_{T}^{2}$ and longitudinal
field squared $F_{z}^{2}$ are given in Ref. \citep{Sheng:2022wsy}.
The figure is taken from Ref. \citep{Sheng:2022wsy}.}
\end{figure}

\begin{figure}
\centering \includegraphics[width=9cm]{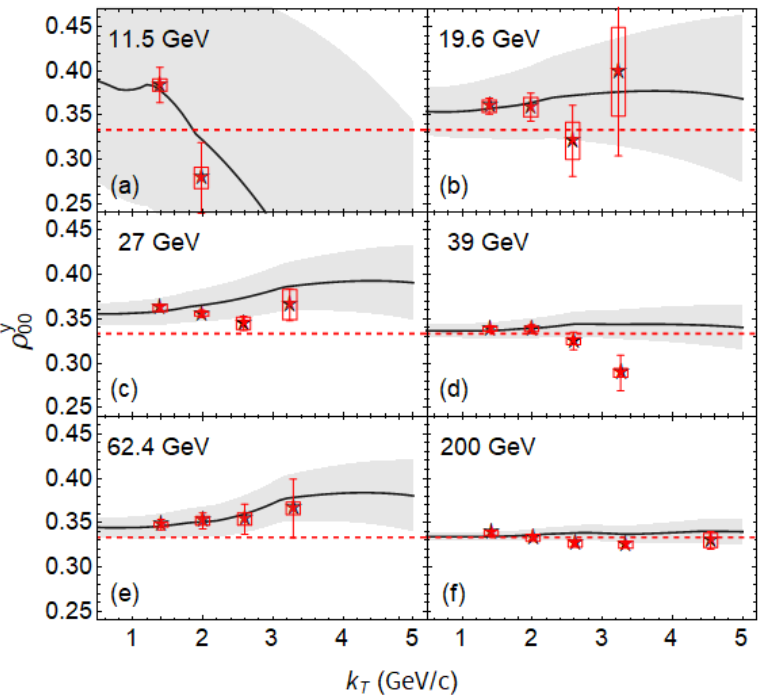} \caption{\label{fig:rho00-pt} Calculated $\rho_{00}^{y}$ for $\phi$ mesons
(solid lines) as functions of transverse momenta in 0-80\% Au+Au collisions
at different colliding energies as compared to STAR data \citep{STAR:2022fan}.
Shaded error bands are from the extracted parameters $F_{T}^{2}$
and $F_{z}^{2}$. The figure is taken from Ref. \citep{Sheng:2022wsy}. }
\end{figure}

The process of fitting this model involves adjusting $G_{s}^{(y)}$,
representing the quadratic form of field strengths multiplied by the
effective coupling constant ($g_{\phi}$). In its specific form~\citep{Sheng:2019kmk}
, $G_{s}^{(y)}$ is defined as 
\begin{eqnarray}
G_{s}^{(y)} & \equiv & g_{\phi}^{2}\left[3\langle B_{\phi,y}^{2}\rangle+\frac{\langle\textbf{p}^{2}\rangle_{\phi}}{m_{s}^{2}}\langle E_{\phi,y}^{2}\rangle-\frac{3}{2}\langle B_{\phi,x}^{2}+B_{\phi,z}^{2}\rangle\right.\nonumber \\
 &  & \left.-\frac{\langle\textbf{p}^{2}\rangle_{\phi}}{2m_{s}^{2}}\langle E_{\phi,x}^{2}+E_{\phi,z}^{2}\rangle\right],
\end{eqnarray}
where $E_{\phi,i}$ and $B_{\phi,i}$ denote the $i^{{\rm th}}$-component
of the analogous electric and magnetic parts of the $\phi$ field,
respectively. Additionally, $m_{s}$ represents the $s$-quark mass,
$\textbf{p}$ represents its momentum in the $\phi$ rest frame, and
$\langle\textbf{p}^{2}\rangle_{\phi}$ denotes the average $\textbf{p}^{2}$
inside the $\phi$ meson's wave function. When applying the model
from Ref. \citep{Sheng:2019kmk} to fit the data in Fig. \ref{fig-rho00_phiKstar},
the resulting free parameter in the fit, denoted as $G_{s}^{(y)}$,
is determined to be $(4.64\pm0.73)\,m_{\pi}^{4}$. The value of $G_{s}^{(y)}$
reflects the strength of local correlation or fluctuation of the $\phi$
field. The non-relativistic model in Ref.~\citep{Sheng:2019kmk}
has been promoted to a more rigorous relativistic transport model
~\citep{Sheng:2022wsy,Sheng:2022ffb,Sheng:2023urn}, which provides
a comprehensive description of STAR's data for the $\phi$ meson's
spin alignment as shown in Figs. \ref{fig:rho00-eng} and \ref{fig:rho00-pt}.

\begin{figure}[ht]
\centering \includegraphics[clip,width=8.5cm]{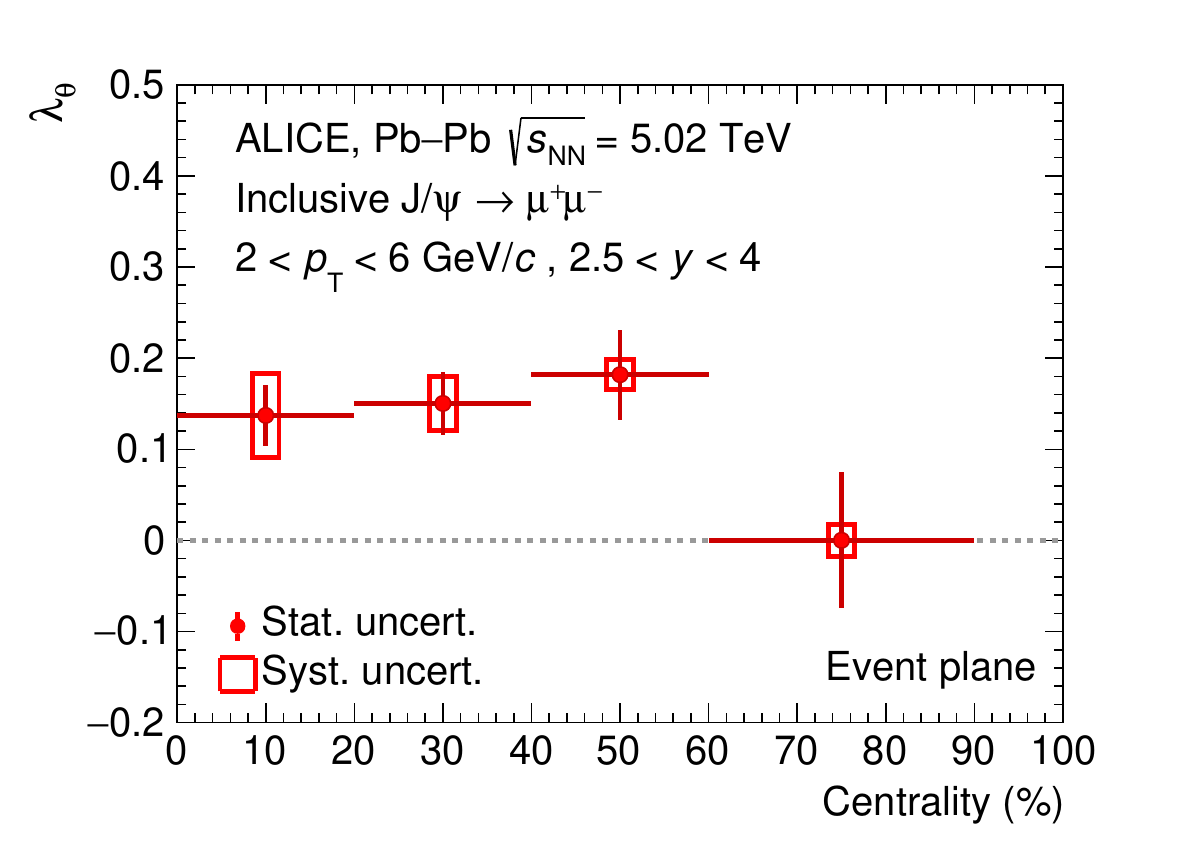}
\caption{Centrality dependence of $\lambda_{\theta}$ for $J/\psi$ as observed
by the ALICE collaboration. For $J/\psi$, the relationship $\lambda_{\theta}\propto(1-3\rho_{00})/(1+\rho_{00})$
holds. The peak value of $\lambda_{\theta}$ for $J/\psi$ (approximately
0.2) implies an associated $\rho_{00}$ value of approximately 0.25.
This figure is sourced from Ref.~\citep{ALICE:2022dyy}. }
\label{fig-rho00_lambdaTheta_vs_centrality_alice} 
\end{figure}

A key factor enabling the theoretical calculation of $\rho_{00}$
for the $\phi$-meson lies in the fact that the two quarks comprising
the $\phi$-meson originate from the same flavor family. This characteristic
also renders the measurement of $\rho_{00}$ for $J/\psi$ intriguing.
The $J/\psi$ particle is composed of $c$ and $\bar{c}$ quarks,
both belonging to the same flavor family. The ALICE collaboration
has conducted a study~\citep{ALICE:2022dyy} on the polarization
of $J/\psi$ particles produced in Pb+Pb collisions at $\sqrt{s_{NN}}=5.02\,\mathrm{TeV}$
in the dimuon channel. The obtained results indicate a deviation of
$-0.08$ from the expected $\rho_{00}=1/3$, as shown in Fig.~\ref{fig-rho00_lambdaTheta_vs_centrality_alice}.

Interestingly, according to the argument of fluctuating strong force
fields, one would anticipate $\rho_{00}$ to be larger than 1/3, which
contrasts with the findings reported by ALICE. However, interpreting
ALICE's results requires consideration of additional complexities.
The measurement was carried out at forward rapidity ($2.5<y<4$),
adding another layer of intricacy. Furthermore, the impact of color
screening and regeneration on the $\rho_{00}$ value of $J/\psi$
remains a topic that warrants thorough investigation.

\section{Summary}


In this review article we have overviewed the most significant advances
in the spin physics in heavy ion collisions up to 2023 both from a
theoretical and experimental standpoint. In the theory part, we put
emphasis on two topics: theoretical models for global and local polarization
in equilibrium and spin alignment of vector mesons. For the first
topic, we reviewed the quantum statistical field theory and the spin
hydrodynamics. In quantum statistical field theory, we discussed the
mean spin vector, the spin density matrix and Wigner functions. Then
we derived the freezeout formula for the spin polarization of fermions
which can be used to calculate spin observables in heavy ion collisions.
For the second topic, we indroduced the spin density matrix, the angular
distribution of decay daughters, and Green's functions for vector
mesons in the CTP formalism. Then we showed how to derive kinetic
equations and the spin Boltzmann equation for vector mesons from Dyson-Schwinger
equations in the CTP formalism. An overview was given on the application
of the spin Boltzmann equation with on-shell approximation to the
spin alignment of the $\phi$ meson. Finally we accounted for the
most recent experimental results in hyperons' global and local spin
polarization and vector mesons' spin alignment.



\section*{Acknowledgments}

M.B. is supported by the U.S. Department of Energy Grants No. DE-SC0023692.
T.N. is supported by JSPS KAKENHI Grant Number JP22K03648. Q.W. is
supported by the National Natural Science Foundation of China (NSFC)
under Grants No. 12135011 and by the Strategic Priority Research Program
of the Chinese Academy of Sciences (CAS) under Grant No. XDB34030102.
S.P. is supported in part by the National Key Research and Development
Program of China under Contract No. 2022YFA1605500; by the Chinese
Academy of Sciences (CAS) under Grants No. YSBR-088; and by NSFC under
Grants No. 12075235. 

\bibliographystyle{h-physrev}
\bibliography{ref-qgp6}

\end{document}